\documentclass[twocolumn]{emulateapj}
\usepackage{graphicx}
\usepackage{dcolumn}

\begin{document}

\title{A Spitzer Spectroscopic Survey of Low Ionization Nuclear Emission-line Regions: Characterization of the Central Source }

\author{R. P. Dudik\altaffilmark{1,2, 3, 4}, S. Satyapal\altaffilmark{1}, D. Marcu\altaffilmark{1}}

\altaffiltext{1}{George Mason University, Department of Physics \& Astronomy, MS 3F3, 4400 University Drive, Fairfax, VA 22030; rachel@physics.gmu.edu}

\altaffiltext{2}{NASA Goddard Space Flight Center, Observational Cosmology Lab, Greenbelt, MD}
\altaffiltext{3}{United States Naval Observatory, 3450 Massachusetts Avenue, NW, Washington, DC 20392-5420; rpdudik@usno.navy.mil}
\altaffiltext{4}{Research Support Instruments, 4325-B Forbes Boulevard, Lanham, MD, 20706 }

\begin{abstract}

We have conducted a comprehensive mid-IR spectroscopic investigation of 67 Low Ionization Nuclear Emission Line Regions (LINERs) using archival observations from the high resolution modules of the Infrared Spectrograph on board the {\it Spitzer Space Telescope}.  Using the [NeV] 14 and 24$\mu$m lines as active galactic nuclei (AGN) diagnostics, we detect active black holes in 39\% of the galaxies in our sample, many of which show no signs of activity in either the optical or X-ray bands.  In particular, a detailed comparison of multi-wavelength diagnostics shows that optical studies fail to detect AGN in galaxies with large far-IR luminosities.  These observations emphasize that the nuclear power source in a large percentage of LINERs is obscured in the optical.  Indeed, the majority of LINERs show mid-IR [NeV]14/[NeV]24$\mu$m flux ratios well below the theoretical low-density limit, suggesting that there is substantial extinction toward even the [NeV]-emitting region . Combining optical, X-ray, and mid-IR diagnostics, we find an AGN detection rate in LINERs of 74\%, higher than previously reported statistics of the fraction of LINERs hosting AGN.   The [NeV]24$\mu$m /[OIV]26$\mu$m mid-IR line flux ratio in “AGN-LINERs” is similar to that of standard AGN, suggesting that the spectral energy distribution (SED) of the intrinsic optical/UV continuum is similar in the two.  This result is in contrast to previous suggestions of a UV deficit in the intrinsic broadband continuum emission in AGN-LINERs.  Consistent with our finding of extinction to the [NeV]-emitting region, we propose that extinction may also be responsible for the observed optical/UV deficit seen in at least some AGN-LINERs. 
\end{abstract}


\section{Introduction}
Low Ionization Nuclear Emission Line Regions (LINERs), the dominant population of active galaxies in our local universe (e.g. Stauffer 1982; Ho, Filippenko and Sargent 1997), remain one of the most captivating and generally elusive subsets of galaxies.  Originally classified using ratios of oxygen emission lines with low ionization potentials, the principal mechanism responsible for their optical spectra has been the subject of debate for almost 30 years (e.g. Heckman 1980a and b, Veilleux and Osterbrock 1987, Ho, Fillipenko and Sargent 1997, Satyapal et al. 2004, Dudik et al. 2005, Gonzalez-Martin et al. 2006).  This is in part because the observed optical spectrum can be explained by a variety of different physical processes (including photoionization from hot stars, shocks, or an active galactic nucleus (AGN)) (Heckman 1980b, Baldwin, Phillips, and Terlevich 1981, Ferland \& Netzer 1983, Halpern \& Steiner 1983, Veilleux and Osterbrock 1987, Dopita \& Sutherland 1995, and Sugai \& Malkan 2000).   In fact, it is clear that the LINER class as a whole is not homogeneous, and only a fraction contains accreting black holes (Heckman 1980, Ho et al. 2001, Nagar et al. 2001, 2002, Dudik et al. 2005, Satyapal et al. 2005).  

While extensive optical spectroscopic studies have been carried out to determine the number of AGNs in LINERs (Heckman 1980a\&b, Ho, Filippenko, \& Sargent 1997a, Ho et al. 1997b; Kim et al. 1995, Veilleux et al. 1995), surprisingly few systematic, multi-wavelength comparisons on a representative set of LINERs have been conducted.  This is a serious shortcoming since we now know that the majority of LINERs are bright in the far-infrared (FIR) and have large FIR to B-band flux ratios, implying that they are heavily obscured (Carrillo et al. 1999, Satyapal et al. 2004, Dudik et al. 2005).  In such cases, the optical emission lines can be dominated by the contribution from star forming regions in the host galaxy.  Thus in order to get a true census of the number of LINERs containing AGN, the investigation should be carried out at wavelengths less sensitive to host galaxy contamination that are also less affected by gas and dust obscuration.  

Mid-IR spectroscopy and X-ray observations are generally considered the ideal tools for uncovering AGN in LINERs. The mid-IR contains a number of high ionization lines that are unambiguous AGN diagnostics.  High ionization lines, such as the [NeV] 14 and 24$\mu$m lines (ionization potential = 97eV)  are prominent in standard AGN and not readily produced in HII regions (Satyapal et al. 2008, Abel \& Satyapal 2008), the ionizing source in star forming regions (Genzel et al. 1998, Sturm et al. 2002, Satyapal et al. 2004).  Detection of either the [NeV] 14 or 24$\mu$m lines provides strong evidence for the presence of an AGN and has been used to detect active black holes in galaxies that show no signatures of them in their optical spectra (Satyapal et al. 2007, Satyapal et al. 2008, Abel \& Satyapal 2008).  Similarly, hard X-rays (2-10keV) are a powerful tool for finding obscured AGN since they are produced in the inner hottest regions of the source, are relatively insensitive to host galaxy contamination, and are not significantly affected by absorption.  Several preliminary investigations employing mid-IR or X-ray observations to search for AGN in relatively small samples of LINERs have been carried out so far (e.g., Ho et al. 2001;  Eracleous et al. 2001; Satyapal, Sambruna \& Dudik 2004, Dudik et al. 2005; Satyapal et al. 2005), but no large-scale study of a representative sample of LINERs that compares the multi-wavelength properties and detection rates has yet been undertaken.

In light of this deficiency, we conducted an exhaustive search for LINERs observed by the high resolution (HR) modules of the Infrared Spectrograph (IRS) on board the {\it Spitzer Space Telescope} in order to determine the fraction that show mid-IR signatures of AGNs and to compare this detection rate with that obtained using other detection methods.  This survey includes 67 LINERs and represents the largest mid-IR spectroscopic survey of LINERs to date.  In addition this is the only study that includes extensive observations of the [NeV] line in LINERs, the only robust AGN indicator in the mid-IR.  The principal goal of our study was to determine the fraction of LINERs showing definitive AGN signatures at mid-IR wavelengths and to compare this fraction with the AGN detection rate obtained using X-ray and optical surveys.

This Paper is structured as follows.  Section 2 summarizes the properties of the {\it Spitzer} sample.  Section 3 describes the data reduction and analysis for the mid-IR observations.  In Section 4 the mid-IR AGN detection rate is presented and in Section 5 this mid-IR detection rate is compared with other multi-band AGN indicators.  In Section 6 the mid-IR line ratios in LINERs are examined and these ratios are compared with others from a sample of standard AGNs.  Finally, a summary of the major conclusions is given in Section 7.

\section{The Sample}

The majority of LINERs are luminous in the far-IR (FIR) and span a wide range of far-IR to optical B-band ratios (IR-brightness ratios, Carrillo et al. 1999, Satyapal et al. 2004, Dudik et al. 2005, Satyapal 2005).  Indeed, using the sample from Carrillo et al. (1999), which represents the largest compilation of LINERs to date, 70\% of LINERs have FIR luminosities in excess of 10$^{43}$ ergs s$^{-1}$ (or $\sim$ 10$^{9}$ L$\odot$.  Therefore, a statistically significant, representative sample of LINERs requires inclusion of both IR-bright and IR-faint LINERs.   The sample of LINERs presented here is derived from the Palomar survey of bright nearby galaxies (containing predominantly IR-faint LINERs, H97) and the Veilleux et al. (1995, hereafter V95) sample of luminous infrared galaxies (containing IR-bright LINERs). For consistency, we excluded all galaxies that did not adhere to all four of the Veilleux et al. (1987)\footnote[1]{The four criteria for LINER classification outlined by Veilleux et al. 1987 are:  [OI]$_{6300}$ $\geq$ 0.17H$\alpha$, [NII]$_{6583}$ $\geq$ 0.6H$\alpha$, [SII]$_{6716, 6731}$ $\geq$ 0.4H$\alpha$, [OIII]$_{5007}$ $<$ 3H$\beta$. See Veilleux et al. 1987 for details of the classification scheme } optical LINER classification criteria, since this scheme is the least sensitive to observational effects associated with optical spectroscopy such as reddening, line blending and abundance sensitivities.

\begin{figure*}[htbp]
\begin{center}
\begin{tabular}{cc}
  \includegraphics[width=0.45\textwidth]{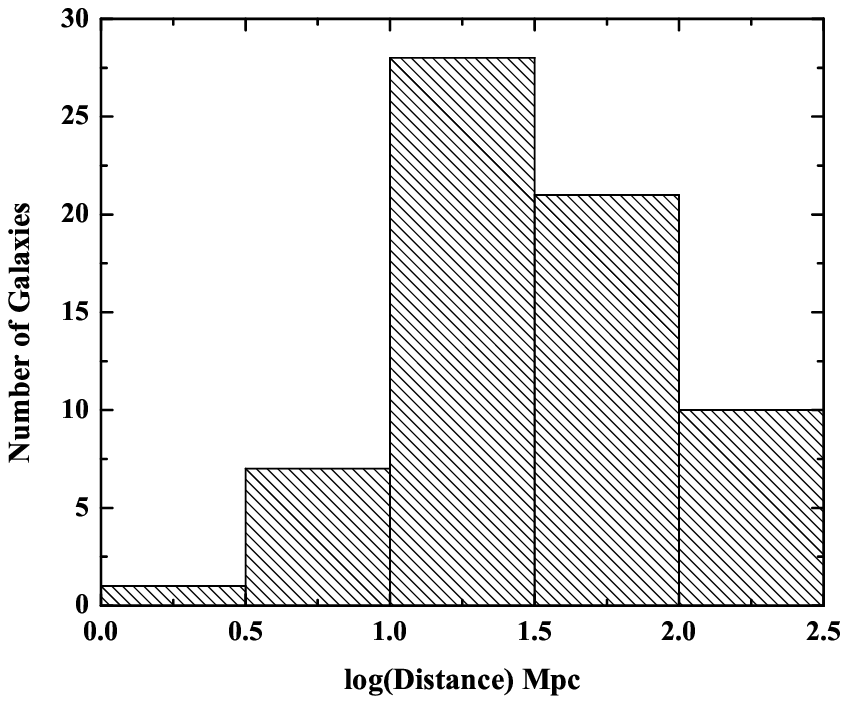} &
  \includegraphics[width=0.45\textwidth]{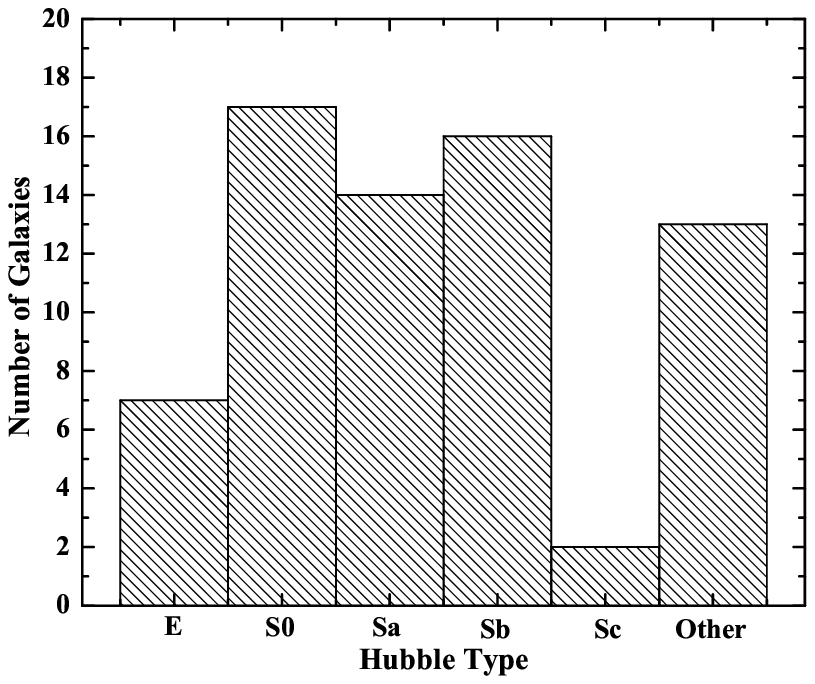} \\
 \includegraphics[width=0.45\textwidth]{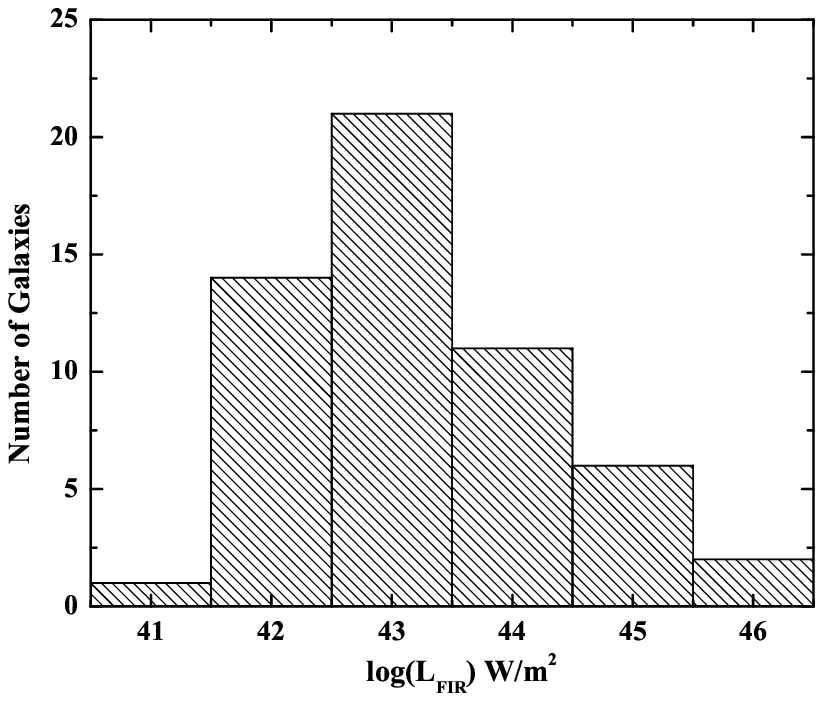} &
  \includegraphics[width=0.45\textwidth]{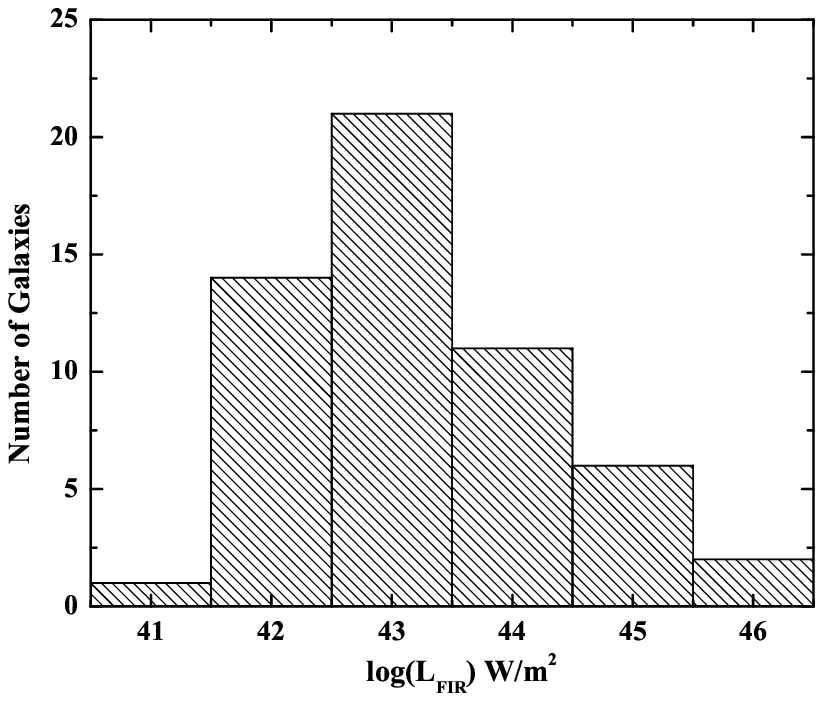} \\
 \end{tabular}
\end{center}
\caption[]{Characteristics of the {\it Spitzer} sample of LINERs.  Most galaxies are nearby and span a wide range of far-IR luminosities, IR-brightness ratios, and Hubble types.}
\end{figure*}

A search of the {\it Spitzer} archive for the subset of LINERs with high-resolution IRS (Houck et al. 2004) observations yielded 67 available galaxies.   The resulting sample of galaxies span a wide range of distances (2 to 276 Mpc; median = 28 Mpc), Hubble types, FIR luminosities (log (L$_{FIR}$) $\sim$ 41 to 45, median = 43, FIR luminosities correspond to the 40-500$\mu$m wavelength interval and were calculated using the IRAS 60 and 100 $\mu$m fluxes according to the prescription: $L_{FIR}$=1.26$\times$10$^{-14}$(2.58f$_{60}$+f$_{100}$) in W m$^{-2}$ (Sanders \& Mirabel 1996)), and IR-brightness ratios (L$_{FIR}$/L$_B$ $\sim$ 0.02 to 158, median = 1).  These basic properties of the sample are given in Figure 1 and Tables 1 and 2.  We also list in Tables 1 and 2 the program identification (PID) numbers corresponding to each observation.  We note that the majority of observations were obtained in programs targeting LINERs, and in couple cases nearby starburst and normal galaxies. Only 1 galaxy was observed as part of a program targeting known AGN (NGC5005, PID 86).  The sample is therefore not biased toward hosts with known AGN.  However, we emphasize that the objects presented here were selected based on the availability of high resolution IRS {\it Spitzer} observations.  The sample therefore should not be viewed as complete in any sense, or representative of the LINER class.  For example, a comparison of the FIR luminosities in our archival sample with that from the Carrillo et al. (1999) sample reveals that this archival sample is slightly deficient in the most FIR luminous LINERs. On the other hand, we have conservatively chosen our LINER sample to include only those sources that adhere to all four Veilleux et al. (1987) selection criteria.  Carrillo et al. (1999) includes LINERs selected using less stringent criteria.  Thus the two samples are not strictly comparable.

\begin{table*}
\fontsize{9pt}{10pt}\selectfont
\begin{center}
\caption{Properties of the H97 Sample}
\begin{tabular}{lccccccccc}
\hline
\multicolumn{1}{c}{Galaxy} & \multicolumn{1}{c}{PID}& Distance & Hubble & {\underline [OIII]} & {\underline [OI]} & {\underline [NII]} & {\underline [SII]} & log & {\underline L$_{FIR}$} \\

\multicolumn{1}{c}{Name} & \multicolumn{1}{c}{} & (Mpc) & Type & H$_{\beta}$
 & H$_{\alpha}$ & H$_{\alpha}$ & H$_{\alpha}$ & L$_{FIR}$ & L$_B$  \\

\multicolumn{1}{c}{(1)} & \multicolumn{1}{c}{(2)} & (3) & (4) & (5) & (6) & (7) & (8) & (9) & (10)\\ 

\hline
NGC1052 & 3674 & 18 & E4 & 2.01 & 0.71 & 1.20 & 1.72 & 42.25 & -0.59\\
NGC1055 & 3124 & 13 & SBb & $\cdots$ & 0.06 & 0.66 & 0.41 & 43.46 & 0.66\\
NGC1961 & 30323 & 53 & SABc & 1.18 & 0.21 & 1.96 & 1.07 & 44.21 & 0.28\\
NGC266 & 3674 & 62 & SBab & 2.38 & 0.28 & 2.36 & 1.27 & 43.50 & -0.11\\
NGC2681 &3124 & 13 & SAB0/a &1.73 & 0.19 & 2.36 & 0.75 & 42.91 & 0.13\\
NGC2787  & 3674 & 13 & SB0 & 1.41 & 0.55 & 1.61 & 1.41 & 41.83 & -0.64\\
NGC2841  & 159 & 12 & SAb & 1.86 & 0.17 & 1.83 & 1.14 & 42.89 & -0.32\\
NGC3166  & 20140 & 22 & SAB0/a  & 2.71 & 0.27 & 2.57 & 1.32 & 43.34 & 0.17\\
NGC3169  & 3674 & 20 & Saa & 2.88 & 0.28 & 2.07 & 1.45 & 43.40 & 0.31\\
NGC3190  & 159 & 22 & SAa  & 2.06 & 0.34 & 3.47 & 1.82 & 43.14 & 0.13\\
NGC3226  & 3674 & 23 & E2 pec & 2.22 & 0.59 & 1.45 & 1.36 & 41.95 & -0.69\\
NGC3368  & 3124 & 8 & SABab  & 1.83 & 0.18 & 1.11 & 0.97 & 42.76 & -0.02\\
NGC3507  & 3237 & 20 & SBb  & 0.98 & 0.18 & 1.10 & 0.94 & $\cdots$ & $\cdots$\\
NGC3521  & 159 & 7 & SABbc  & 1.00 & 0.23 & 0.64 & 0.88 & 43.27 & 0.39\\
NGC3642  & 3237 & 28 & SAbc  & 1.32 & 0.18 & 0.71 & 1.04 & 42.98 & -0.20\\
NGC3884  & 3237 & 92 & SA0/a  & 2.24 & 0.58 & 2.07 & 1.12 & $\cdots$ & $\cdots$\\
NGC3998  & 3237 & 22 & SA0 & 1.98 & 0.53 & 1.08 & 0.99 & 42.24 & -0.71\\
NGC4036  & 3237 & 25 & SA0 & 1.85 & 0.46 & 2.25 & 1.81 & 42.43 & -0.64\\
NGC404  & 3237 & 2 & SA0 & 1.26 & 0.17 & 0.44 & 0.99 & 40.94 & -0.34\\
NGC4143  & 3674 & 17 & SAB0 & 2.15 & 0.71 & 1.75 & 1.39 & $\cdots$ & $\cdots$\\
NGC4203  & 3674 & 10 & SAB0 & 1.57 & 1.22 & 1.84 & 1.42 & 41.70 & -0.52\\
NGC4261  & 3674 & 35 & E2 & 2.44 & 0.49 & 2.60 & 1.29 & 42.45 & -0.98\\
NGC4278  & 3237 & 10 & E1 & 1.31 & 0.41 & 1.22 & 1.48 & 41.66 & -0.81\\
NGC4314  & 3674 & 10 & SBa  & 0.72 & 0.18 & 1.01 & 0.57 & 42.38 & -0.01\\
NGC4394  & 20140 & 17 & SBb  & 2.80 & 0.44 & 1.08 & 0.98 & 42.53 & -0.20\\
NGC4438  & 3237 & 17 & SA0/a  & 1.59 & 0.27 & 1.64 & 1.51 & 42.99 & -0.15\\
NGC4450  & 159 & 17 & SAab  & 2.12 & 0.67 & 1.94 & 2.15 & 42.73 & -0.31\\
NGC4457  & 3237 & 17 & SAB0/a  & 0.87 & 0.19 & 1.04 & 0.90 & 42.98 & 0.24\\
NGC4486  & 3237 & 17 & E0  & 1.89 & 0.30 & 2.12 & 1.30 & 41.77 & -1.77\\
NGC4548  & 3674 & 17 & SBb  & 1.12 & 0.23 & 1.38 & 1.06 & 42.89 & -0.13\\
NGC4594  & 3674 & 20 & SAa  & 1.57 & 0.18 & 2.19 & 1.07 & 43.31 & -0.82\\
NGC4596  & 3674 & 17 & SB0 & 1.00 & 0.27 & 1.48 & 1.16 & 42.03 & -0.71\\
NGC4736  & 159 & 4 & SAab  & 1.47 & 0.24 & 2.15 & 1.39 & 42.92 & 0.27\\
NGC474  & 20140 & 33 & SA0 & 2.41 & 0.21 & 0.47 & 1.01 & 41.87 & -1.25\\
NGC5005  & 86 & 21 & SABbc  & 2.27 & 0.65 & 4.94 & 3.31 & 43.92 & 0.46\\
NGC5195  & 159 & 9 & IA0 & 1.22 & 0.55 & 5.43 & 2.00 & 42.96 & 0.29\\
NGC5353  & 20140 & 38 & SA0 & 1.22 & 0.19 & 1.71 & 0.83 & 42.66 & -0.58\\
NGC5371  & 3237 & 38 & SABbc  & 2.07 & 0.28 & 2.36 & 1.48 & 43.84 & 0.27\\
NGC5850  & 20140 & 29 & SBb  & 2.38 & 0.22 & 1.69 & 1.43 & 42.89 & -0.35\\
NGC5982  & 20140 & 39 & E3 & 1.08 & 0.49 & 2.42 & 0.71 & 42.12 & -1.12\\
NGC5985  & 20140 & 39 & SABb  & 2.65 & 0.30 & 3.08 & 1.51 & 43.32 & -0.20\\
NGC6500  & 3237 & 40 & SAab  & 1.40 & 0.23 & 0.73 & 0.96 & 43.00 & -0.09\\
NGC7626  & 3674 & 46 & E  & 1.62 & 0.22 & 2.35 & 1.28 & 42.33 & -1.05\\ 
\hline
\end{tabular}
\end{center}
\tablecomments{Columns Explanation:
Col(1):Common Source Names; 
Col(2):  Program ID Number;
Col(3): Distance,  = taken from  H97 ; 
Col(4): Hubble Type;
Col(5):  [OIII] to H$_\beta$ ratio taken from H97;  
Col(6):  [OI] to H$_\alpha$ ratiotaken from H97; 
Col(7):  [NII] to H$_\alpha$ ratio taken from H97;
Col(8):  [SII] to H$_\alpha$ ratio taken from H97; 
Col(9):  Log of the far-IR luminosity in units of ergs s$^{-1}$, Far-infrared luminosities correspond to the 40-500$\mu$m wavelength interval and were calculated using the IRAS 60 and 100 $\mu$m fluxes according to the prescription: $L_{FIR}$=1.26$\times$10$^{-14}$(2.58f$_{60}$+f$_{100}$) in W m$^{-2}$ (Sanders \& Mirabel 1996);
Col(10):  IR-brightness Ratio (L$_{FIR}$/L$_B$) 
}

\end{table*}
\begin{table*}
\fontsize{9pt}{10pt}\selectfont
\begin{center}
\caption{Properties of the V95 Sample}
\begin{tabular}{lccccccccc}
\hline
\multicolumn{1}{c}{Galaxy} & \multicolumn{1}{c}{PID}& Distance & Hubble & {\underline [OIII]} & {\underline [OI]} & {\underline [NII]} & {\underline [SII]} & log & {\underline L$_{FIR}$} \\

\multicolumn{1}{c}{Name} & \multicolumn{1}{c}{} & (Mpc) & Type & H$_{\beta}$
 & H$_{\alpha}$ & H$_{\alpha}$ & H$_{\alpha}$ & L$_{FIR}$ & L$_B$  \\

\multicolumn{1}{c}{(1)} & \multicolumn{1}{c}{(2)} & (3) & (4) & (5) & (6) & (7) & (8) & (9) & (10)\\ 
\hline
UGC556 & 3237 & 62 & S & 0.63 & 0.14 & 0.59 & 0.49 & 44.16 & 1.47\\
IR01364-1042 & 20549 & 193 & n/a  & 1.62 & 0.35 & 0.95 & 0.59 & 45.13 & $\cdots$\\
NGC660 & 14 & 11 & SBa & 2.19 & 0.08 & 1.00 & 0.43 & 43.74 & 1.32\\
IIIZw35 & 3237 & 110 & n/a & 1.17 & 0.21 & 1.02 & 0.49 & 44.94 & $\cdots$\\
IR02438+2122 & 3237 & 93 & n/a  & 2.57 & 0.20 & 1.35 & 0.29 & 44.46 & $\cdots$\\
NGC1266 & 159 & 29 & SB0 & 1.20 & 0.35 & 3.72 & 1.62 & 43.82 & 1.74\\
ugc5101 & 105 & 157 & S & 2.29 & 0.11 & 1.32 & 0.35 & 45.27 & 1.95\\
NGC4666 & 3124 & 20 & SABc & 1.20 & 0.06 & 1.29 & 0.60 & 44.06 & 0.83\\
UGC8387 & 3237 & 93 & Im & 0.69 & 0.11 & 0.66 & 0.40 & 44.95 & 1.86\\
NGC5104 & 3237 & 74 & Sa & 1.10 & 0.09 & 0.85 & 0.48 & 44.41 & 1.37\\
NGC5218 & 3237 & 39 & SBb & 0.44 & 0.13 & 0.85 & 0.47 & 43.86 & 0.93\\
Mrk273 & 105 & 151 & Ring & 2.82 & 0.14 & 1.02 & 0.58 & 45.44 & 1.82\\
Mrk848 & 3605 & 161 & S0 & 1.15 & 0.09 & 0.76 & 0.45 & 45.11 & $\cdots$\\
NGC5953 & 59 & 26 & SAa & 1.82 & 0.06 & 0.78 & 0.40 & 43.72 & $\cdots$\\
IR15335-0513 & 3237 & 109 & n/a & 0.56 & 0.20 & 1.17 & 0.46 & 44.60 & $\cdots$\\
IR16164-0746 & 3237 & 109 & n/a & 1.02 & 0.25 & 1.00 & 0.44 & 44.85 & $\cdots$\\
NGC6240 & 105 & 98 & I0 & 1.17 & 0.37 & 1.23 & 1.10 & 45.09 & 1.50\\
NGC6286 & 3237 & 73 & Sb & 0.56 & 0.15 & 0.83 & 0.60 & 44.58 & 1.54\\
ESO593-IG008 & 30323 & 195 & Double & 0.65 & 0.10 & 2.82 & 0.59 & 45.17 & $\cdots$\\
NGC7130 & 30577 & 65 & Sa & 2.40 & 0.26 & 0.91 & 0.48 & 44.64 & 1.26\\
ZW453.062 & 3237 & 100 & n/a & 1.07 & 0.14 & 1.20 & 0.48 & 44.66 & 1.82\\
NGC7591 & 3237 & 66 & SBbc & 1.10 & 0.10 & 0.95 & 0.39 & 44.37 & 1.03\\
IR23365+3604 & 105 & 258 & Sba & 0.50 & 0.07 & 0.66 & 0.43 & 45.45 & 2.19\\
IR04259-0440 & 20577 & 62 & E-S0 & 1.66 & 0.14 & 0.72 & 0.51 & 43.91 & $\cdots$\\
\hline
\end{tabular}
\end{center}
\tablecomments{Columns Explanation:
Col(1):Common Source Names; 
Col(2):  Program ID Number;
Col(3): Distance,  = taken from  H97 ; 
Col(4): Hubble Type;
Col(5):  [OIII] to H$_\beta$ ratio taken from H97;  
Col(6):  [OI] to H$_\alpha$ ratiotaken from H97; 
Col(7):  [NII] to H$_\alpha$ ratio taken from H97;
Col(8):  [SII] to H$_\alpha$ ratio taken from H97; 
Col(9):  Log of the far-IR luminosity in units of ergs s$^{-1}$, Far-infrared luminosities correspond to the 40-500$\mu$m wavelength interval and were calculated using the IRAS 60 and 100 $\mu$m fluxes according to the prescription: $L_{FIR}$=1.26$\times$10$^{-14}$(2.58f$_{60}$+f$_{100}$) in W m$^{-2}$ (Sanders \& Mirabel 1996);
Col(10):  IR-brightness Ratio (L$_{FIR}$/L$_B$) 
}
\end{table*}

The majority of IR-bright LINERs in our sample are from the V95 survey which includes galaxies that are, on average, somewhat more distant than those in the H97 survey.  Using a FIR luminosity cut off of 10$^{43}$ erg s$^{-1}$, the average distance for the IR-faint vs. IR-bright galaxies is 19 Mpc and 75 Mpc respectively.  Larger extraction regions for the most distant objects may increase the contamination of the optical spectrum from star formation and shocks , resulting in a greater fraction of LINERs dominated by star formation with distance.  We point out that such distance-based aperture effects  are applicable to all LINER samples (both IR-bright and IR-faint) and therefore unavoidable.  Indeed galaxies from the Palomar survey have been shown to change classification when higher spatial resolution {\it HST} observations are employed (Shields et al. 2007). 

\section{Observational Details and Data Reduction}
	The galaxies in this sample were observed by both the short-wavelength, high-resolution (SH, 4.7''$\times$11.3'', $\lambda$  =  9.9-19.6$\mu$m) and long-wavelength, high-resolution (LH, 11.1''$\times$22.3'', $\lambda$ = 18.7-37.2$\mu$m) modules of IRS.  That data were preprocessed by the IRS pipeline (version 15.3) prior to download.  For more information about data preprocessing, see {\it Spitzer} Observers Manual, Chapter 7.\footnote[2]{Calibration and Preprocessing:  http://ssc.spitzer.caltech.edu/documents/som/som8.0.irs.pdf}  The {\it Spitzer} data were further processed using the SMART v. 6.2.5 analysis package (Higdon et al. 2004) and the corresponding version of the calibration files (v.1.4.8).  The SH and LH modules are too small for interactive background subtraction to take place, and separate SH or LH background observations do not exist for most observations.   In the few cases where dedicated, off-source, background files were available, the background observations were subtracted from the nuclear spectrum.   For both high and low resolution spectra, the ends of each order were manually clipped from the rest of the spectrum and bad pixels missed in preprocessing were removed.  

	The 67 observations presented in this work are archived from various programs, and therefore contain both mapping and staring observations.   All of the staring observations were centered on the nucleus of the galaxy.  The radio or {\it 2MASS} coordinates of the nucleus were used to verify the coordinates of each observation.  The SH and LH staring observations include data from two slit positions overlapping by one third of a slit.   Because the slits in staring observations occupy distinctly different regions of the sky, averaging the spectra from two slits cannot occur unless the emission originates from a compact source that is contained entirely in each slit.  Therefore the procedure of Dudik et al. (2007) was adopted for estimating the nuclear flux of each line: 1) If the fluxes measured from the two slits differed by no more than the calibration error of the instrument, then the fluxes were averaged; otherwise, the slit with the highest measured line flux was chosen.  2) If an emission line was detected in one slit, but not in the other, then the detection was selected. 

In order to isolate the nuclear region in the mapping observations, fluxes were extracted from a single slit coinciding with the radio or {\it 2MASS} nuclear coordinates.  The surrounding slits were also checked to ensure that detections of some of the high ionization lines presented here were not missed either because of incorrect nuclear coordinates or because of low signal to noise (S/N) characteristic of many the mapping observations.  In all cases, the final fluxes presented from mapping observations result from the single, nuclear slit.

\begin{table*}
\fontsize{9pt}{10pt}\selectfont
\begin{center}
\caption{Mid-IR Line Fluxes: H97}
\begin{tabular}{lccccc}
\hline
\multicolumn{1}{c}{Galaxy} & \multicolumn{1}{c}{[NeII]} & [NeIII] & [NeV] & [NeV] & [OIV] \\

\multicolumn{1}{c}{Name} & \multicolumn{1}{c}{12$\mu$m} & 15$\mu$m & 14$\mu$m
 & 24$\mu$m & 26$\mu$m \\

\multicolumn{1}{c}{(1)} & \multicolumn{1}{c}{(2)} & (3) & (4) & (5) & (6) \\ 

\hline
NGC1052 & 21.0$\pm$0.3 & 12.4$\pm$0.2 & 0.54$\pm$0.15 & 2.5$\pm$0.4 & 2.1$\pm$0.5\\
NGC1055 & 28.3$\pm$0.3 & 2.7$\pm$0.2 & $<$0.12 & $<$0.37 & $<$0.4\\
NGC1961 & 23.8$\pm$0.7 & 2.1$\pm$0.3 & $<$0.13 & $<$0.61 & $<$0.3\\
NGC266 & 1.5$\pm$0.1 & 1.1$\pm$0.1 & $<$0.03 & $<$0.16 & $<$0.2\\
NGC2681 & 8.5$\pm$0.5 & 3.4$\pm$0.2 & $<$0.1 & $<$0.27 & 2.3$\pm$0.4\\
NGC2787 & 1.5$\pm$0.1 & 1.0$\pm$0.1 & $<$0.03 & $<$0.12 & $<$0.3\\
NGC2841 & 4.5$^a$ & 5.4$^a$ & $<$0.4 & $<$0.31 & 0.7$^a$\\
NGC3166 & 7.9$\pm$0.4 & 3.8$\pm$0.4 & $<$0.52 & 0.75$\pm$0.22 & 1.4$\pm$0.3\\
NGC3169 & 29.1$\pm$0.8 & 4.8$\pm$0.1 & $<$0.08 & $<$0.24 & 1.5$\pm$0.2\\
NGC3190 & 5.7$\pm$0.3 & 3.5$\pm$0.5 & $<$0.57 & $<$0.75 & $<$1.3\\
NGC3226 & 2.0$\pm$0.2 & 1.9$\pm$0.1 & $<$0.14 & $<$1.4 & $<$1.4\\
NGC3368 & 5.3$\pm$0.2 & 2.9$\pm$0.2 & 0.38$\pm$0.12 & $<$2.0 & $<$0.4\\
NGC3507 & 3.9$\pm$0.1 & 1.3$\pm$0.1 & 0.09$\pm$0.02 & $<$0.14 & $<$0.2\\
NGC3521 & 2.9$\pm$0.4 & 2.5$\pm$0.2 & 0.38$\pm$0.12 & $<$1.6 & $<$1.9\\
NGC3642 & 9.8$\pm$0.4 & 2.3$\pm$0.1 & $<$0.07 & 0.90$\pm$0.17 & 1.2$\pm$0.3\\
NGC3884 & 2.1$\pm$0.1 & 1.1$\pm$0.1 & $<$0.04 & $<$0.11 & $<$0.1\\
NGC3998 & 12.2$\pm$0.1 & 7.8$\pm$0.2 & $<$0.10 & $<$1.8 & $<$1.5\\
NGC4036 & 5.1$\pm$0.1 & 3.0$\pm$0.1 & $<$0.02 & 0.90$\pm$0.16 & 0.9$\pm$0.2\\
NGC404 & 3.7$\pm$0.1 & 1.8$\pm$0.1 & $<$0.05 & $<$0.18 & 0.8$\pm$0.2\\
NGC4143 & 3.2$\pm$0.1 & 2.3$\pm$0.1 & $<$0.04 & $<$0.15 & 0.6$\pm$0.2\\
NGC4203 & 1.8$\pm$0.1 & 2.1$\pm$0.1 & $<$0.03 & $<$0.30 & $<$0.5\\
NGC4261 & 5.4$\pm$0.2 & 2.6$\pm$0.2 & $<$0.03 & $<$0.23 & 1.2$\pm$0.4\\
NGC4278 & 5.6$\pm$0.1 & 4.0$\pm$0.1 & $<$0.03 & 0.70$\pm$0.14 & 1.6$\pm$0.1\\
NGC4314 & 11.3$\pm$0.3 & 1.9$\pm$0.2 & $<$0.05 & $<$0.33 & $<$0.4\\
NGC4394 & 1.7$\pm$0.3 & 1.2$\pm$0.3 & $<$0.16 & $<$0.11 & 0.7$\pm$0.1\\
NGC4438 & 22.2$\pm$0.8 & 8.6$\pm$0.5 & $<$0.21 & 1.1$\pm$0.19 & 3.5$\pm$0.5\\
NGC4450 & 3.8$\pm$0.9 & 2.5$\pm$0.6 & $<$0.71 & $<$1.46 & 1.8$\pm$0.7\\
NGC4457 & 8.9$\pm$0.2 & 3.2$\pm$0.1 & $<$0.03 & $<$0.15 & 2.1$\pm$0.5\\
NGC4486 & 7.8$\pm$0.1 & 7.9$\pm$0.2 & $<$0.10 & $<$0.94 & 2.4$\pm$0.4\\
NGC4548 & 2.5$\pm$0.2 & 1.5$\pm$0.1 & $<$0.05 & $<$3.5 & 2.3$\pm$0.4\\
NGC4594 & 9.5$^a$ & 8.2$^a$ & $<$0.32 & $<$0.40 & 2.4$^a$\\
NGC4596 & 1.1$\pm$0.1 & 1.7$\pm$0.1 & $<$0.03 & $<$0.32 & $<$0.2\\
NGC4736 & 11.9$\pm$2.6 & 6.7$\pm$0.4 & 1.1$\pm$0.32 & $<$1.7 & 3.1$\pm$1.0\\
NGC474 & $<$0.5 & $<$0.2 & $<$0.09 & $<$0.37 & $<$0.7\\
NGC5005 & 41.8$\pm$0.7 & 13.3$\pm$0.7 & $<$0.14 & $<$0.39 & $<$0.8\\
NGC5195 & 21.7$\pm$3.4 & 5.4$\pm$0.4 & $<$1.14 & $<$3.4 & $<$6.2\\
NGC5353 & 3.3$\pm$0.4 & 2.0$\pm$0.3 & $<$0.39 & 0.44$\pm$0.13 & 0.5$\pm$0.1\\
NGC5371 & 2.0$\pm$0.1 & 1.1$\pm$0.04 & 0.07$\pm$0.02 & 0.56$\pm$0.17 & 0.6$\pm$0.2\\
NGC5850 & 7.0$\pm$0.3 & 1.8$\pm$0.4 & $<$0.54 & $<$0.10 & $<$0.1\\
NGC5982 & 4.5$\pm$0.4 & $<$0.6 & $<$0.15 & $<$0.10 & $<$0.1\\
NGC5985 & 5.1$\pm$0.5 & 0.8$\pm$0.2 & $<$0.45 & $<$0.08 & $<$0.1\\
NGC6500 & 5.4$\pm$0.1 & 2.6$\pm$0.1 & $<$0.02 & $<$0.21 & $<$0.2\\
NGC7626 & 1.0$\pm$0.1 & 0.5$\pm$0.1 & $<$0.03 & $<$0.27 & $<$0.4\\
\hline
\end{tabular}
\end{center}
\tablecomments{Columns Explanation:
Col(1): Common Source Names; 
Col(2) - (6):  Fluxes in units of 10$^{-21}$ W cm$^{-2}$. 3 $\sigma$ upper limits are reported for nondetections.}
\tablerefs{ $^a$ Dale et al. 2006, $^b$ Armus et al. 2004 \& 2006, $^c$ Dudik et al. 2007, $^d$ Farrah et al. 2007}

\end{table*}

\begin{table*}
\fontsize{9pt}{10pt}\selectfont
\begin{center}
\caption{Mid-IR Line Fluxes: V95}
\begin{tabular}{lccccc}
 \hline
\multicolumn{1}{c}{Galaxy} & \multicolumn{1}{c}{[NeII]} & [NeIII] & [NeV] & [NeV] & [OIV] \\

\multicolumn{1}{c}{Name} & \multicolumn{1}{c}{12$\mu$m} & 15$\mu$m & 14$\mu$m
 & 24$\mu$m & 26$\mu$m \\

\multicolumn{1}{c}{(1)} & \multicolumn{1}{c}{(2)} & (3) & (4) & (5) & (6) \\ 

\hline
UGC556 & 35.3$\pm$1.3 & 8.0$\pm$0.1 & $<$0.18 & $<$0.30 & $<$0.4\\
IR01364-1042 & 9.3$\pm$0.2 & 2.2$\pm$0.2 & $<$1.18 & $<$1.4 & $<$1.5\\
NGC660 & 367 $\pm$5.1 & 36.8$\pm$0.8 & 3.1$\pm$0.56 & $<$10.0 & 19.8$\pm$4.6\\
IIIZw35 & 3.6$\pm$0.1 & 0.6$\pm$0.1 & $<$0.49 & $<$2.4 & $<$3.7\\
IR02438+2122 & 18.6$\pm$0.6 & 1.6$\pm$0.2 & $<$1.63 & $<$2.2 & $<$1.8\\
NGC1266 & 20.9$\pm$0.8 & 5.3$\pm$0.7 & $<$0.80 & $<$2.9 & $<$6.6\\
ugc5101 & 55.2$\pm$2.5$^b$ & 23.9$\pm$1.4$^b$ & 5.2$\pm$0.70$^b$ & 4.9$\pm$1.0$^b$ & 5.5$\pm$1.4$^b$\\
NGC4666 & 38.3$\pm$0.7 & 8.3$\pm$0.4 & $<$0.15 & 2.4$\pm$0.4 & 5.4$\pm$0.5\\
UGC8387  & 105$\pm$0.9 & 18.9$\pm$0.7 & 1.43$\pm$0.33 & 11.4$\pm$2.4 & 11.2$\pm$1.8\\
NGC5104 & 44.3$\pm$0.5 & 5.4$\pm$0.5 & 1.77$\pm$0.36 & 3.8$\pm$0.5 & 2.9$\pm$0.6\\
NGC5218 & 145.0$\pm$1.8 & 10.6$\pm$0.5 & $<$0.18 & $<$0.44 & $<$1.6\\
Mrk273 & 43.2$\pm$0.4$^c$ & 34.3$\pm$0.5$^c$ & 10.6$\pm$0.47$^c$ & 27.4$\pm$1.9$^c$ & 56.9$\pm$1.8$^c$\\
Mrk848 & 61.4$\pm$0.4 & 14.3$\pm$0.3 & $<$0.09 & $<$0.92 & $<$0.8\\
NGC5953 & 56.0$\pm$0.7 & 16.8$\pm$0.2 & 1.2$\pm$0.26 & 4.7$\pm$1.1 & 17.5$\pm$1.4\\
IR15335-0513 & 35.1$\pm$0.4 & 6.0$\pm$0.2 & 0.23$\pm$0.04 & $<$0.83 & 2.7$\pm$0.5\\
IR16164-0746 & 45.4$\pm$0.4 & 13.7$\pm$0.2 & 1.8$\pm$0.22 & 3.8$\pm$0.7 & 10.2$\pm$1.4\\
NGC6240 & 193$\pm$3.7$^b$ & 70.4$\pm$2.4$^b$ & 5.1$\pm$0.90$^b$ & $<$3.9 & 27.2$\pm$0.7$^b$\\
NGC6286  & 18.5$\pm$0.2 & 3.1$\pm$0.1 & 0.33$\pm$0.11 & 0.99$\pm$0.20 & 0.9$\pm$0.2\\
ESO 593-IG008 & 38.4$\pm$1.3 & 8.2$\pm$0.2 & 0.50$\pm$0.11 & $<$0.42 & 1.2$\pm$0.3\\
NGC7130 & 58.8$\pm$0.8 & 20.6$\pm$0.5 & 2.12$\pm$0.31 & 6.6$\pm$1.0 & 13.3$\pm$1.2\\
ZW453.062 & 25.3$\pm$0.2 & 6.5$\pm$0.1 & 7.2$\pm$0.64 & 2.6$\pm$0.4 & 10.1$\pm$2.0\\
NGC7591 & 34.8$\pm$0.4 & 4.1$\pm$0.4 & $<$0.30 & $<$0.76 & $<$0.4\\
IR23365+3604 & 8.6$^d$ & 0.7$^d$ & $<$0.80$^d$ & $<$0.54$^d$ & $<$2.0$^d$\\
IR04259-0440 & 12.3$\pm$5.3 & $<$1.0 & $<$0.73 & $<$8.5 & $<$5.2\\
\hline
\end{tabular}
\end{center}

\tablecomments{Columns Explanation:
Col(1): Common Source Names; 
Col(2) - (6):  Fluxes in units of 10$^{-21}$ W cm$^{-2}$. 3 $\sigma$ upper limits are reported for nondetections.}
\tablerefs{ $^a$ Dale et al. 2006, $^b$ Armus et al. 2004 \& 2006, $^c$ Dudik et al. 2007, $^d$ Farrah et al. 2007}
\end{table*}

\begin{figure*}[htbp]
\begin{center}
\begin{tabular}{cc}
  \includegraphics[width=0.45\textwidth]{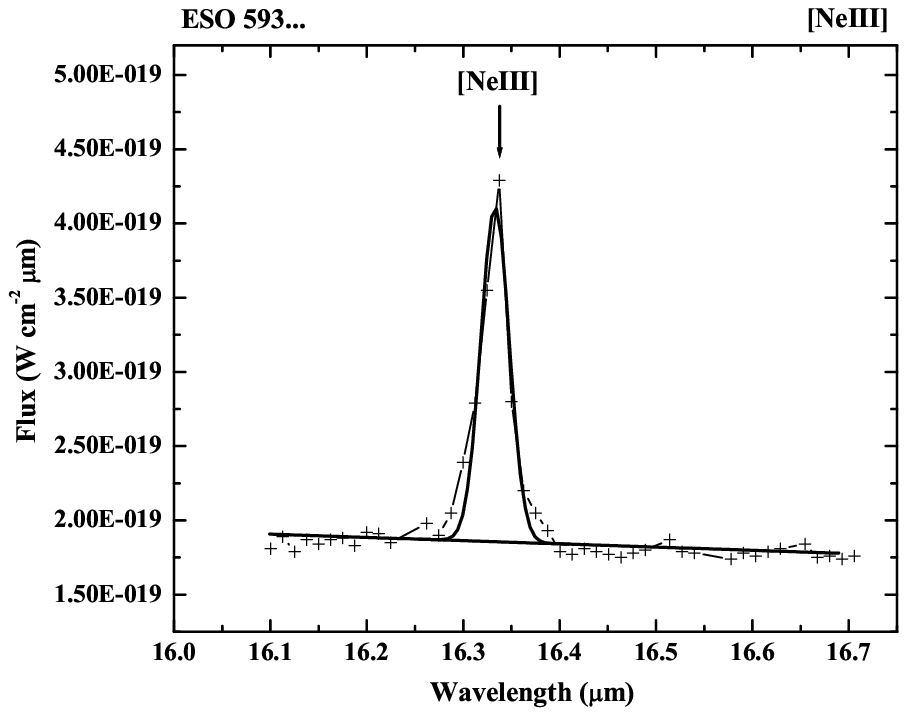} &
  \includegraphics[width=0.45\textwidth]{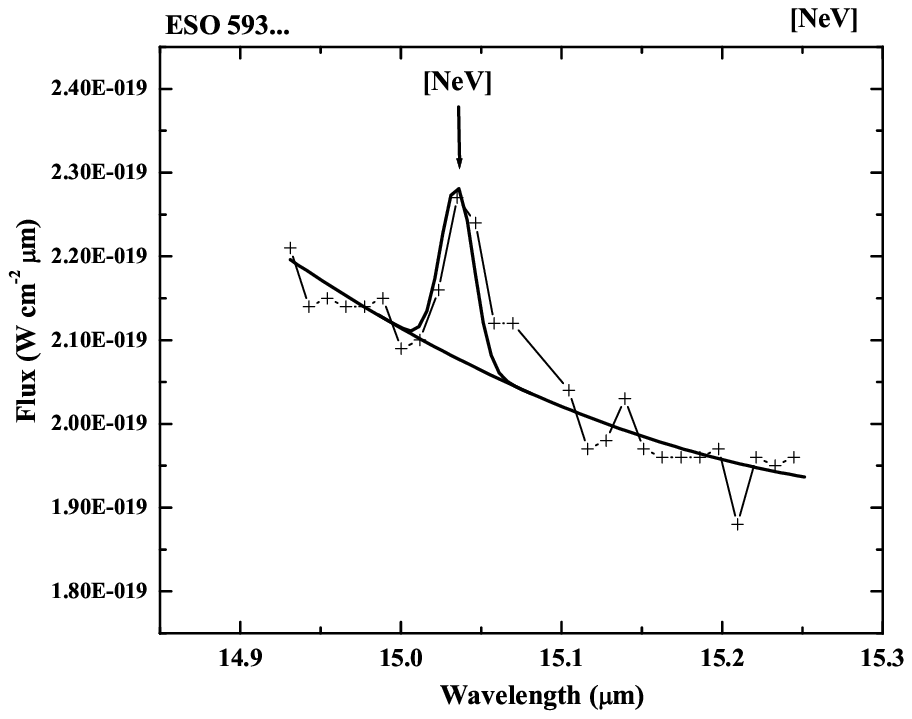} \\
  \includegraphics[width=0.45\textwidth]{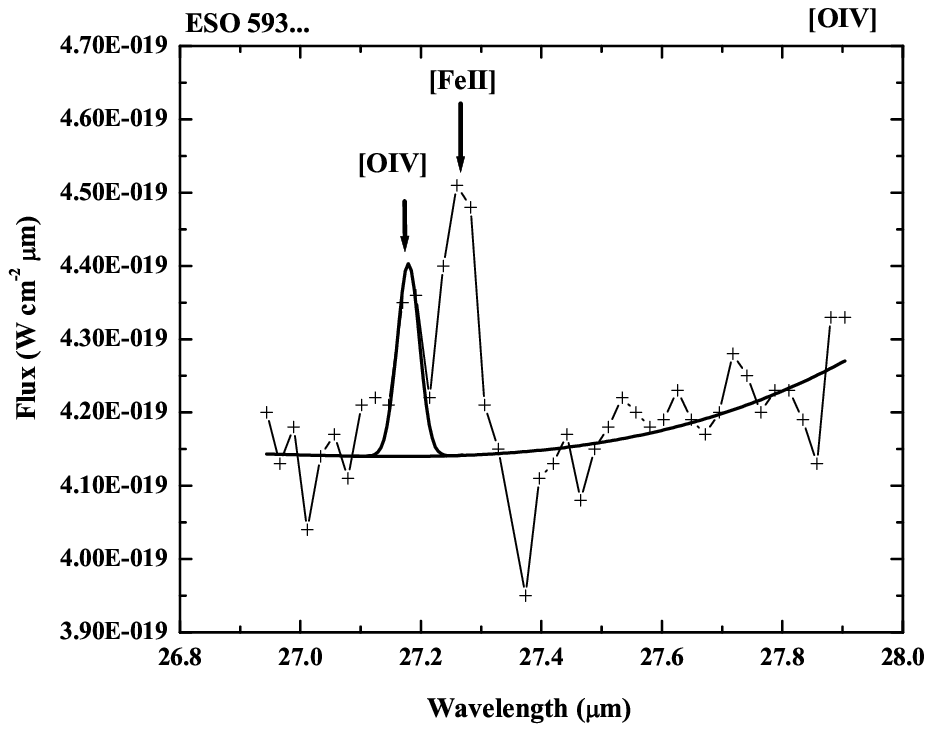} &
 \includegraphics[width=0.45\textwidth]{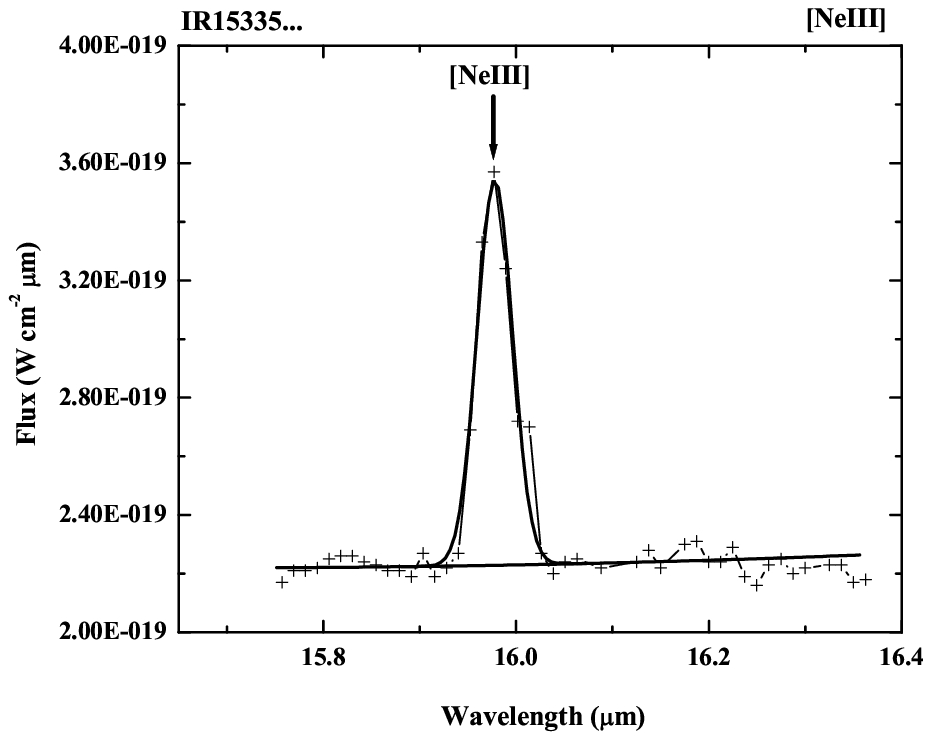}\\
\includegraphics[width=0.45\textwidth]{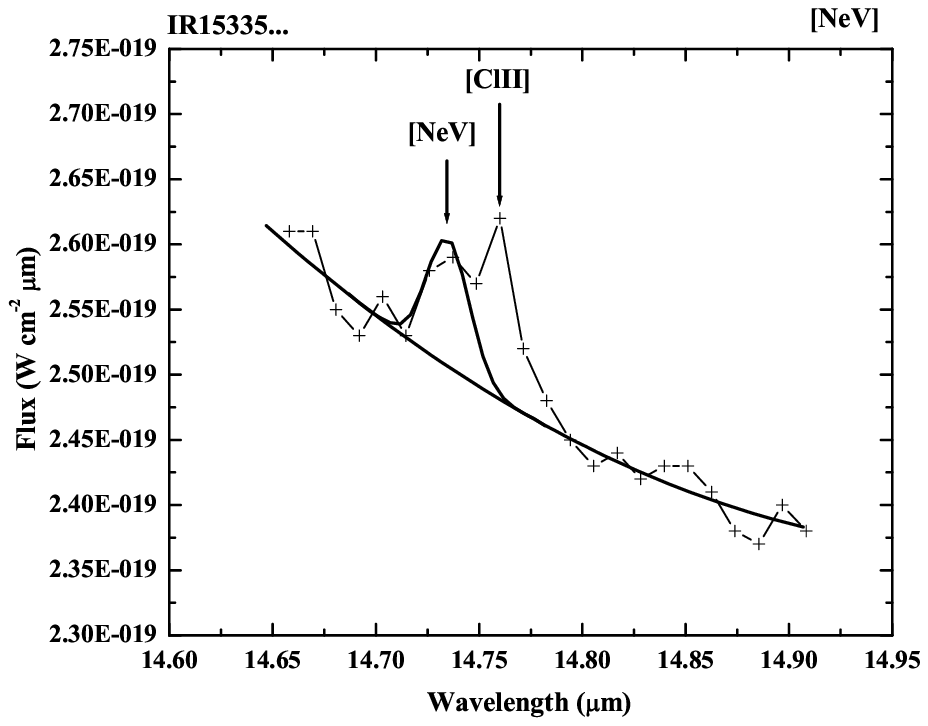} &
  \includegraphics[width=0.45\textwidth]{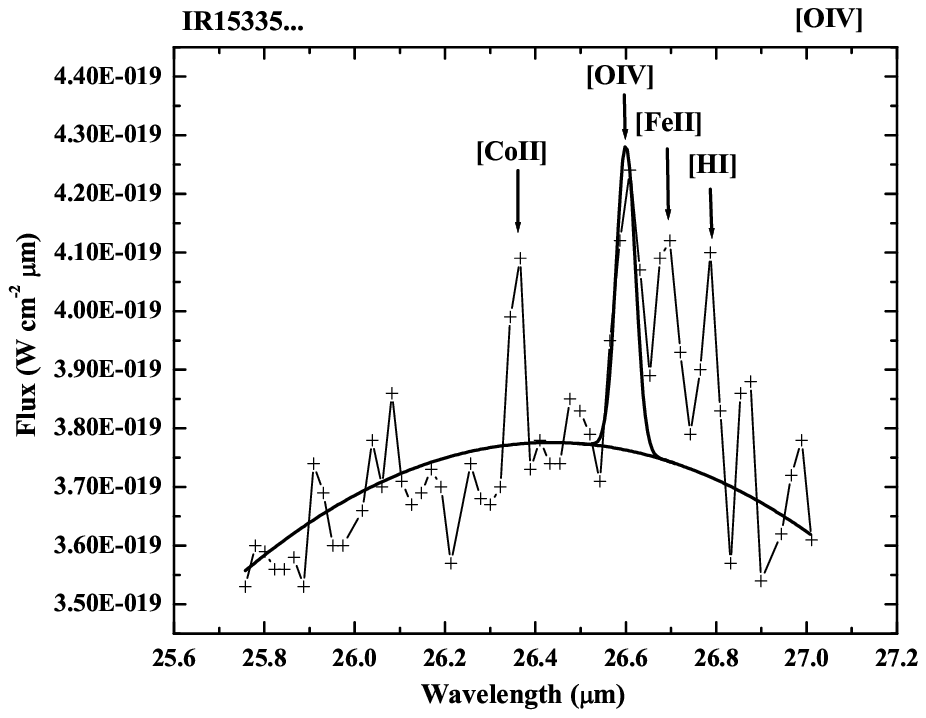} \\ 
 
\end{tabular}
\end{center}
\caption[]{[NeIII]15, [NeV]14, [NeV]24, and [OIV]26 $\mu$m spectra for the [NeV] emitting galaxies in our sample.} 
\end{figure*}
\clearpage
\begin{center}
\begin{tabular}{cc}
\includegraphics[width=0.45\textwidth]{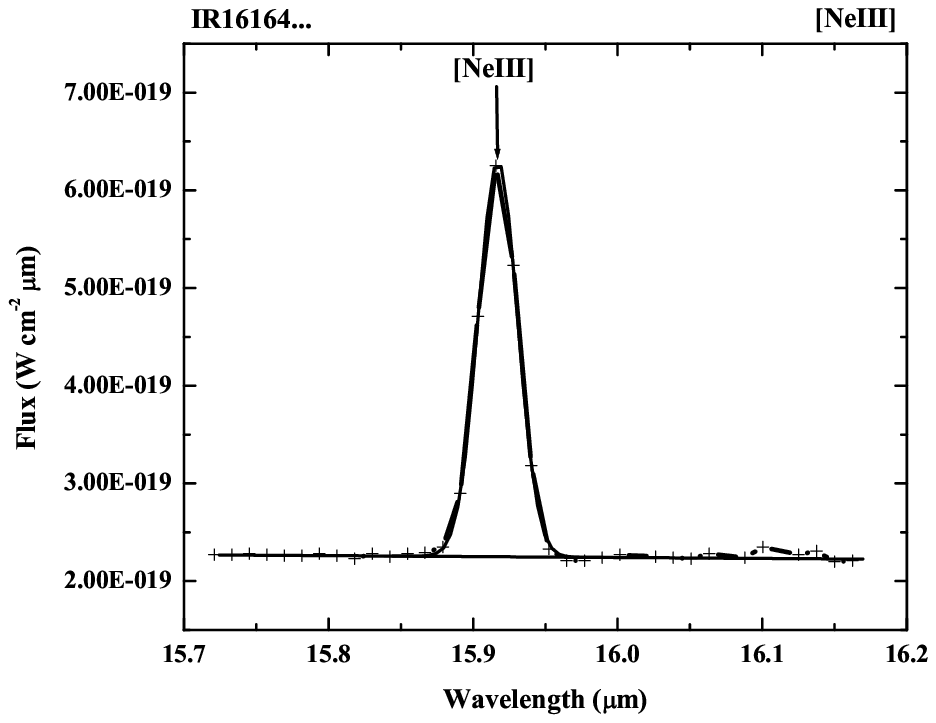} &
  \includegraphics[width=0.45\textwidth]{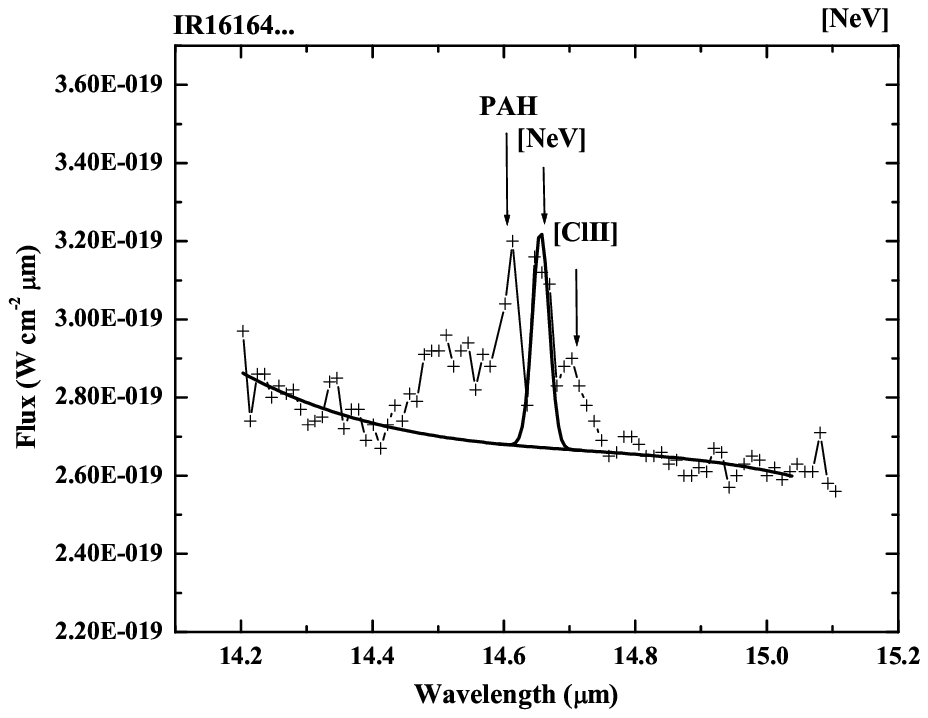} \\
  \includegraphics[width=0.45\textwidth]{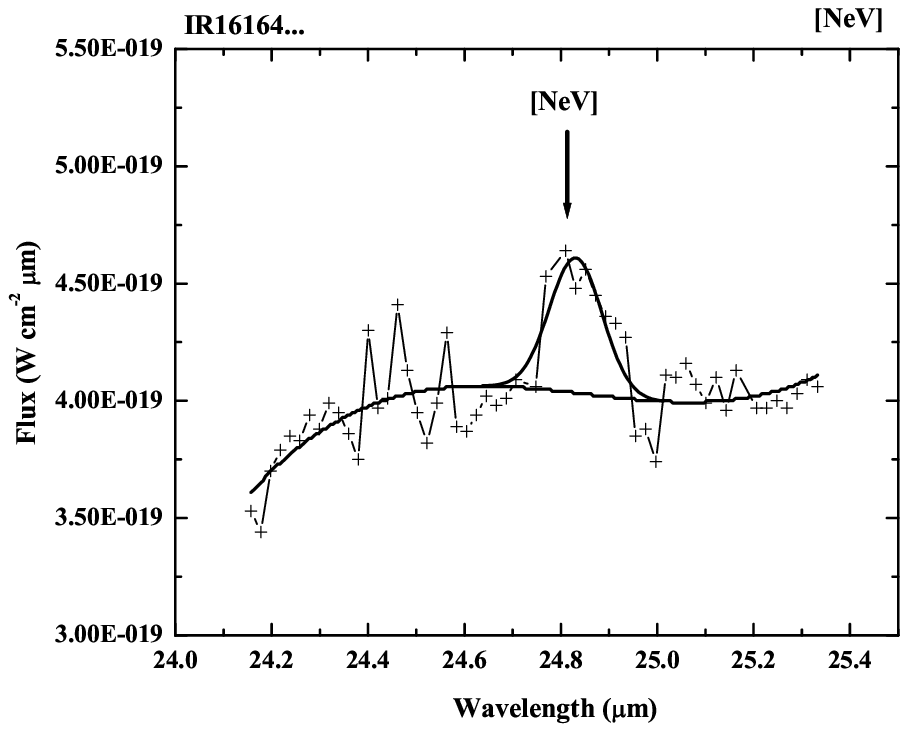} &
 \includegraphics[width=0.45\textwidth]{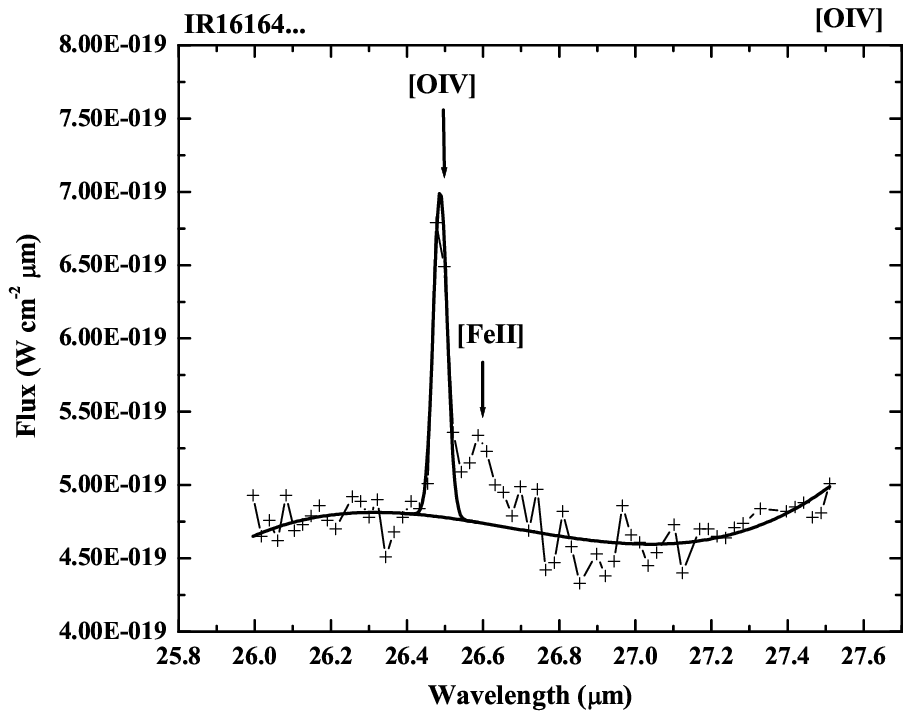} \\
\includegraphics[width=0.45\textwidth]{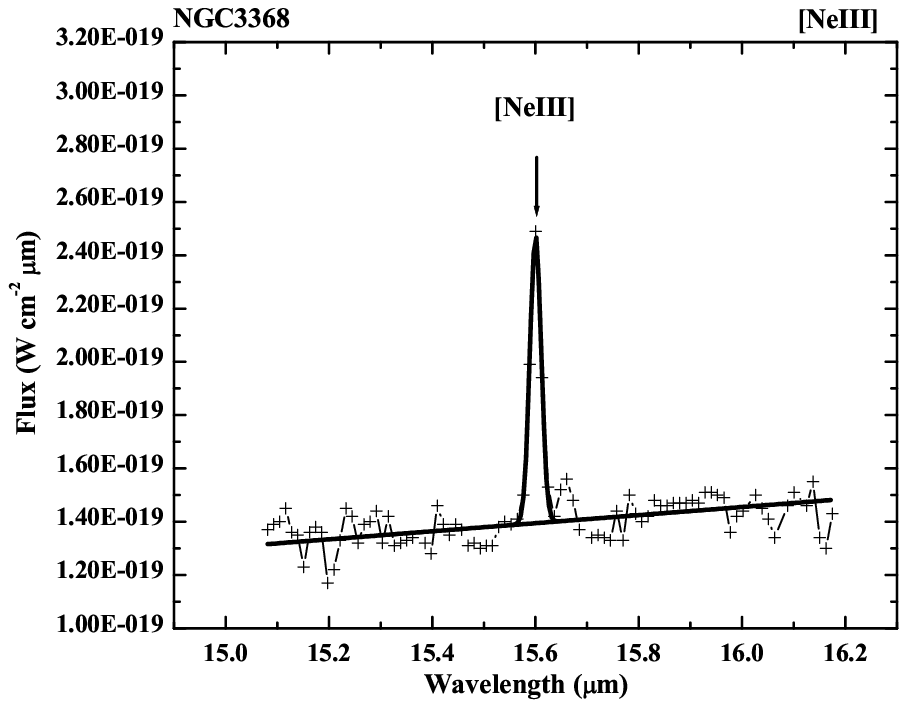} &
  \includegraphics[width=0.45\textwidth]{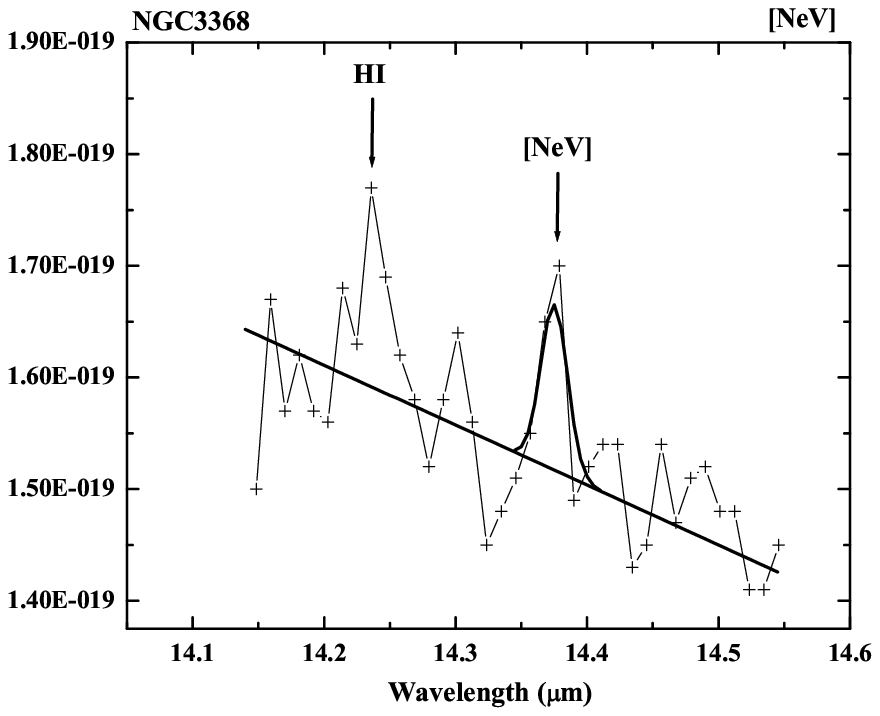} \\

\multicolumn{2}{c}{Fig. 2-- Spectra Continued.}\\
 \end{tabular}
\end{center}
\clearpage
\begin{center}
\begin{tabular}{cc}
\includegraphics[width=0.45\textwidth]{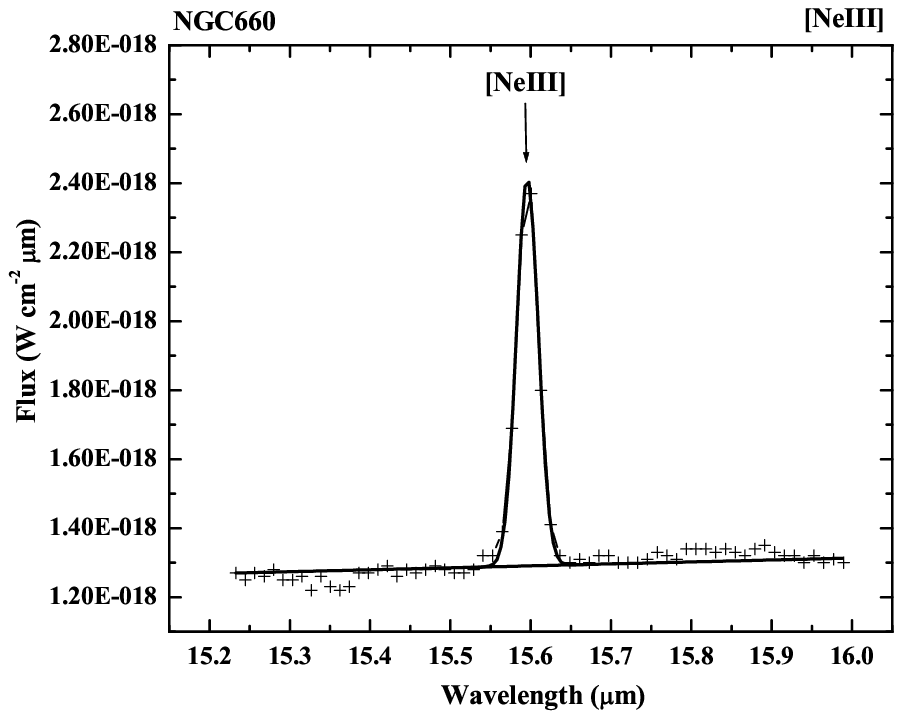} &
  \includegraphics[width=0.45\textwidth]{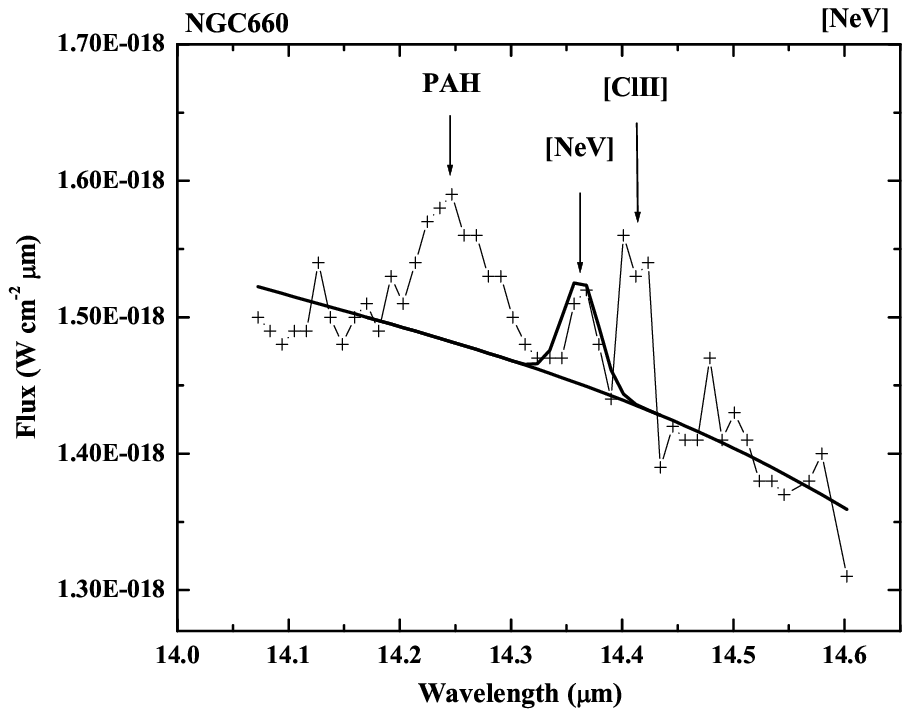} \\
  \includegraphics[width=0.45\textwidth]{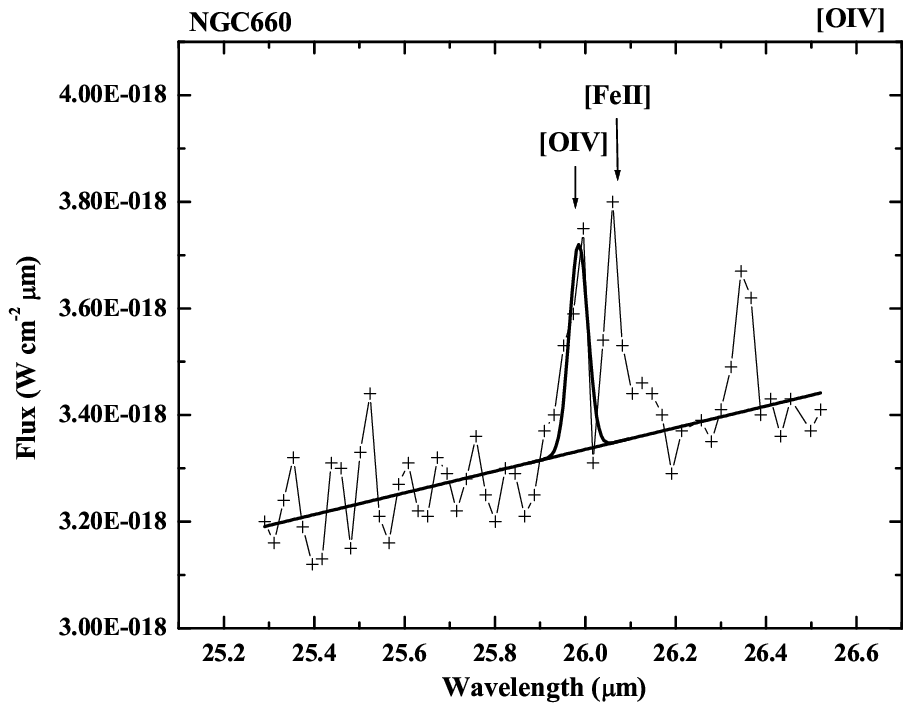} &
 \includegraphics[width=0.45\textwidth]{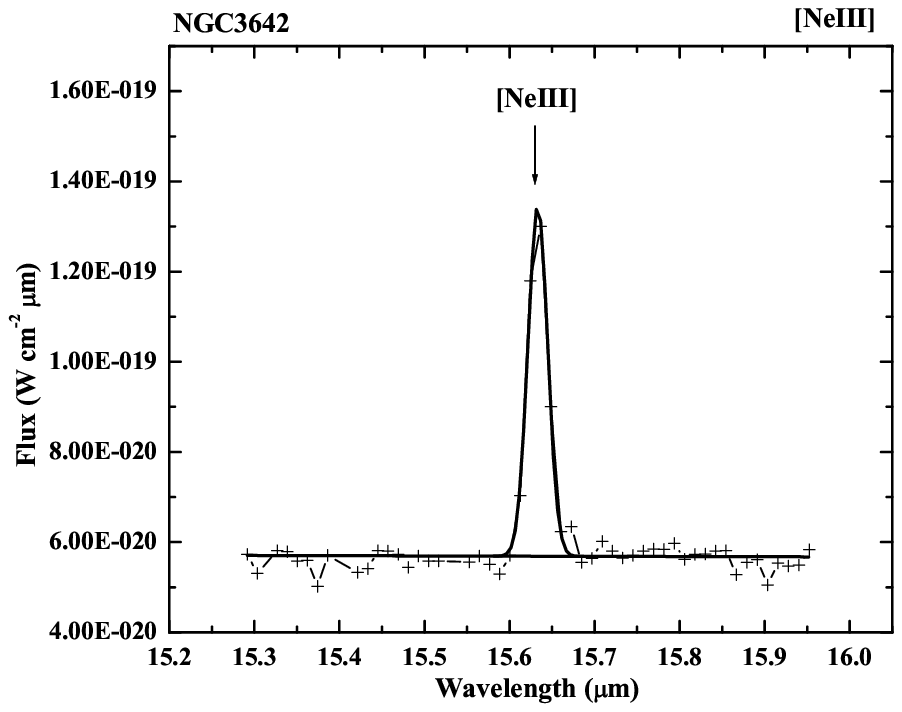} \\
\includegraphics[width=0.45\textwidth]{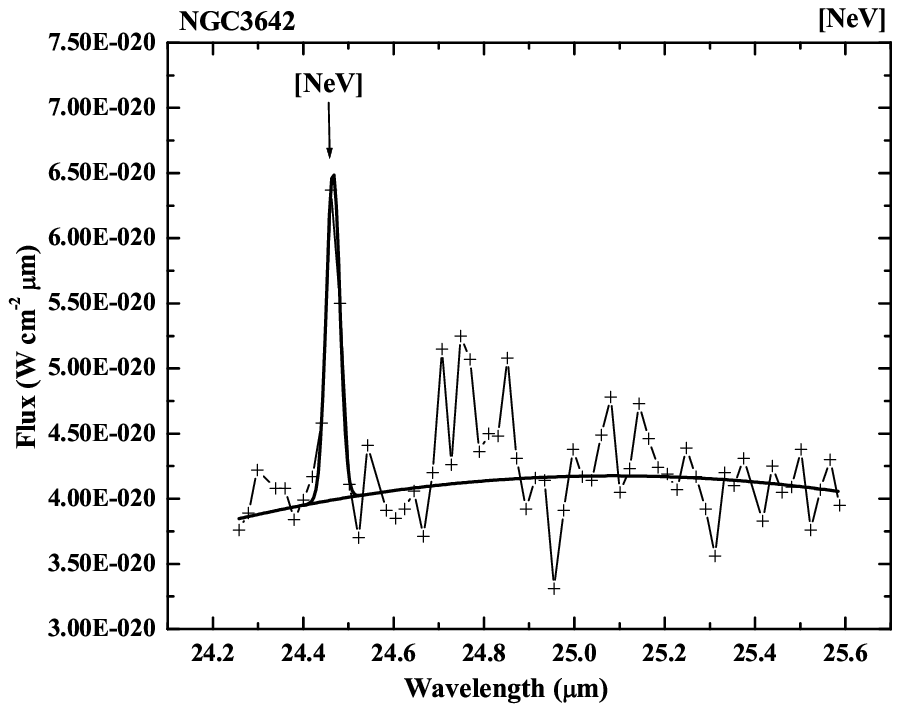} &
  \includegraphics[width=0.45\textwidth]{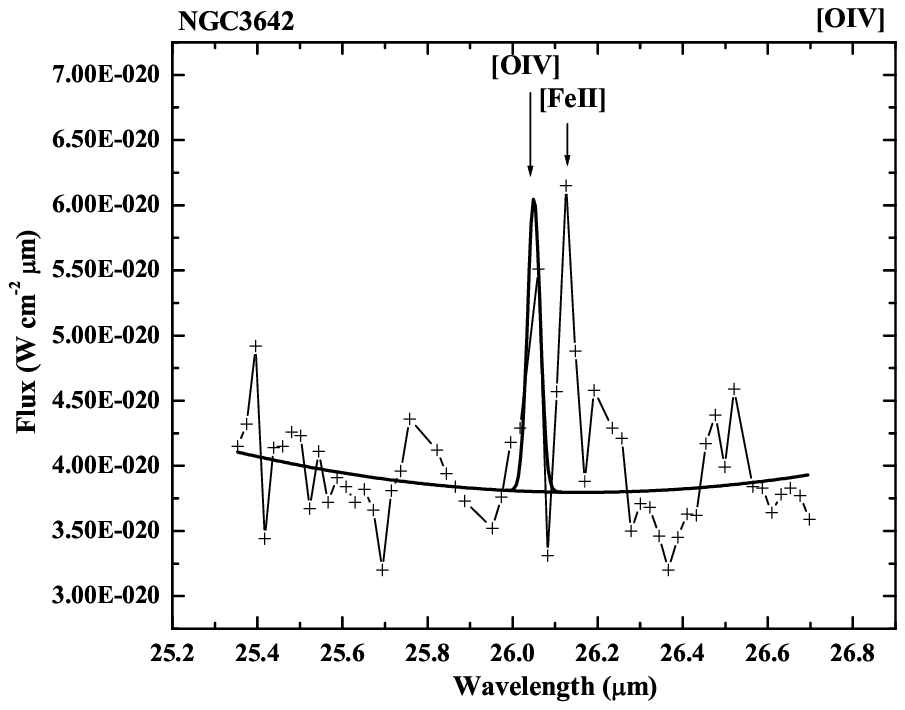}\\
\multicolumn{2}{c}{Fig. 2-- Spectra Continued.}\\
 \end{tabular}
\end{center}
\clearpage
\begin{center}
\begin{tabular}{cc}
 \includegraphics[width=0.45\textwidth]{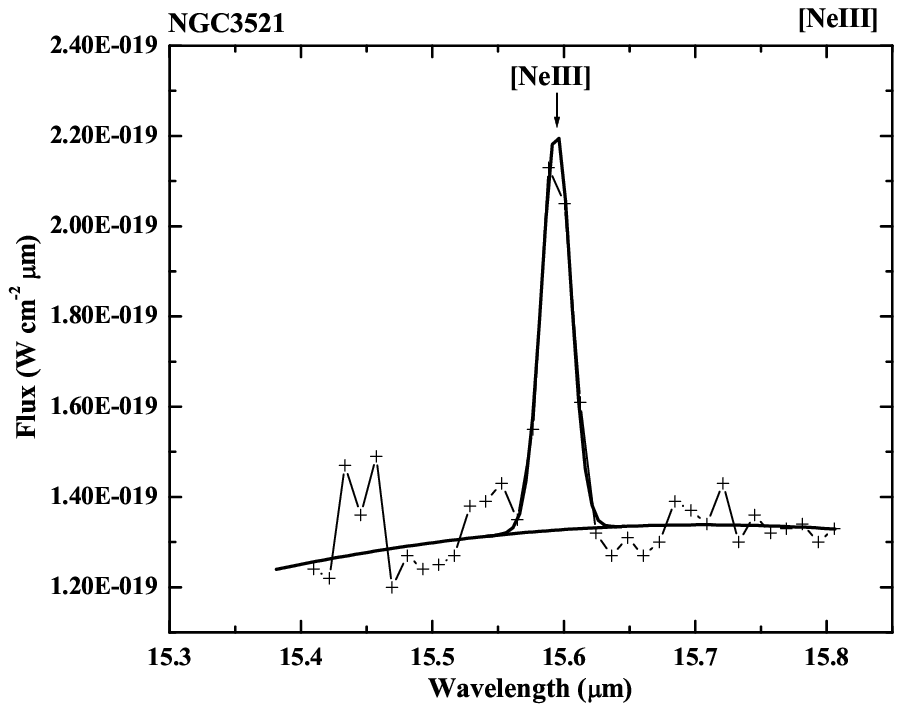} &
  \includegraphics[width=0.45\textwidth]{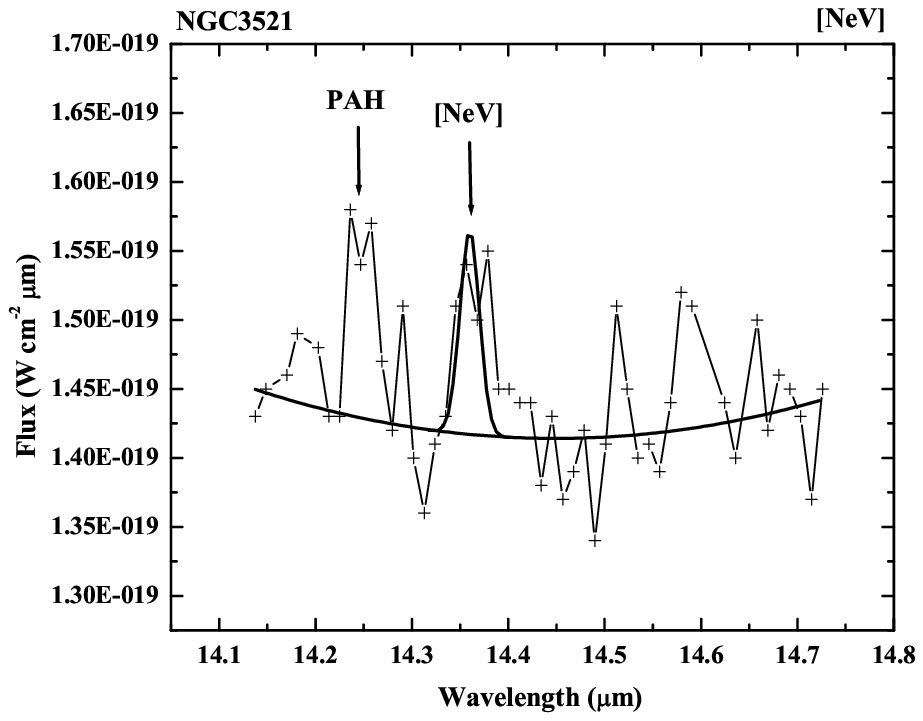} \\
  \includegraphics[width=0.45\textwidth]{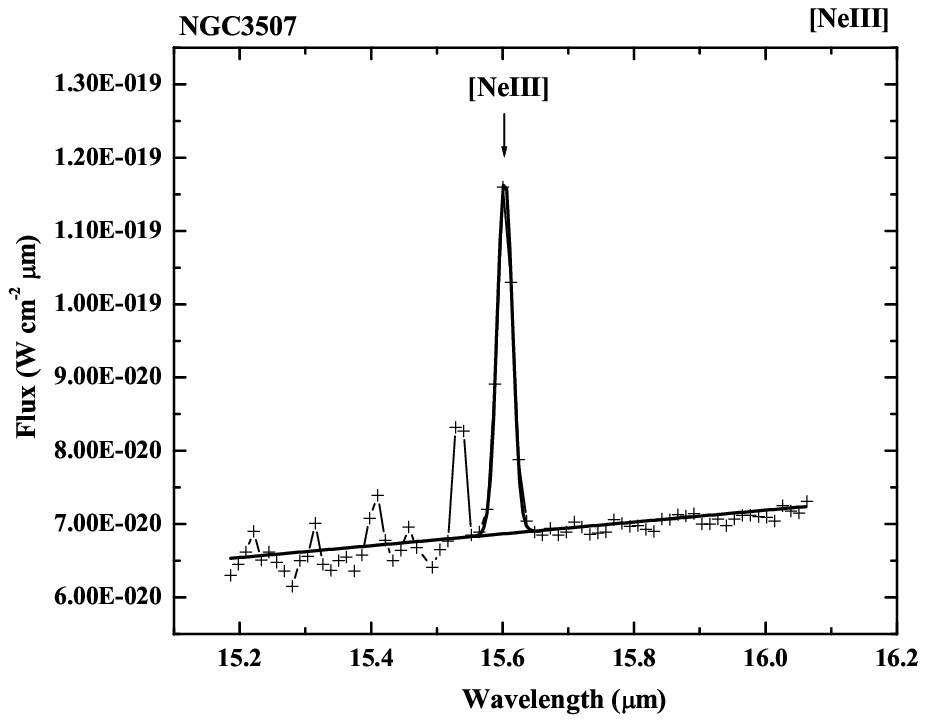} &
 \includegraphics[width=0.45\textwidth]{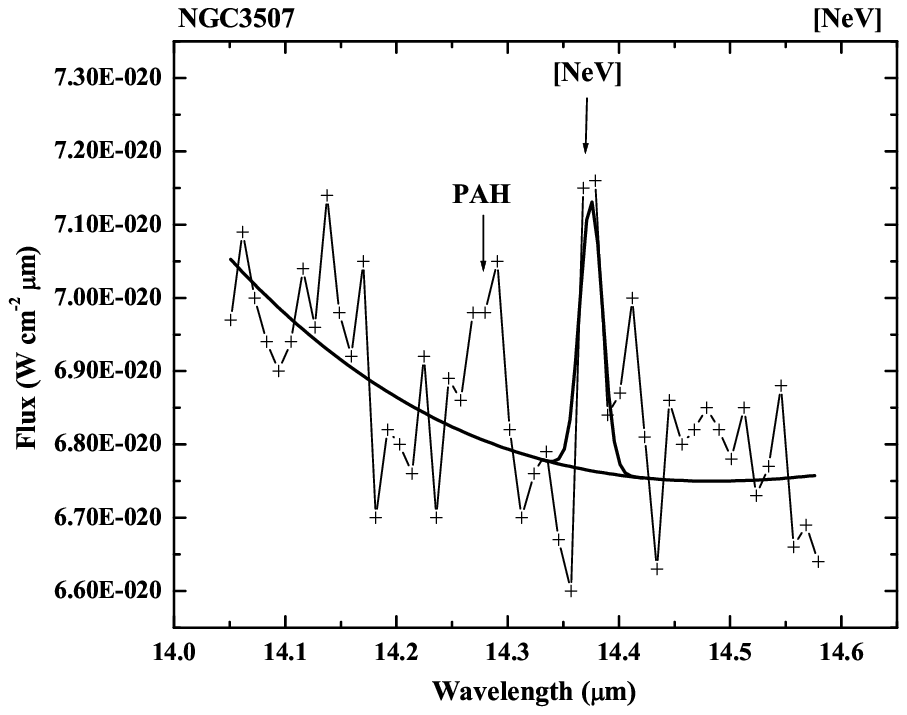} \\
  \includegraphics[width=0.45\textwidth]{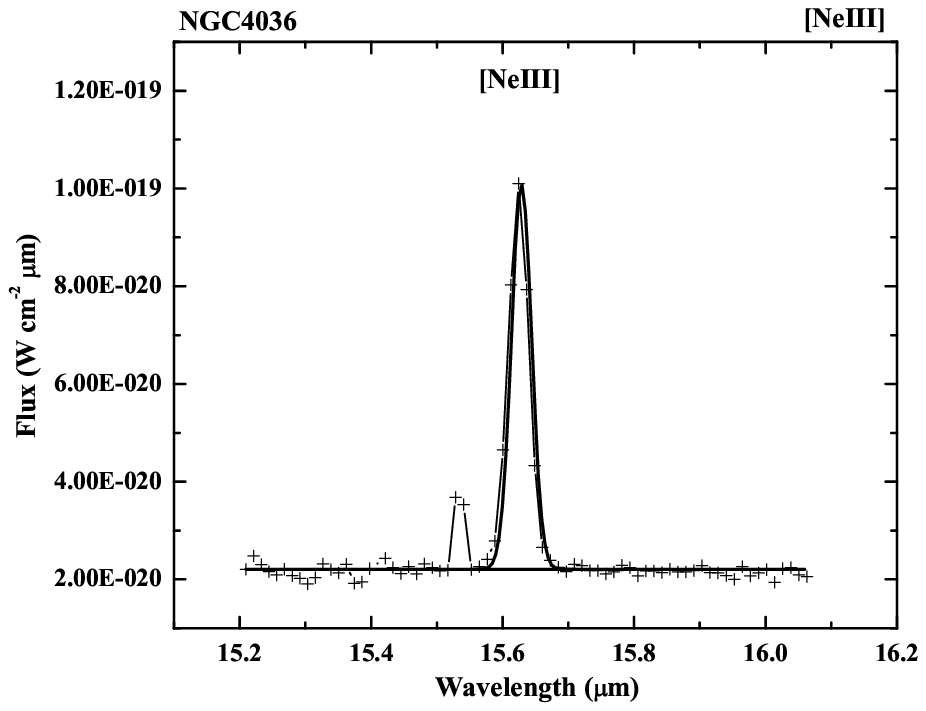} &
  \includegraphics[width=0.45\textwidth]{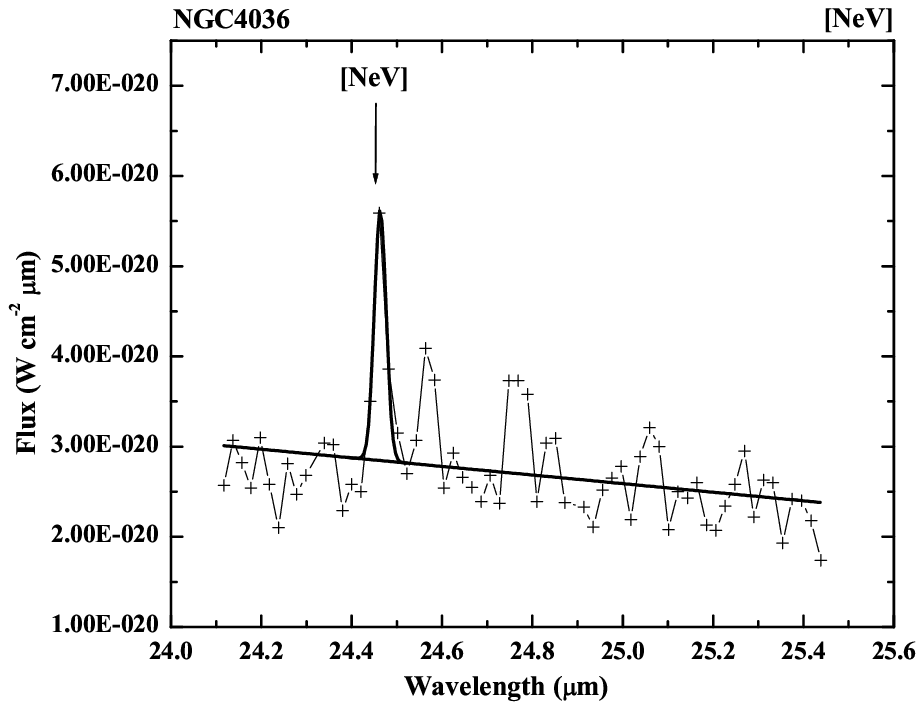} \\
\multicolumn{2}{c}{Fig. 2-- Spectra Continued.}\\
 \end{tabular}
\end{center}
\clearpage
\begin{center}
\begin{tabular}{cc}
  \includegraphics[width=0.45\textwidth]{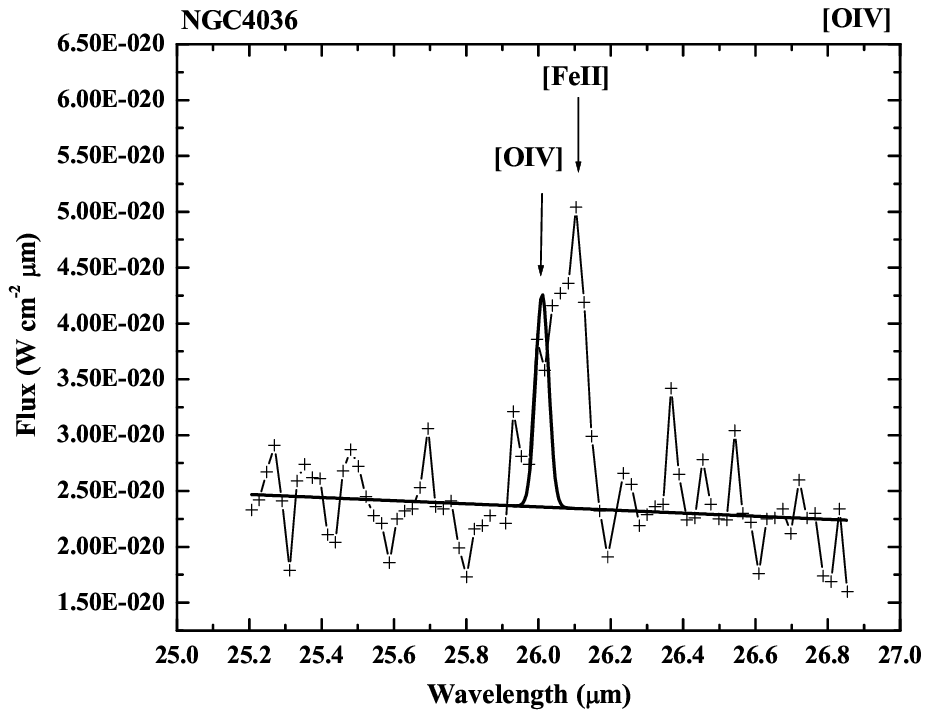} &
 \includegraphics[width=0.45\textwidth]{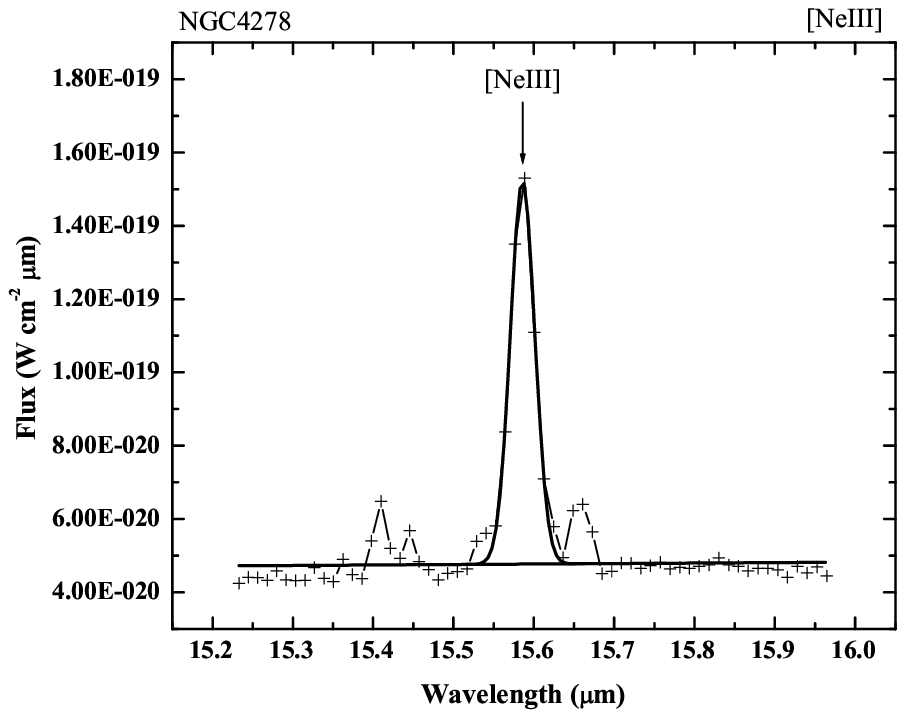} \\
\includegraphics[width=0.45\textwidth]{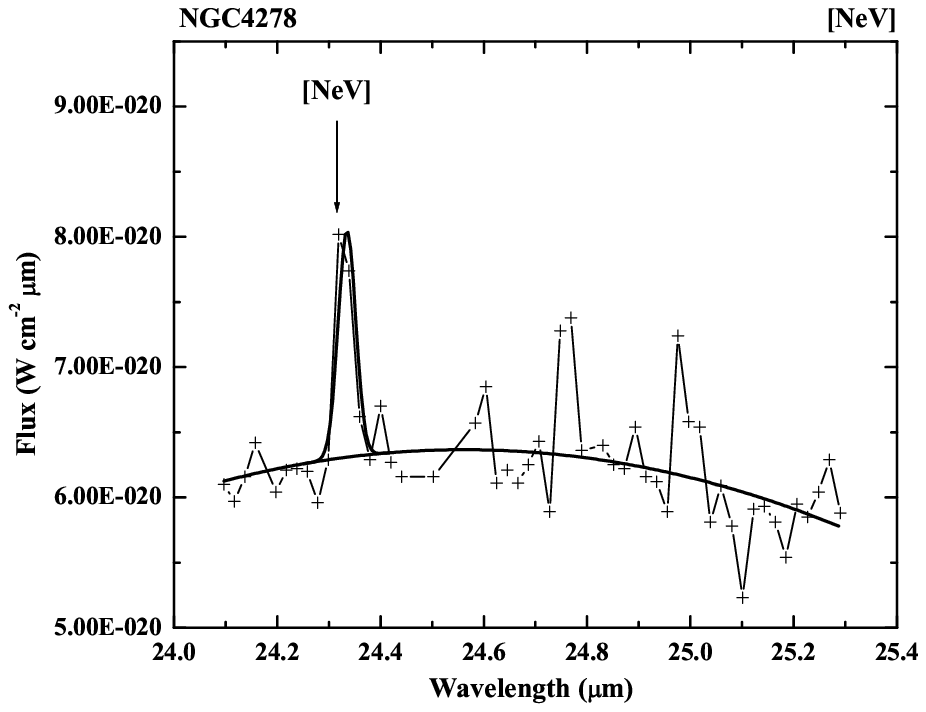} &
  \includegraphics[width=0.45\textwidth]{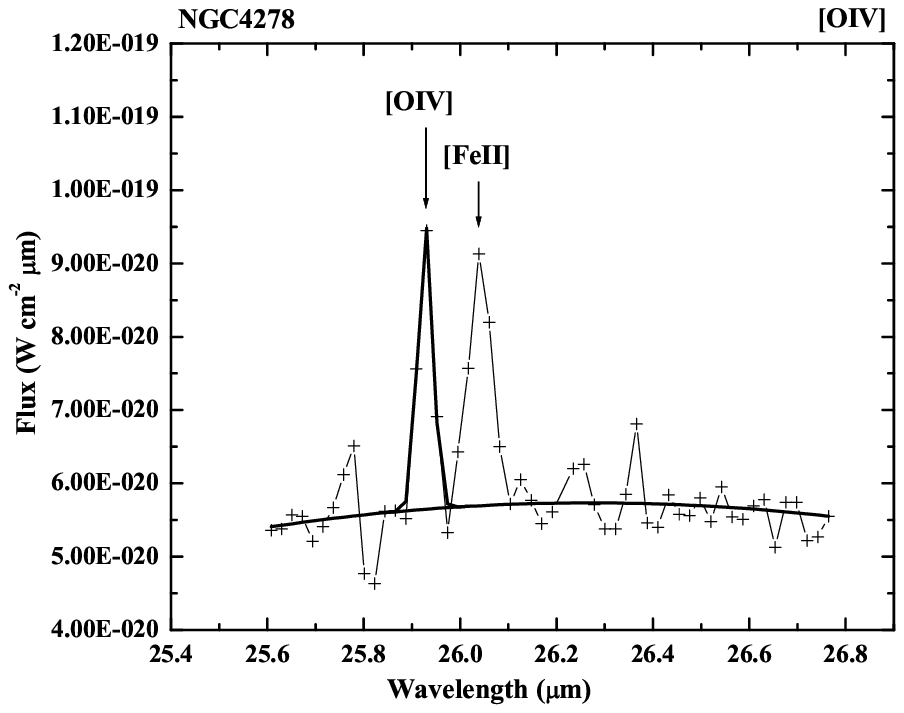} \\
 \includegraphics[width=0.45\textwidth]{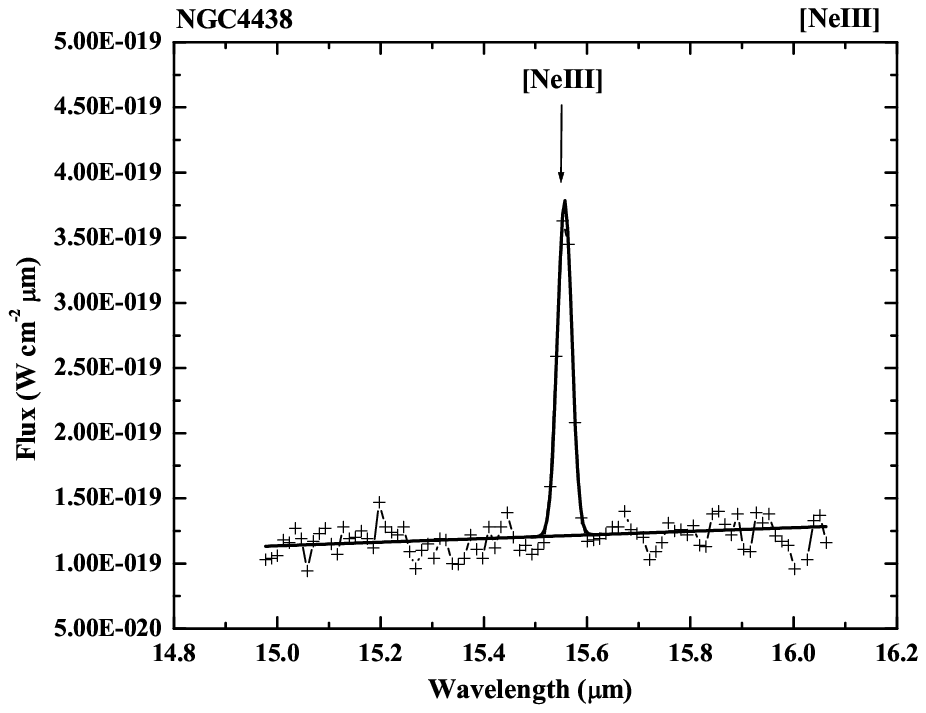} &
  \includegraphics[width=0.45\textwidth]{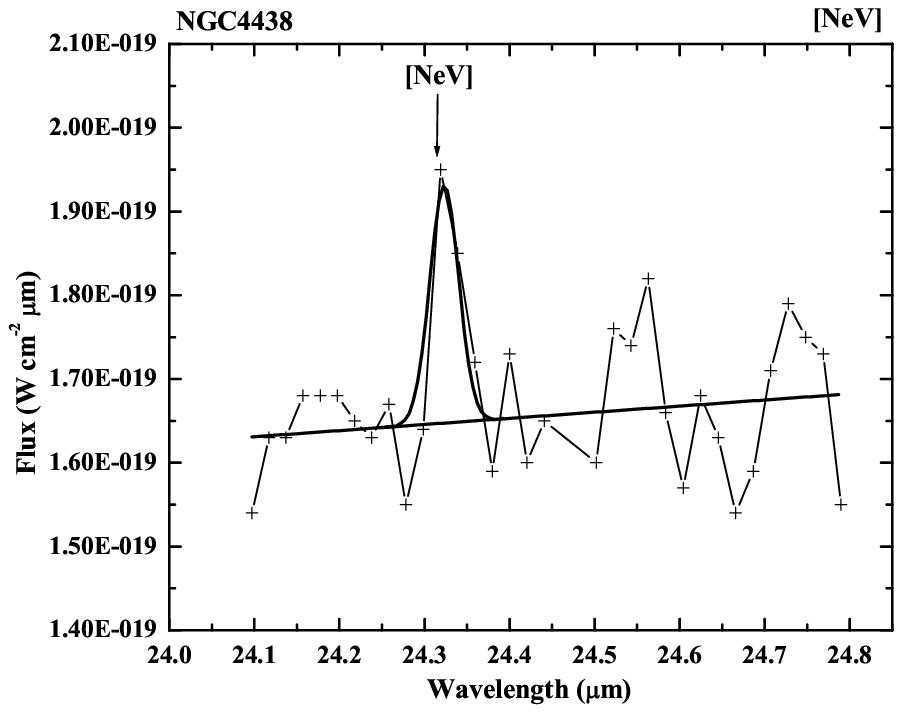} \\
\multicolumn{2}{c}{Fig. 2-- Spectra Continued.}\\
 \end{tabular}
\end{center}
\clearpage
\begin{center}
\begin{tabular}{cc}
 \includegraphics[width=0.45\textwidth]{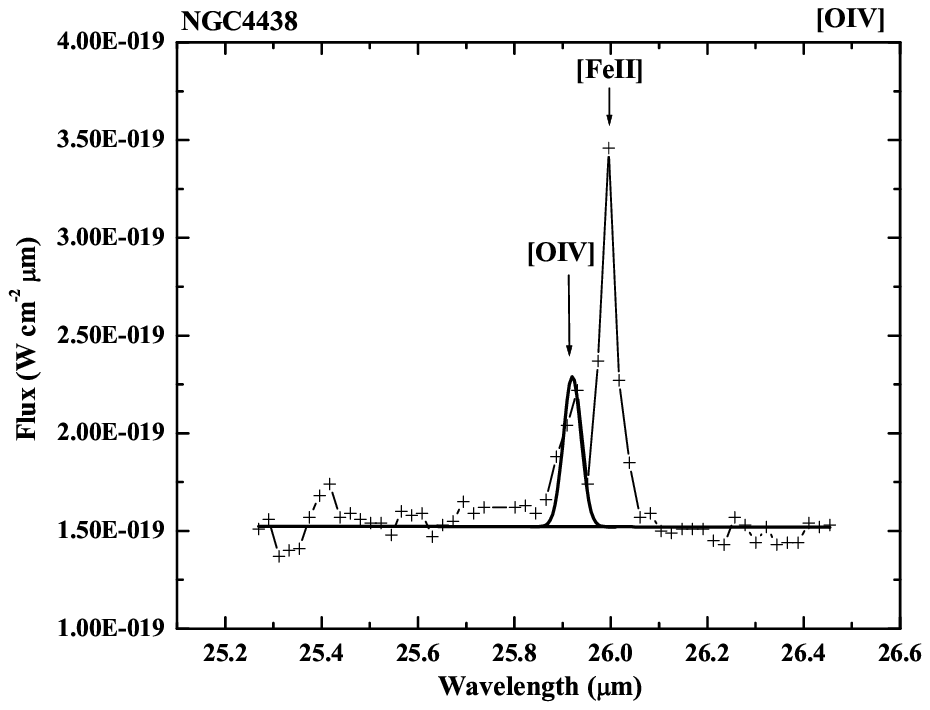} &
 \includegraphics[width=0.45\textwidth]{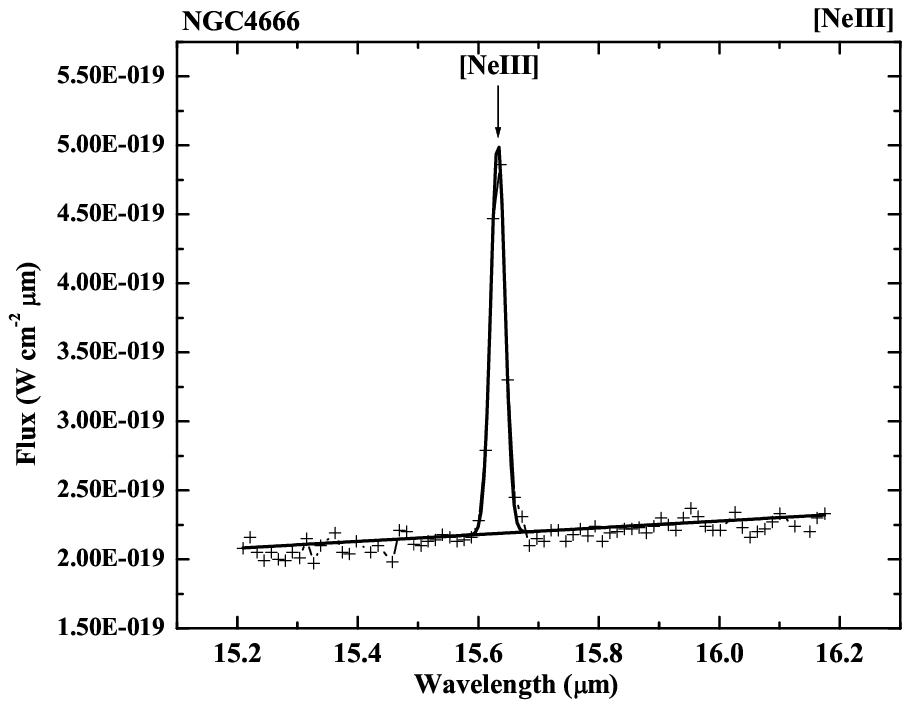} \\
 \includegraphics[width=0.45\textwidth]{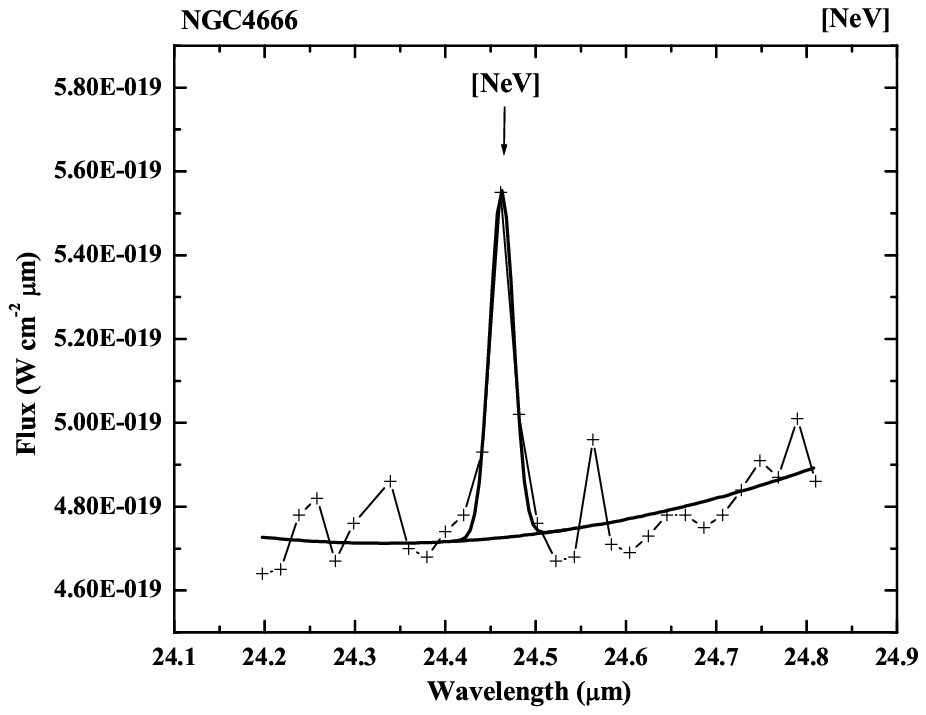} &
  \includegraphics[width=0.45\textwidth]{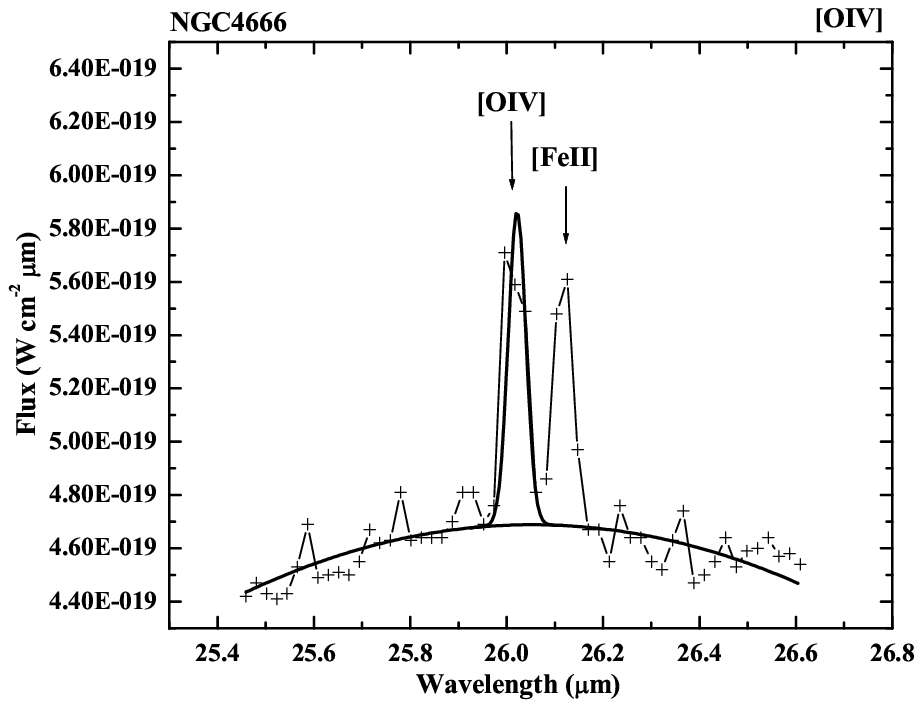} \\
  \includegraphics[width=0.45\textwidth]{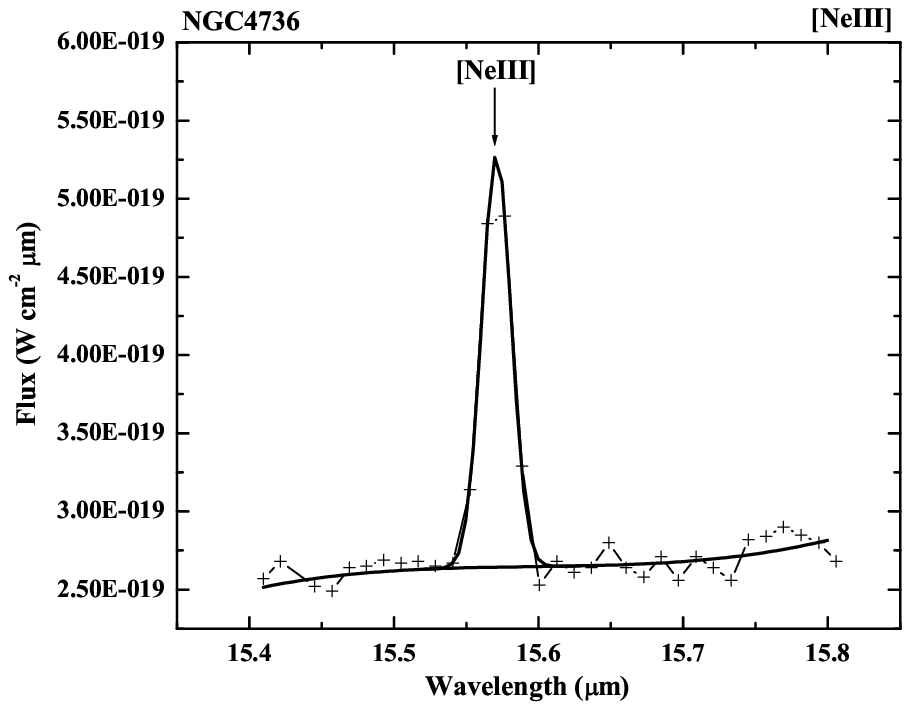} &
  \includegraphics[width=0.45\textwidth]{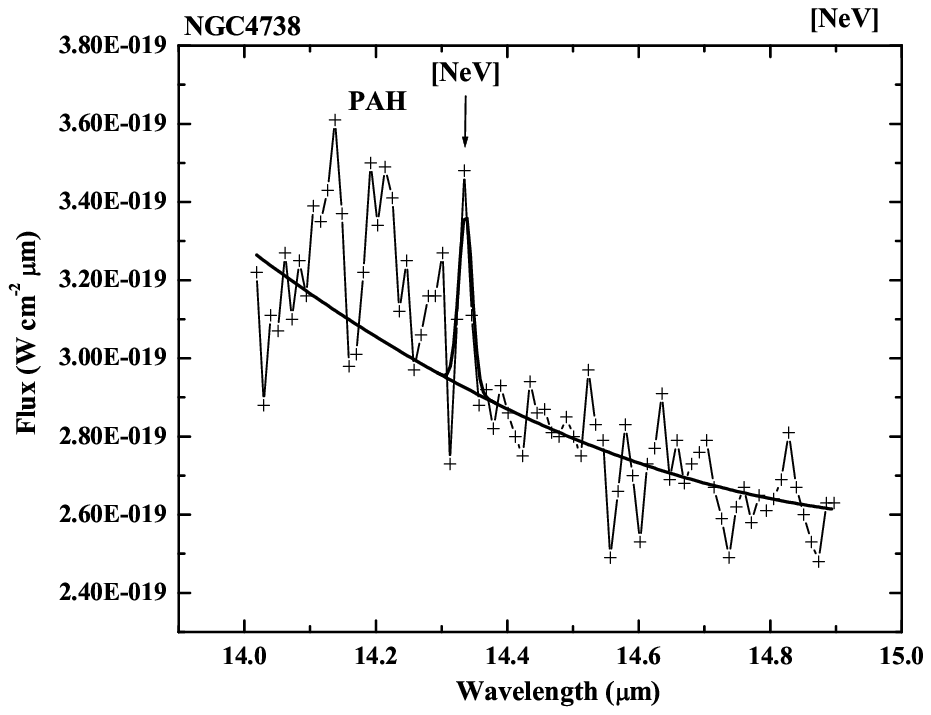} \\
\multicolumn{2}{c}{Fig. 2-- Spectra Continued.}\\
 \end{tabular}
\end{center}
\clearpage
\begin{center}
\begin{tabular}{cc}
\includegraphics[width=0.45\textwidth]{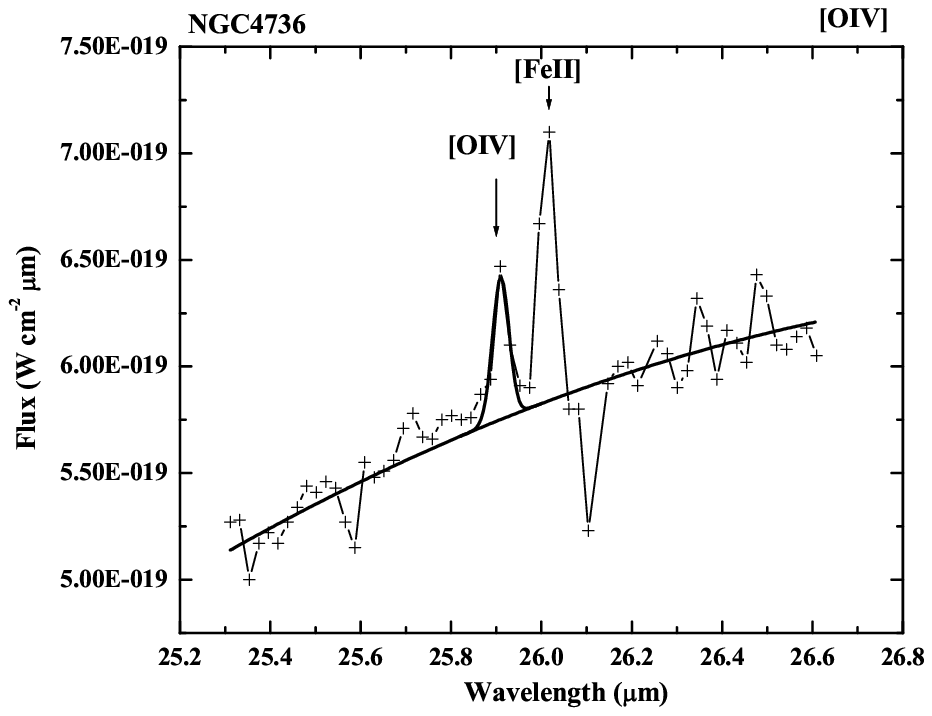} &
 \includegraphics[width=0.45\textwidth]{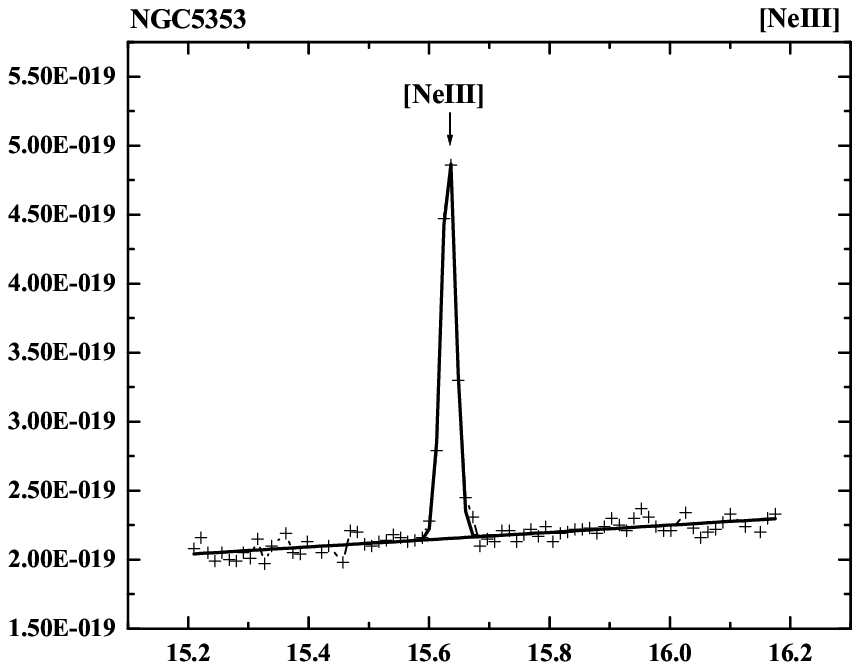} \\
\includegraphics[width=0.45\textwidth]{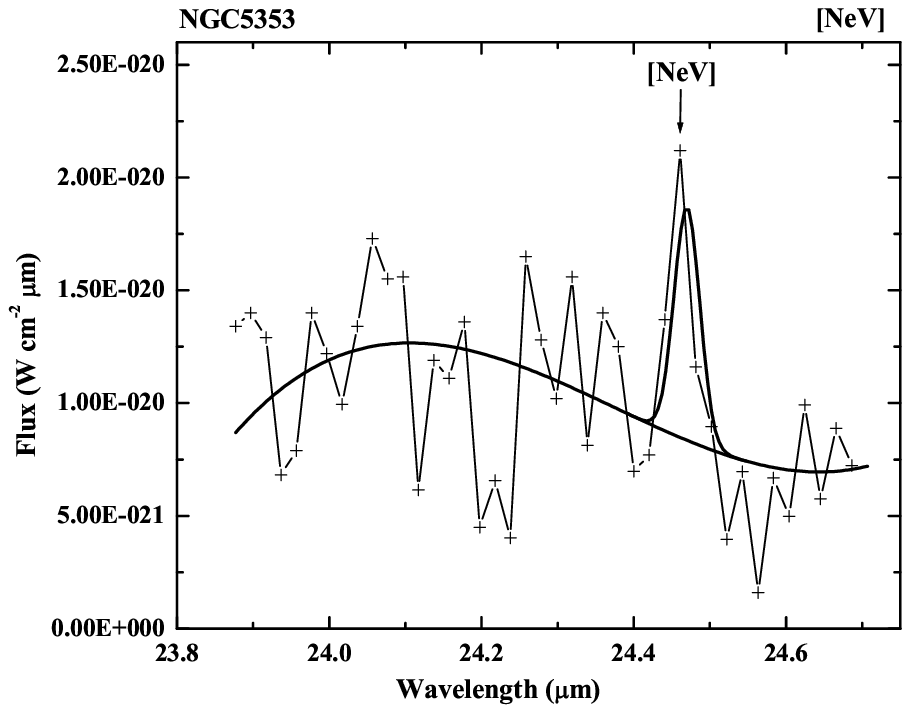} &
  \includegraphics[width=0.45\textwidth]{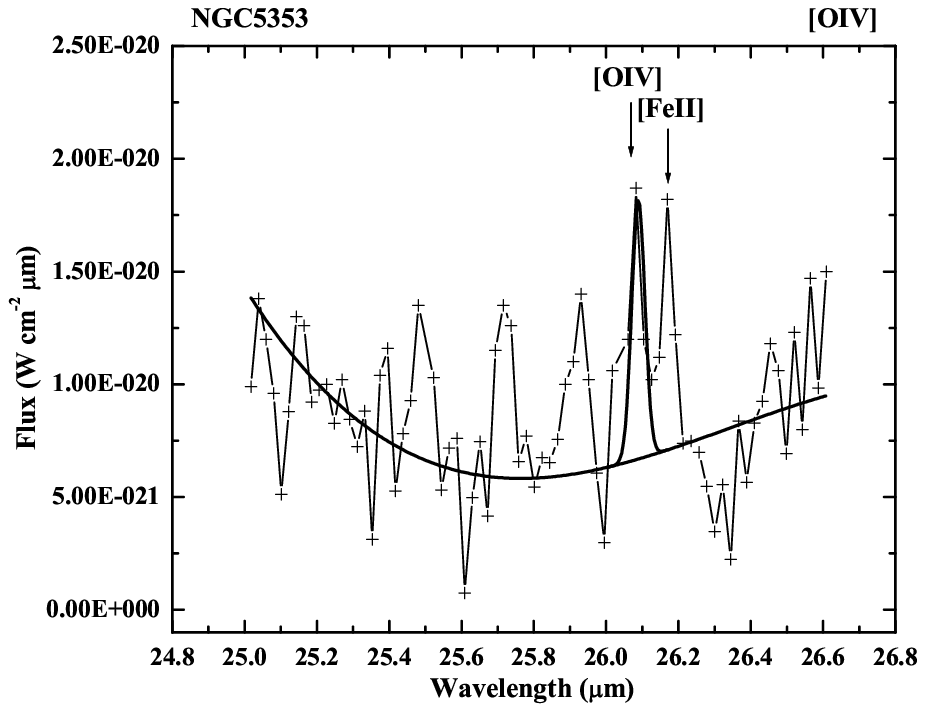} \\
\includegraphics[width=0.45\textwidth]{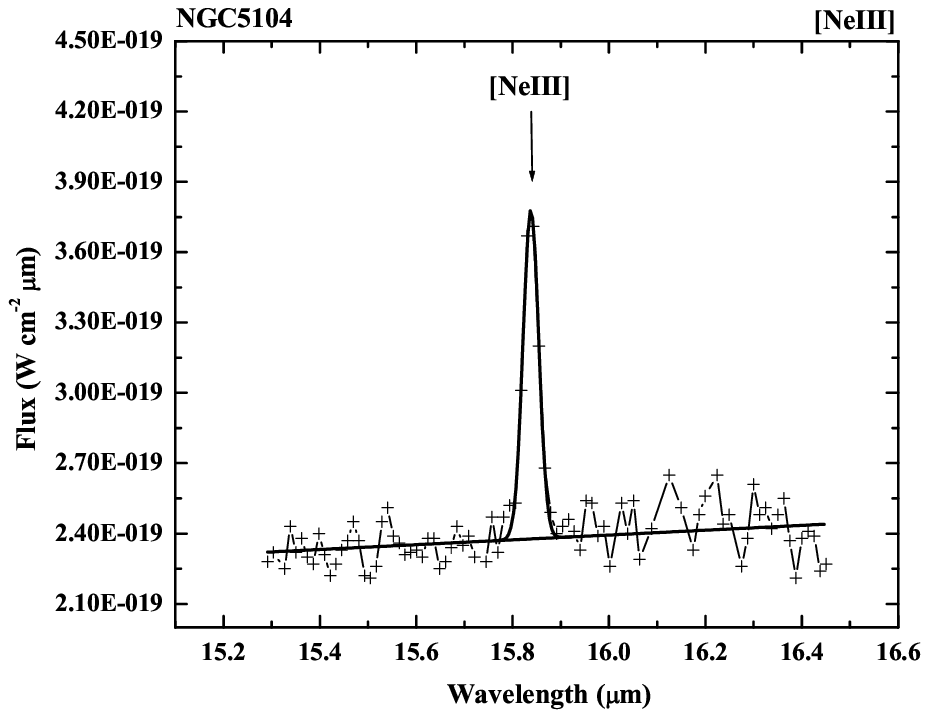} &
 \includegraphics[width=0.45\textwidth]{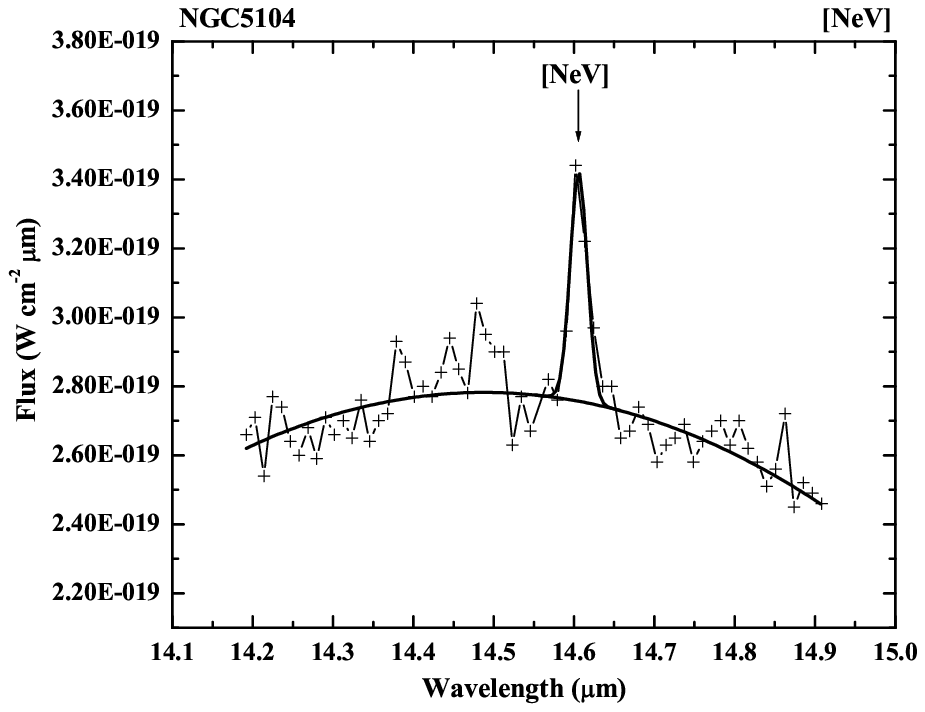} \\
\multicolumn{2}{c}{Fig. 2-- Spectra Continued.}\\
 \end{tabular}
\end{center}
\clearpage
\begin{center}
\begin{tabular}{cc}
\includegraphics[width=0.45\textwidth]{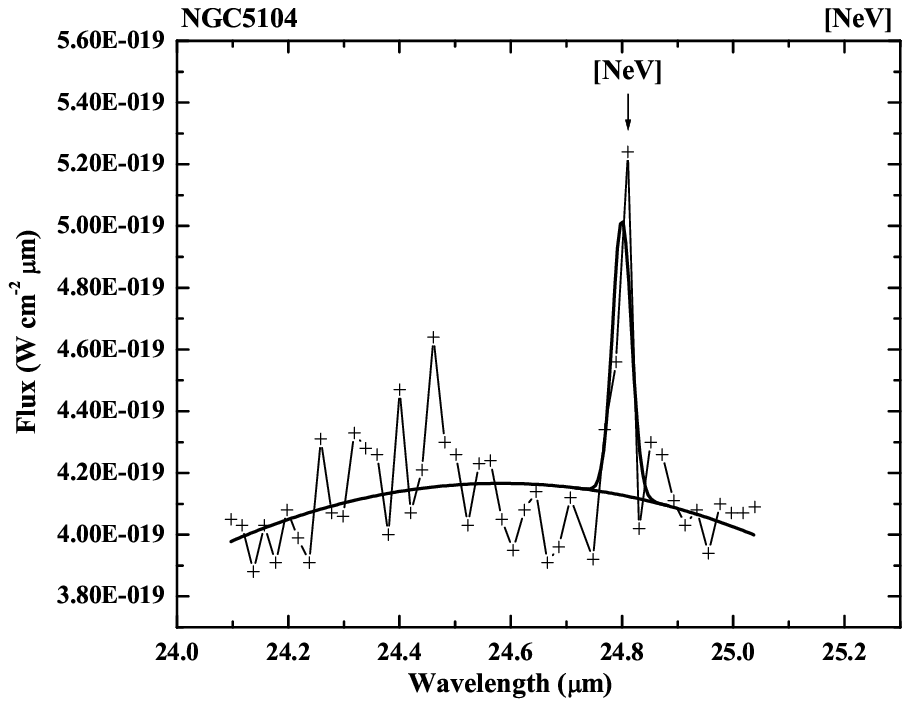} &
 \includegraphics[width=0.45\textwidth]{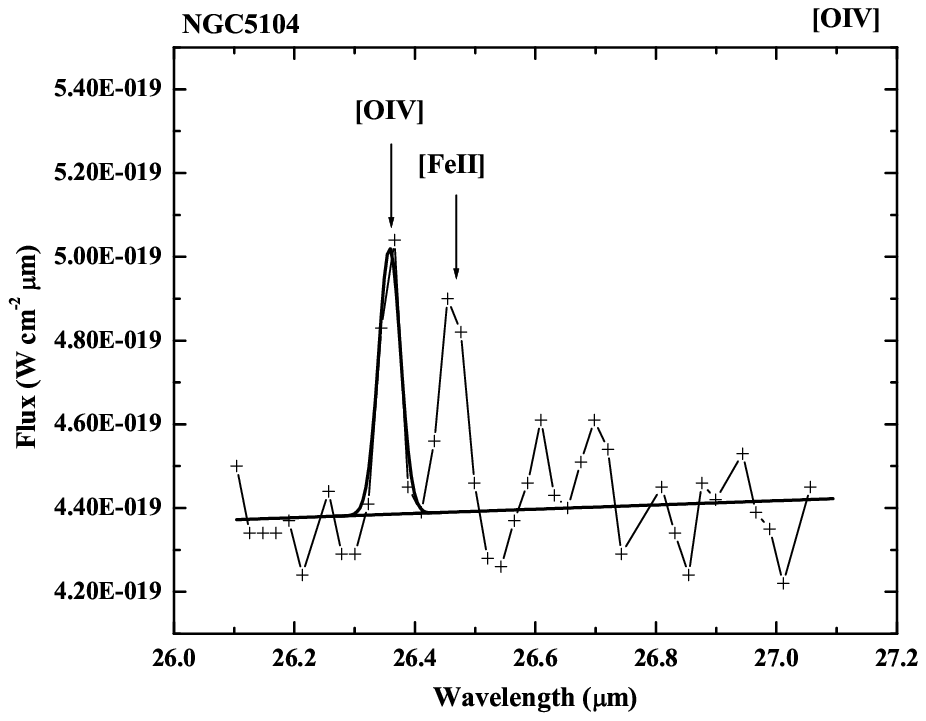} \\
\includegraphics[width=0.45\textwidth]{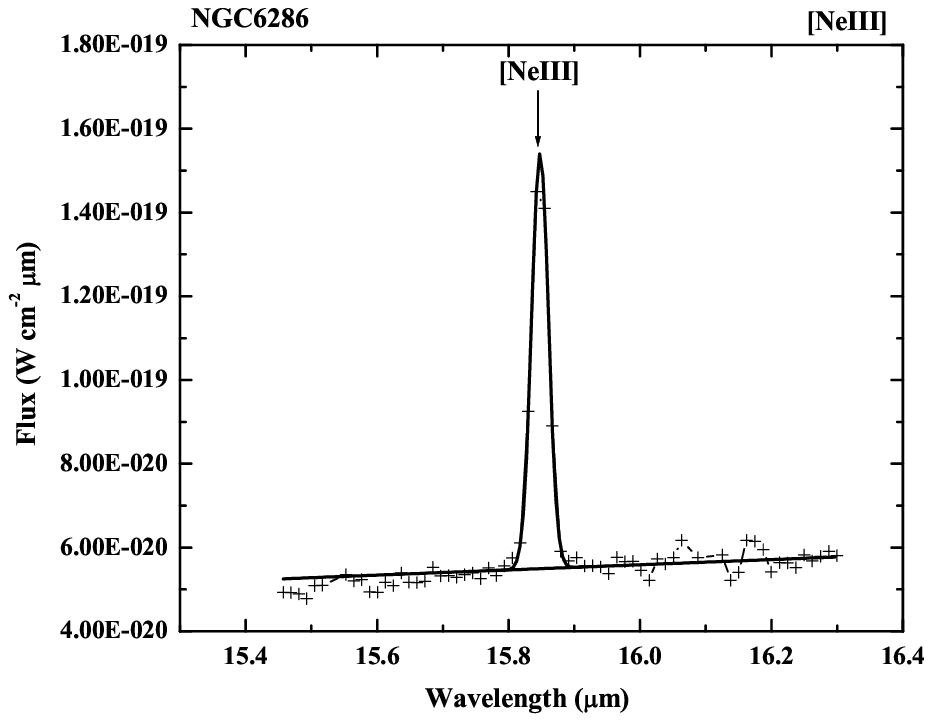} &
 \includegraphics[width=0.45\textwidth]{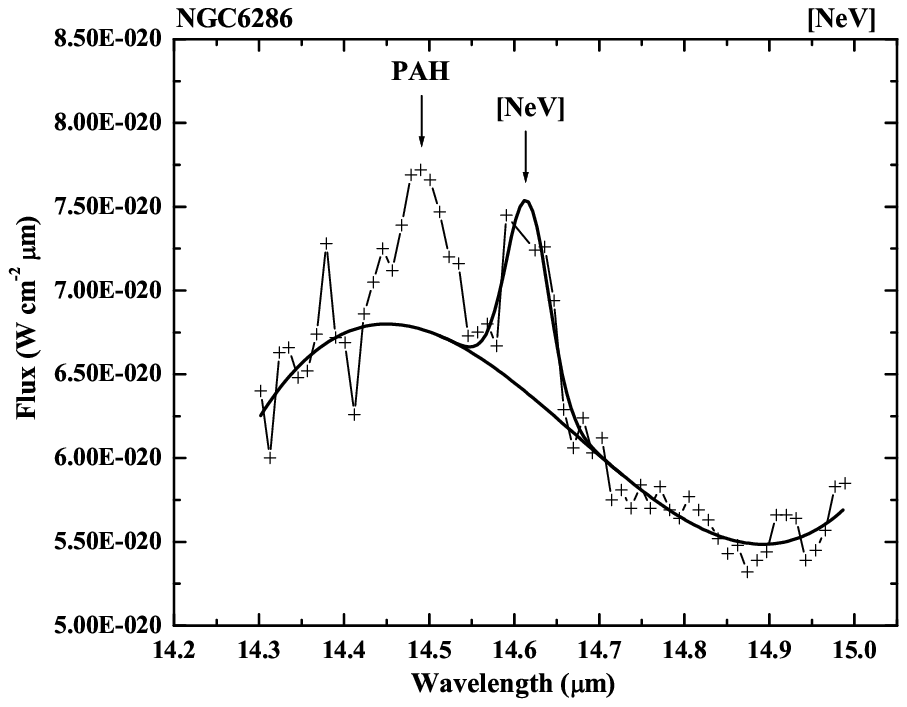} \\
\includegraphics[width=0.45\textwidth]{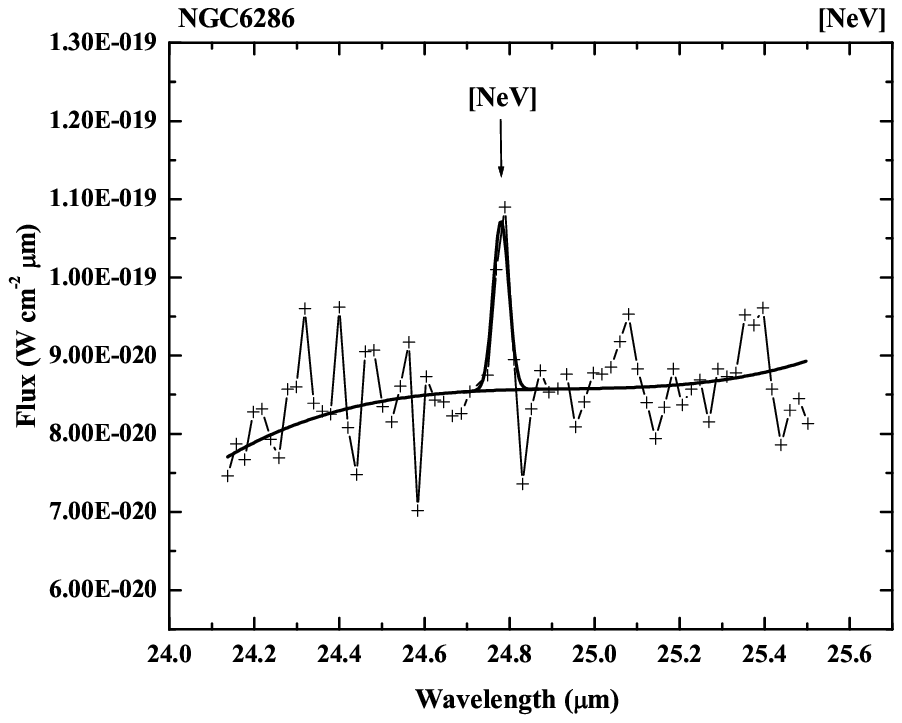} &
 \includegraphics[width=0.45\textwidth]{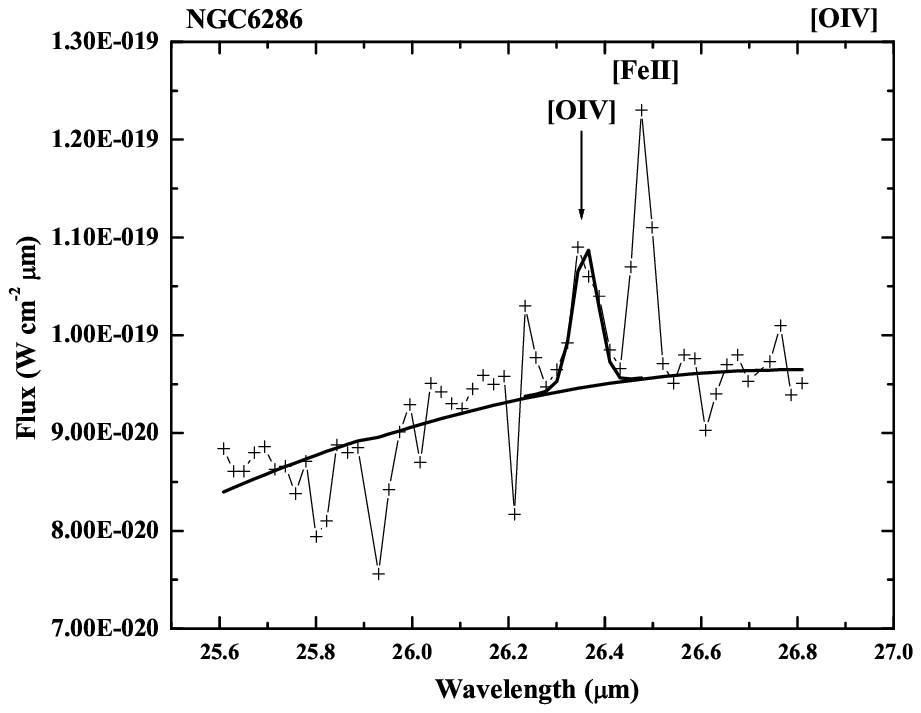} \\
\multicolumn{2}{c}{Fig. 2-- Spectra Continued.}\\
 \end{tabular}
\end{center}
\clearpage
\begin{center}
\begin{tabular}{cc}
\includegraphics[width=0.45\textwidth]{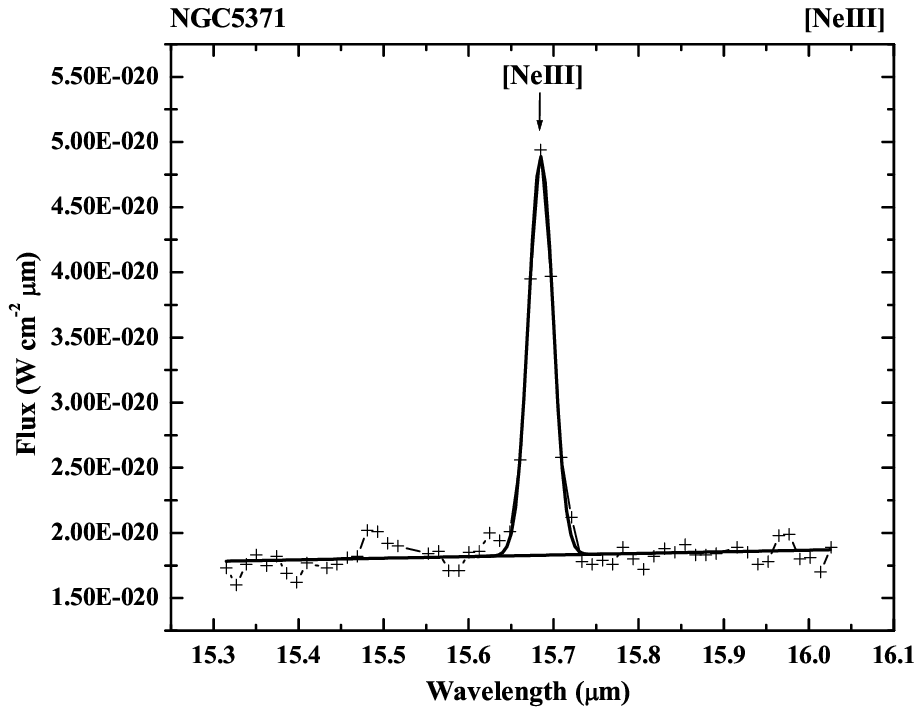} &
  \includegraphics[width=0.45\textwidth]{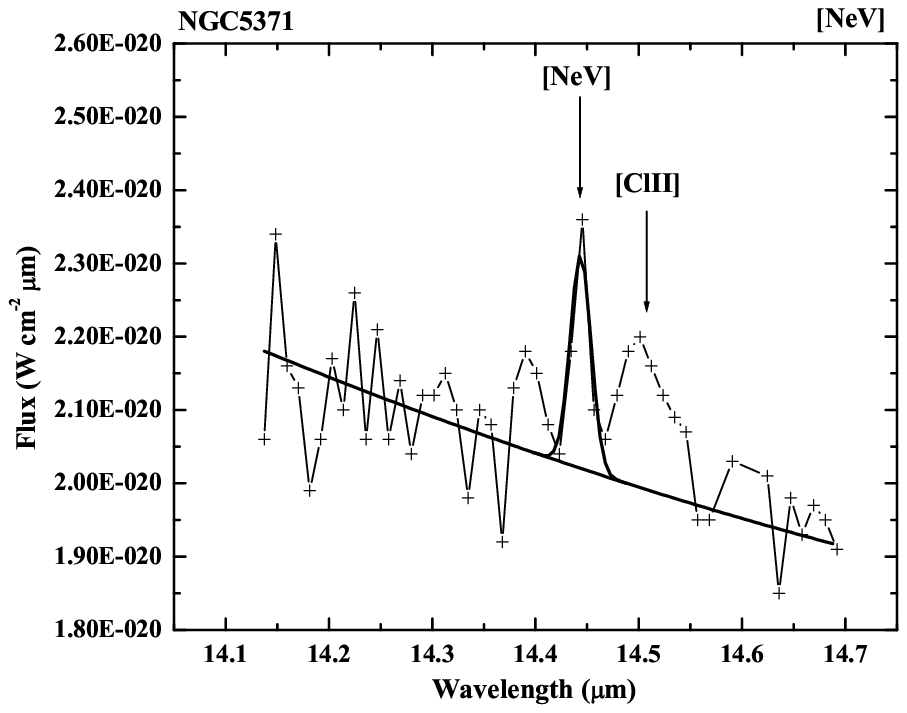} \\
  \includegraphics[width=0.45\textwidth]{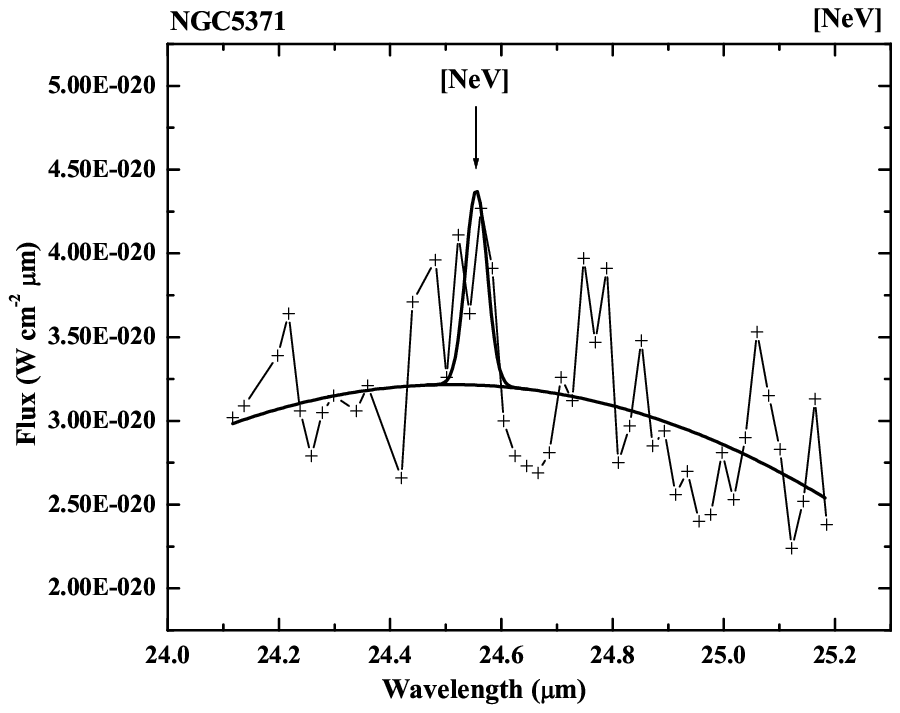} &
 \includegraphics[width=0.45\textwidth]{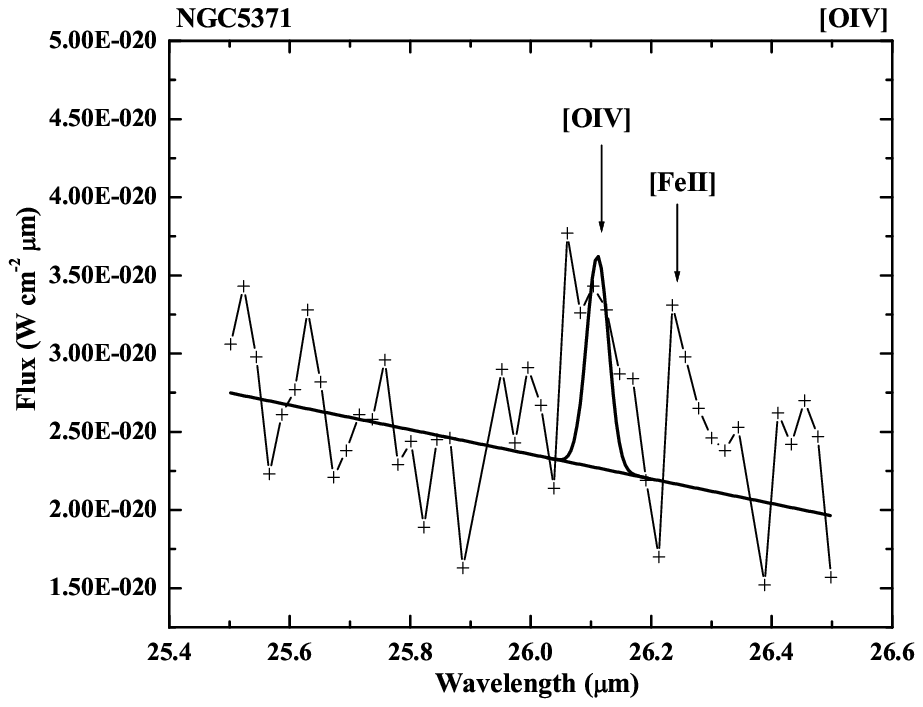} \\
\includegraphics[width=0.45\textwidth]{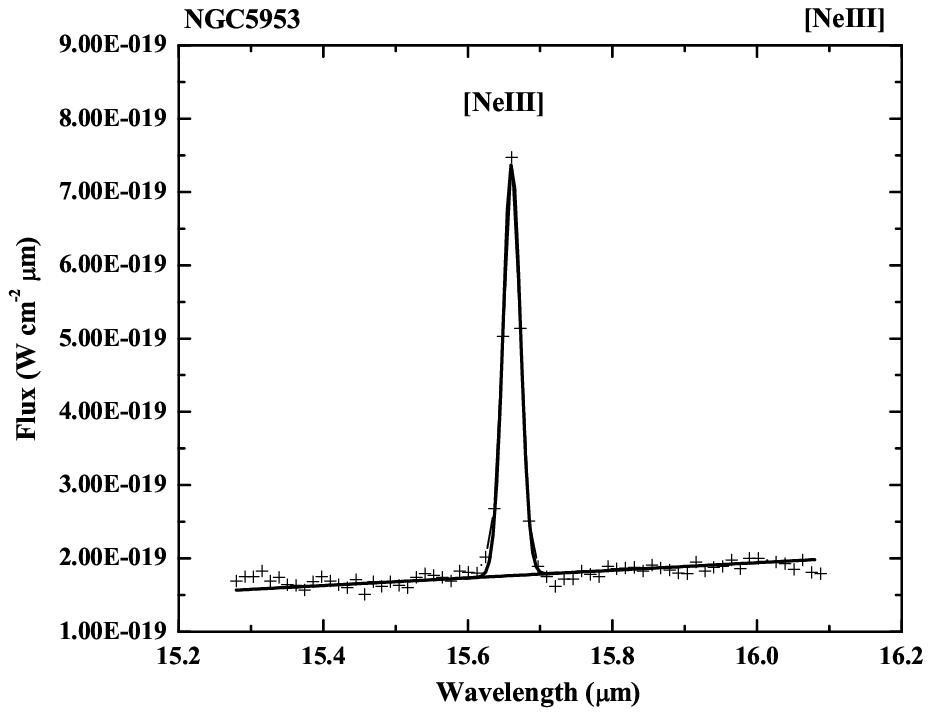} &
  \includegraphics[width=0.45\textwidth]{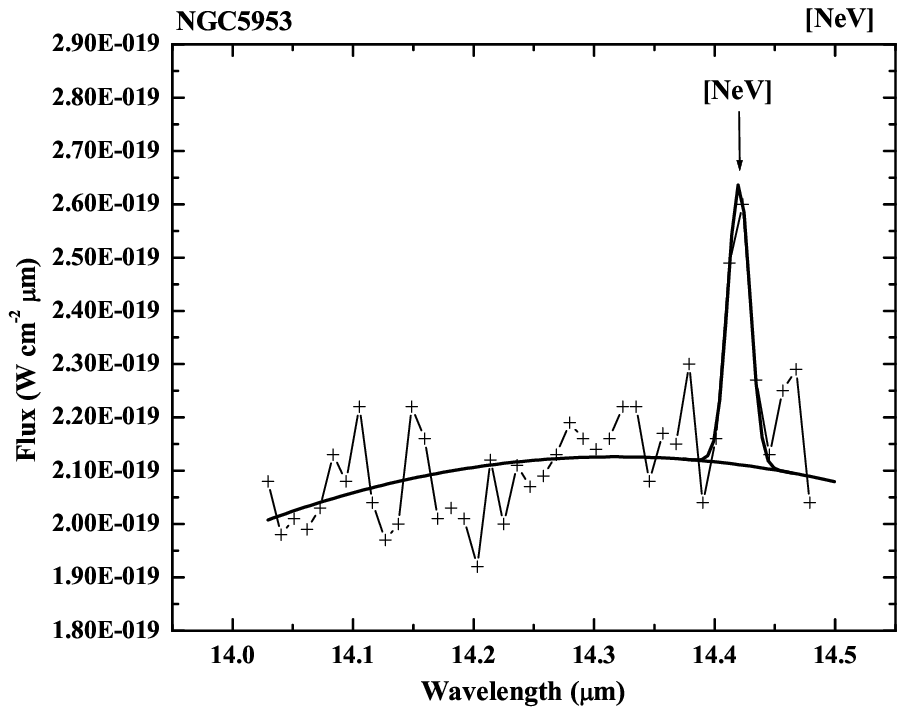} \\
\multicolumn{2}{c}{Fig. 2-- Spectra Continued.}\\
 \end{tabular}
\end{center}
\clearpage
\begin{center}
\begin{tabular}{cc}
 \includegraphics[width=0.45\textwidth]{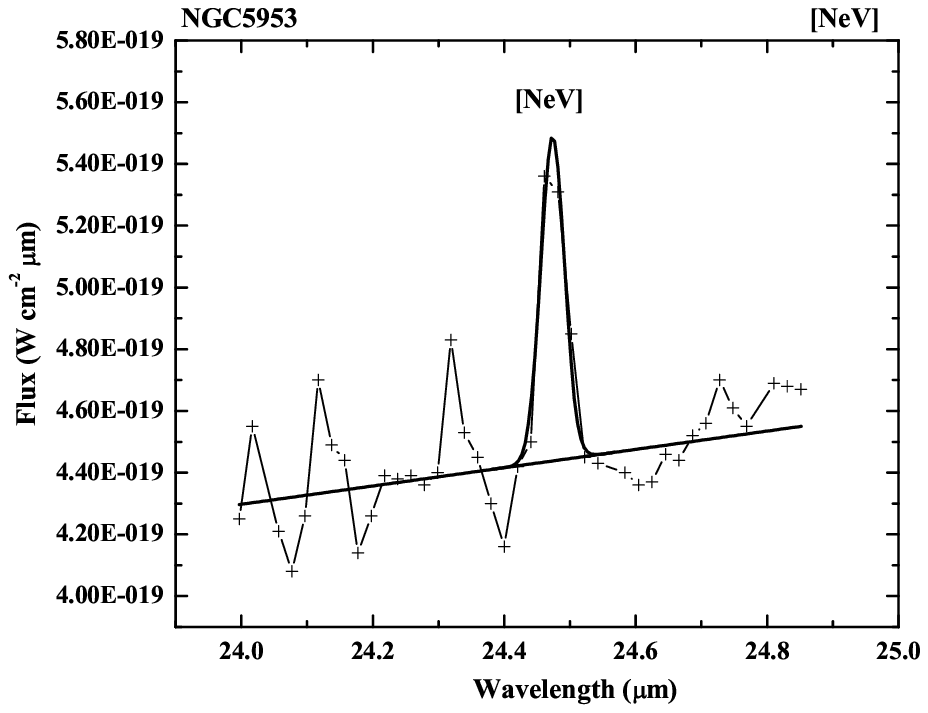} &
 \includegraphics[width=0.45\textwidth]{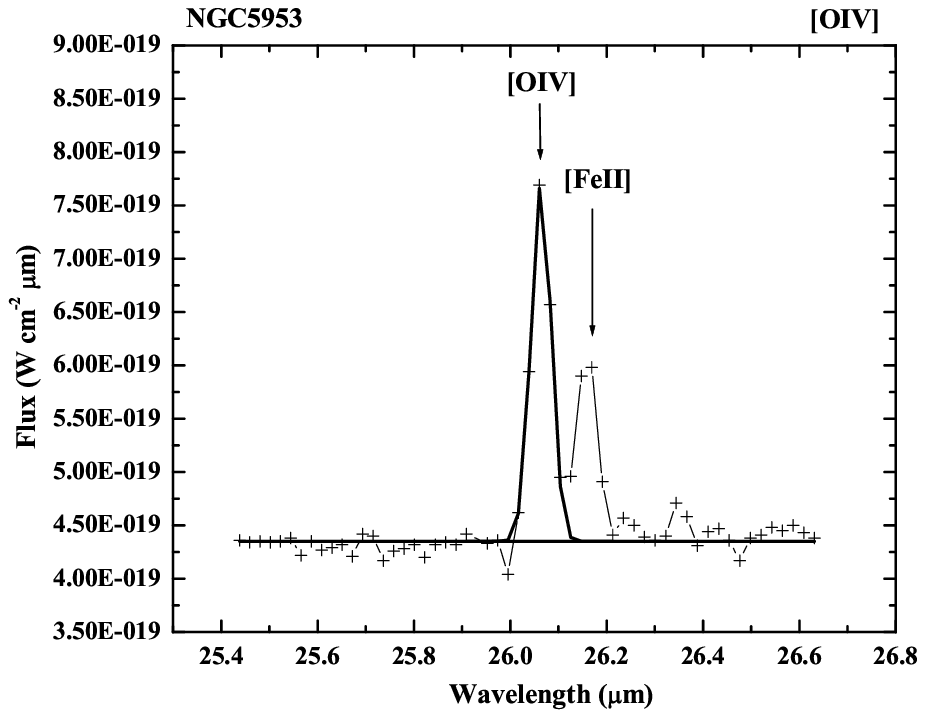} \\
\includegraphics[width=0.45\textwidth]{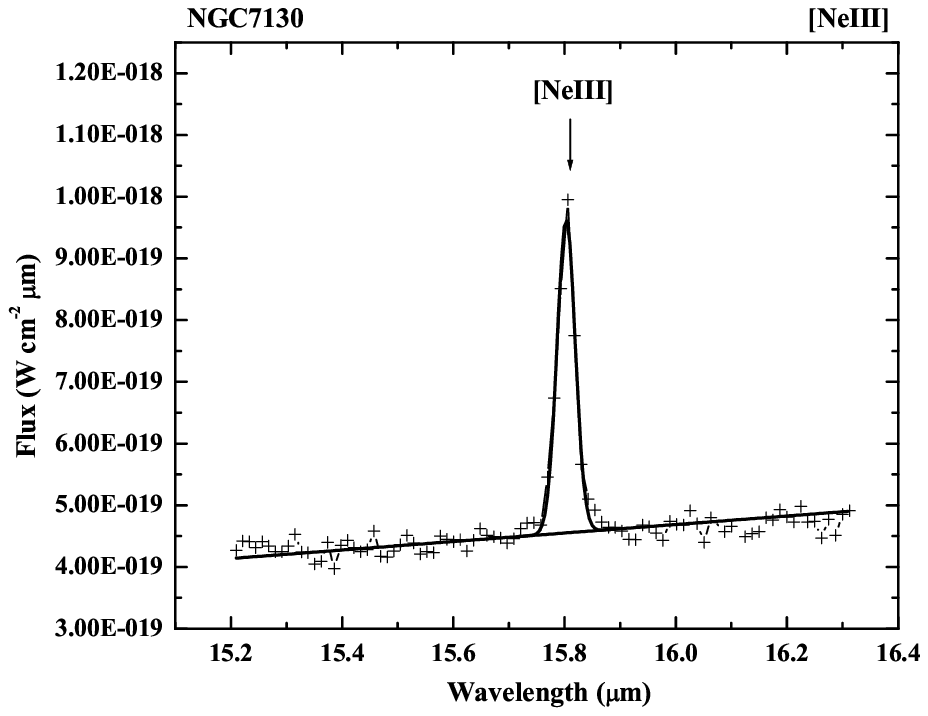} &
 \includegraphics[width=0.45\textwidth]{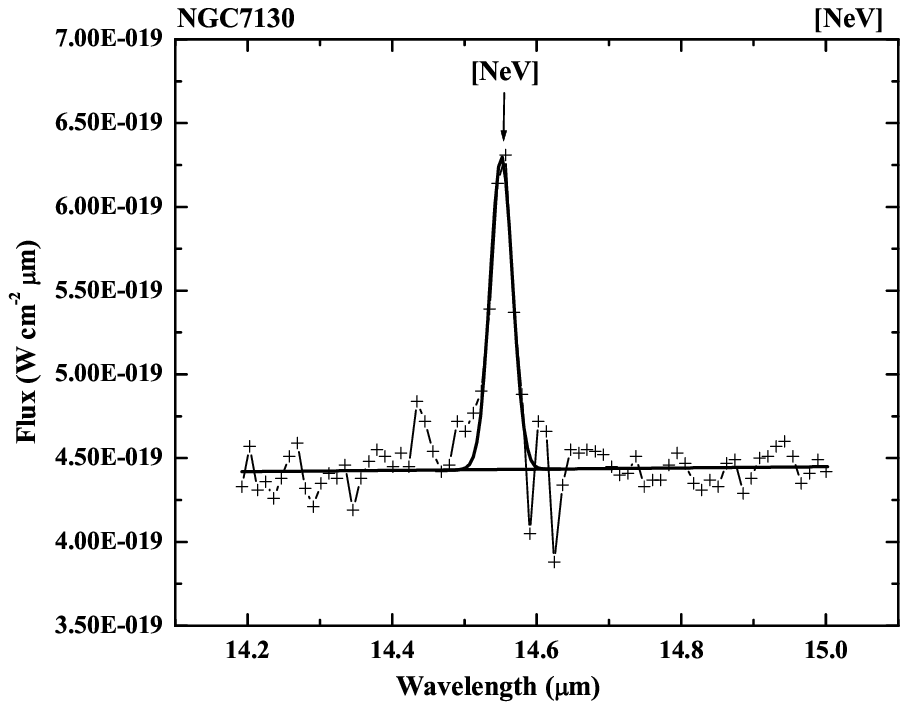}\\
\includegraphics[width=0.45\textwidth]{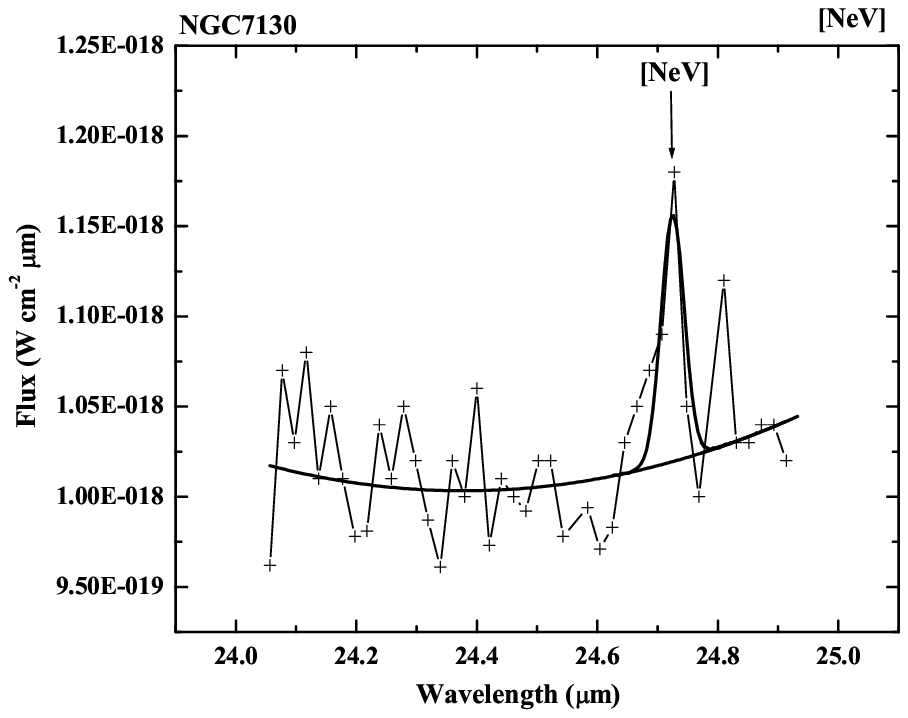} &
 \includegraphics[width=0.45\textwidth]{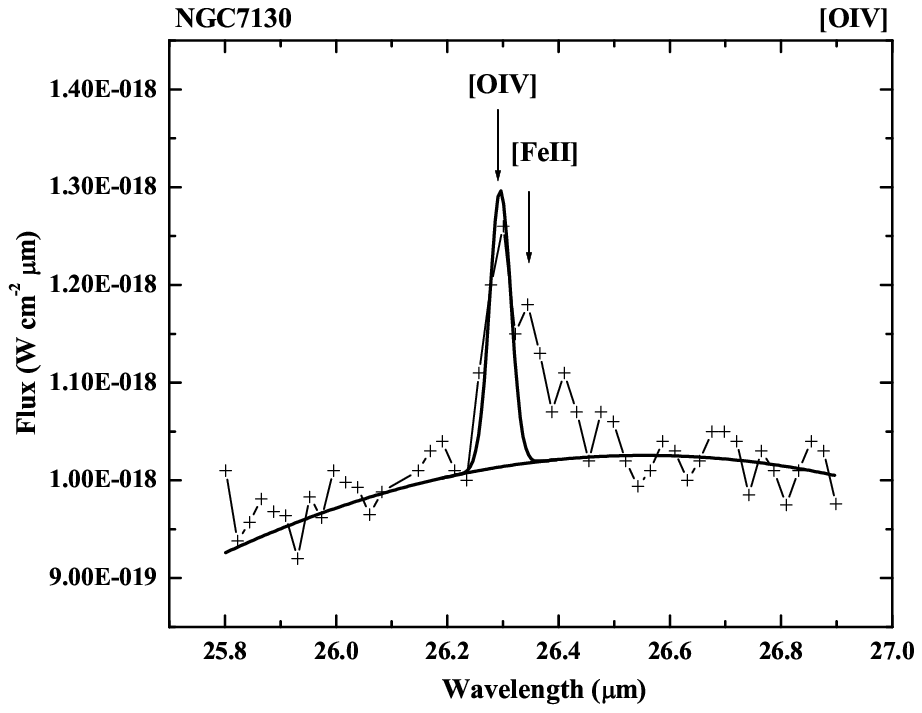} \\
 \multicolumn{2}{c}{Fig. 2-- Spectra Continued.}\\
 \end{tabular}
\end{center}
\clearpage
\begin{center}
\begin{tabular}{cc}
\includegraphics[width=0.45\textwidth]{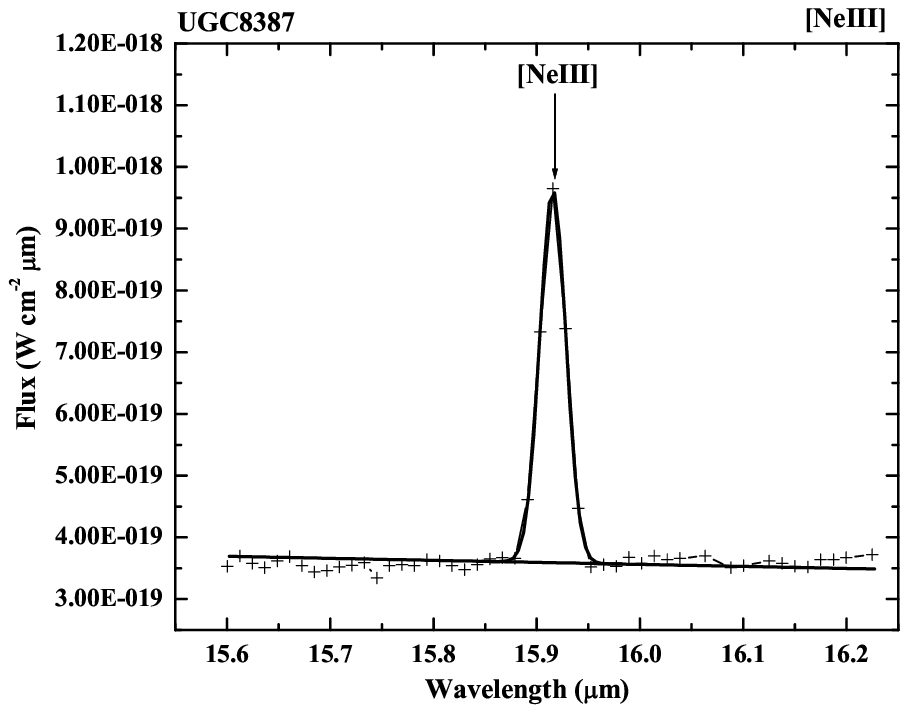} &
  \includegraphics[width=0.45\textwidth]{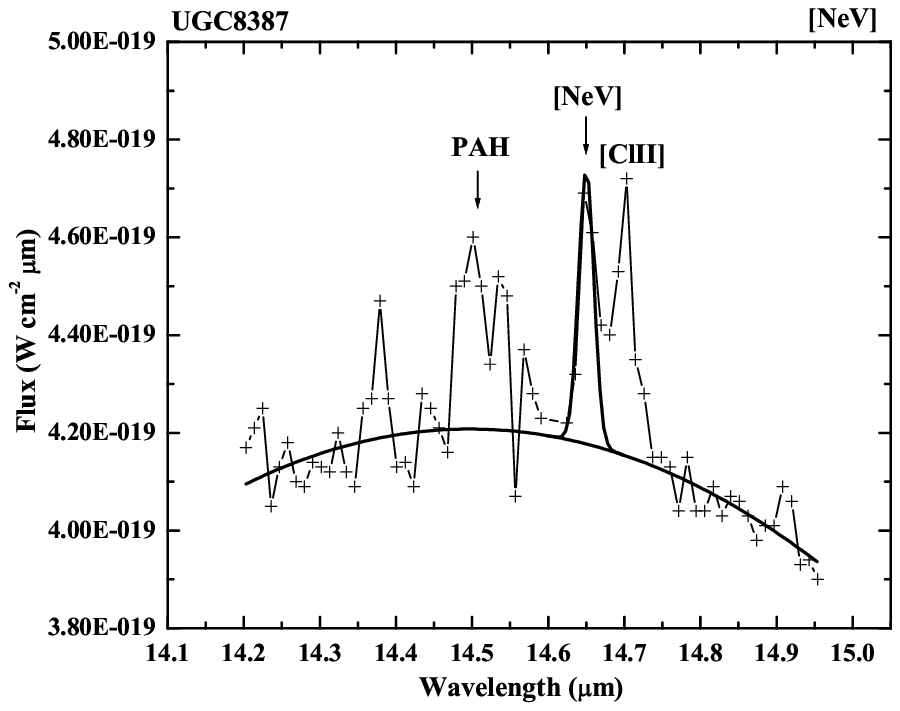} \\
  \includegraphics[width=0.45\textwidth]{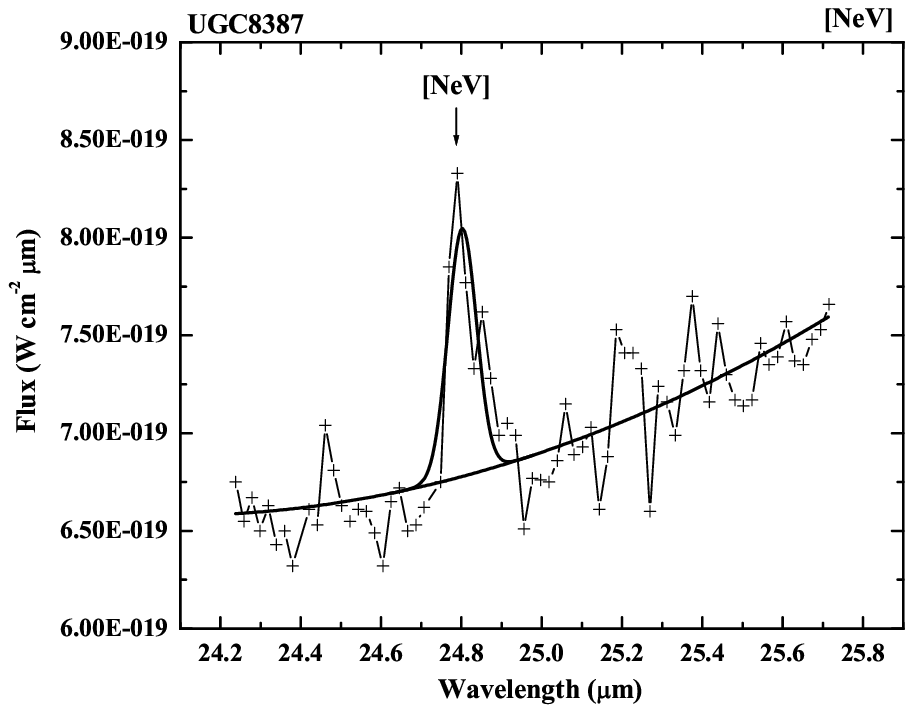} &
  \includegraphics[width=0.45\textwidth]{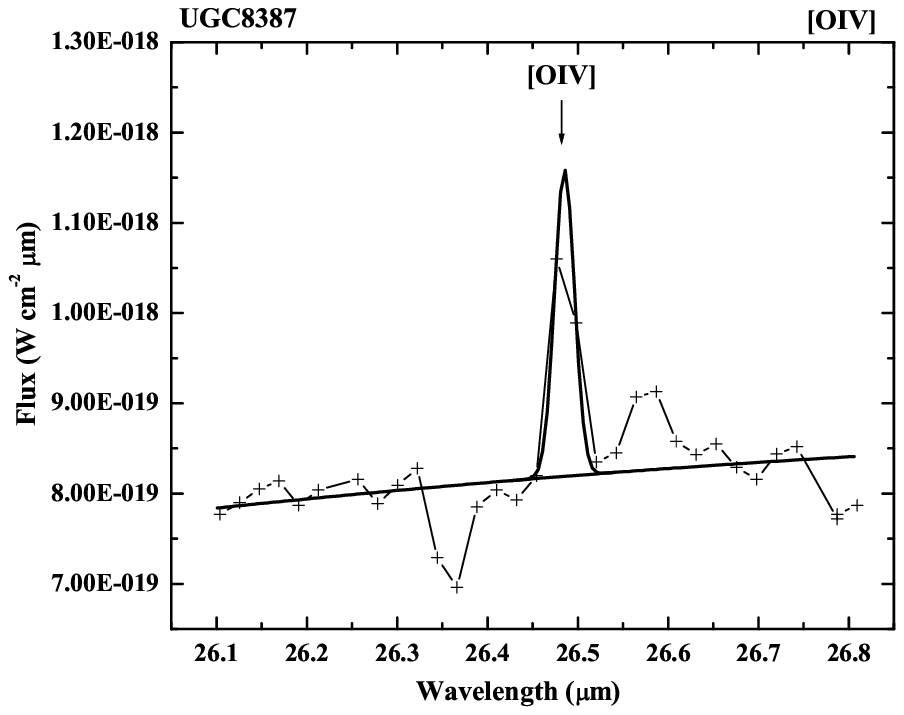} \\
 \includegraphics[width=0.45\textwidth]{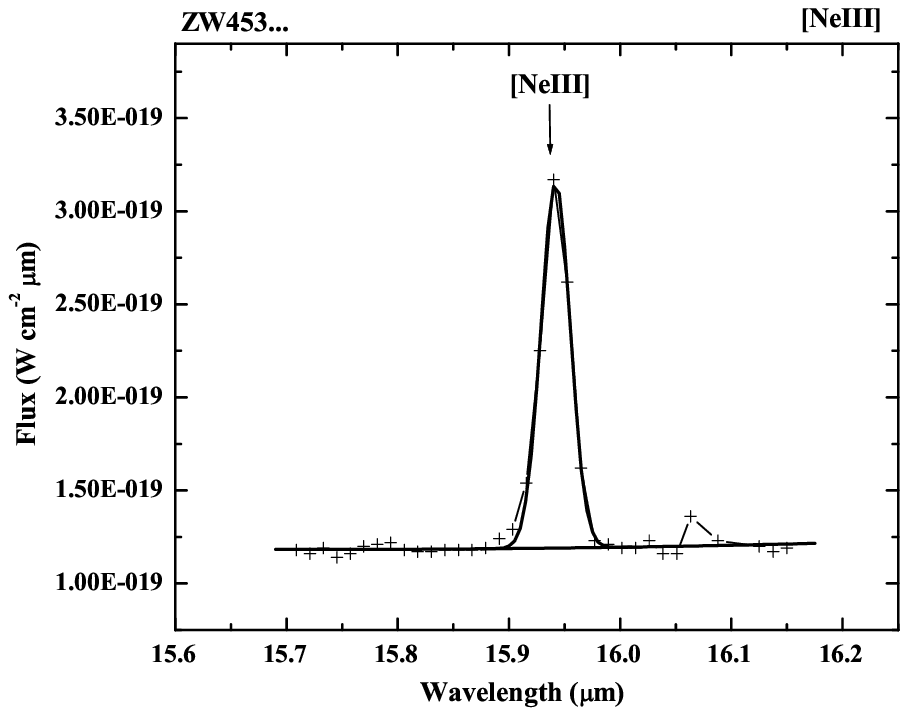} &
  \includegraphics[width=0.45\textwidth]{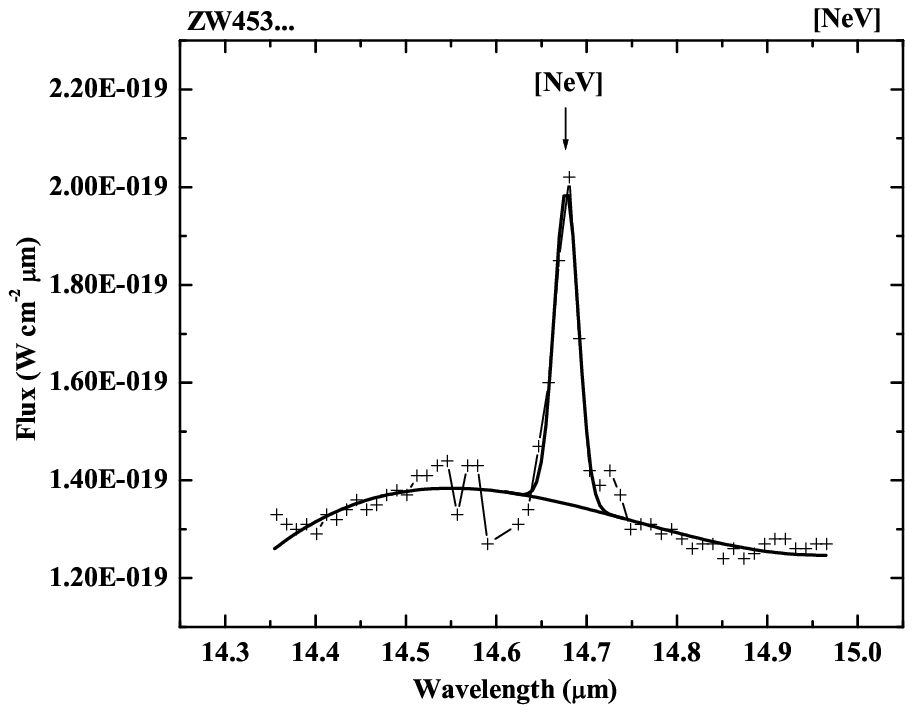} \\
\multicolumn{2}{c}{Fig. 2-- Spectra Continued.}\\
 \end{tabular}
\end{center}
\clearpage
\begin{figure*}[h]
\begin{center}
\begin{tabular}{cc}
 \includegraphics[width=0.45\textwidth]{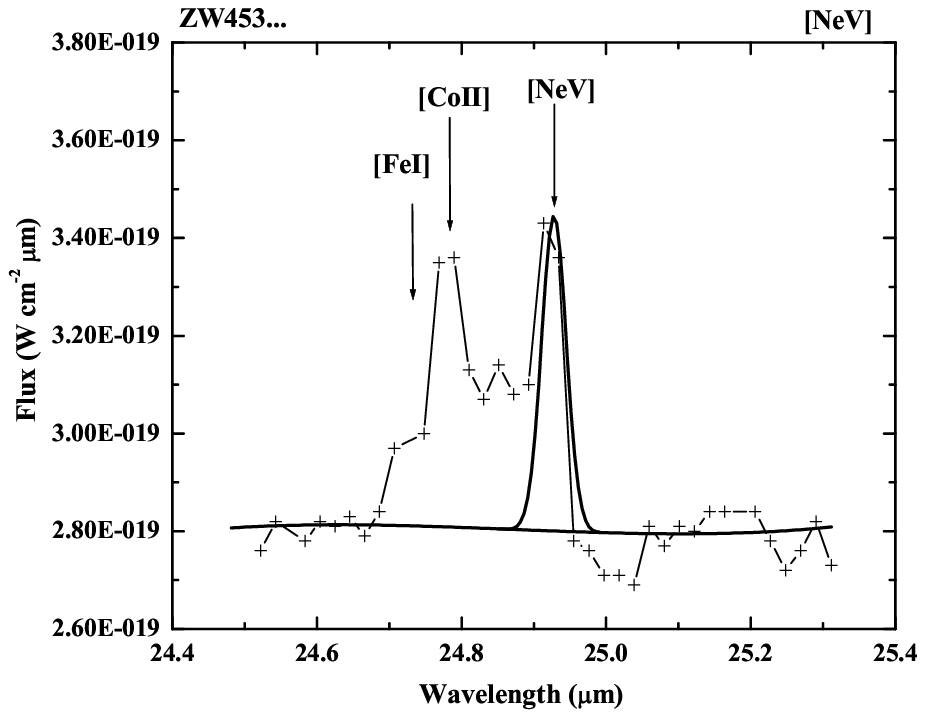} &
 \includegraphics[width=0.45\textwidth]{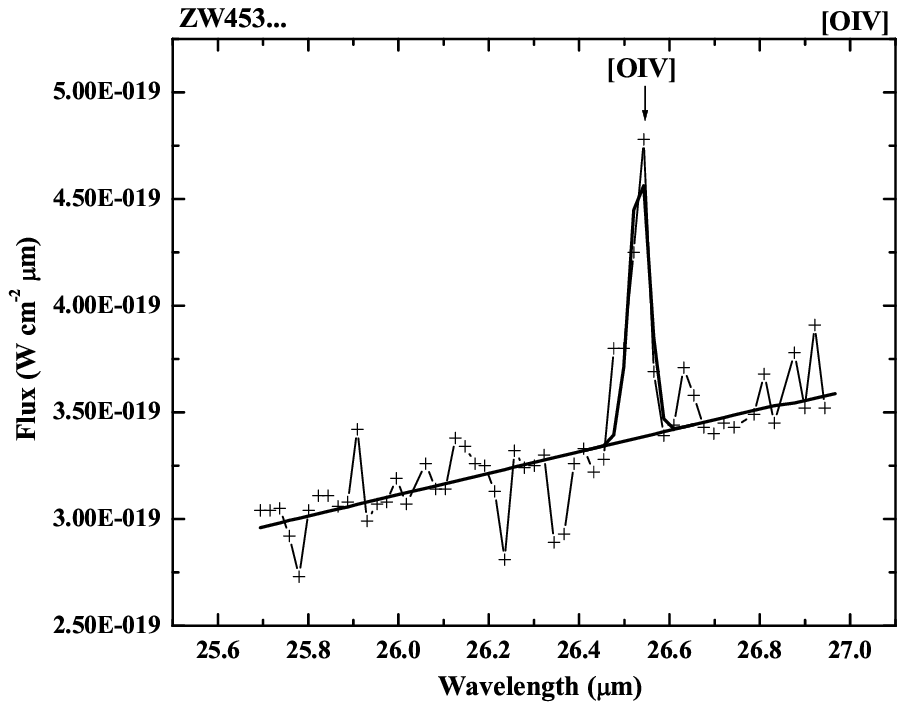}\\
\multicolumn{2}{c}{Fig. 2-- Spectra Continued.}\\
 \end{tabular}
\end{center}
\end{figure*}

In Tables 3 and 4 we list the line fluxes, statistical errors, and upper limits from the SH and LH observations for the [NeII] 12$\mu$m, [NeV] 14$\mu$m, [NeIII] 15$\mu$m,  [NeV] 24$\mu$m and the [OIV] 26$\mu$m lines.  In all cases detections were defined when the line flux was at least 3$\sigma$.  For the absolute photometric flux uncertainty we conservatively adopt 15\%, based on the assessed values given by the {\it Spitzer} Science Center (SSC) over the lifetime of the mission.\footnote[3]{See {\it Spitzer} Observers Manual, Chapter 7, (http://ssc.spitzer.caltech.edu/documents/som/som8.0.irs.pdf and IRS Data Handbook (http://ssc.spitzer.caltech.edu/irs/dh/dh31\_v1.pdf, Chapter 7.2 }.   The dominant component of the total calibration error arises from the uncertainty in the stellar models used in calibration at mid-IR wavelengths and is systematic rather than Gaussian in nature.  For all galaxies with previously published fluxes, we list the published flux values and the references in the column explanation.  Our values agree with the published values to within the adopted calibration error of 15\%.  These differences likely arise from the differences in the pipeline used for preprocessing.

\section{AGN Detection Rate in LINERs Based on Mid-IR Diagnostics }

The Mid-IR band is abundant with emission lines from ions with a wide range of ionization potentials.   Most of these lines have relatively low ionization potentials (IP, [NeII] 12.8$\mu$m, IP= 21.6 eV; [NeIII] 15.6$\mu$m, IP =41 eV; [SIII] 18.7$\mu$m and 33.5$\mu$m, IP=23.3 eV) and are characteristic of prototypical starbursts as well as AGN (Genzel et al. 1998, Alexander et al. 1999, Lutz 2002, Sturm et al. 2002, Satyapal et al. 2004).  The most powerful starburst can even produce emission from the [OIV] 25.89$\mu$m line (IP = 55eV, e.g. M82, Verma et al. 2003), a high ionization line that is generally very prominent in galaxies hosting AGN (Genzel et al. 1998, Sturm et al. 2002).  However, with an ionization potential of 97eV, the [NeV] 14.3 and 24.3$\mu$m lines are not readily produced by the HII regions surrounding young stars--the dominant energy source in starbursts.  This is because even hot massive stars emit very few photons with energy sufficient for the production of Ne$^{4+}$.  In light of this, we use detection of the mid-IR [NeV] 14 \& 24 $\mu$m emission lines as means to uncover buried AGNs in this sample of LINERs.

We detect [NeV] emission in 26 of the 67 galaxies in this sample, yielding an AGN detection rate in LINERs of 39\%.  The fluxes of the 14 and 24$\mu$m [NeV] detections are given in Tables 3 and 4.  The [NeV] spectra, as well as the [NeIII] 15.6$\mu$m and [OIV]$\mu$m spectra, when available, are shown in Figure 2.  Because the sample comprises a random assortment of archival observations with variable S/N and, in some cases, limited sensitivity, the AGN detection rate of 39\% could increase with more sensitive observations.  We investigated this possibility by exploring the dependence of the detection rate on the sensitivity of the observations.

Figure 3 shows the detection rate as a function of the [NeV]14$\mu$m 3$\sigma$ sensitivity.  Some of the galaxies in the sample show [NeV]24$\mu$m emission but not 14$\mu$m emission.  In these cases, the 14 $\mu$m [NeV] flux is estimated from the 24 $\mu$m [NeV] flux using the relationship between the two emission line fluxes found in standard AGNs (Satyapal et al. 2008).  We calculated the detection rate assuming a limiting [NeV]14$\mu$m line sensitivity of log(L[NeV]$_{14}$) = 38 erg s$^{-1}$, the lowest luminosity line detected in a sample of standard AGNs from Dudik et al. (2007).  The detection rate for a log(L[NeV]$_{14}$) sensitivity of $\sim$ 38 erg s$^{-1}$ or higher is 33\% (13 out of 39 galaxies), less than the over-all detection rate for the entire sample.  This suggests that the detection rate would not increase given more sensitive observations.  

\begin{figure*}[htbp]
\begin{center}
 \includegraphics[width=0.60\textwidth]{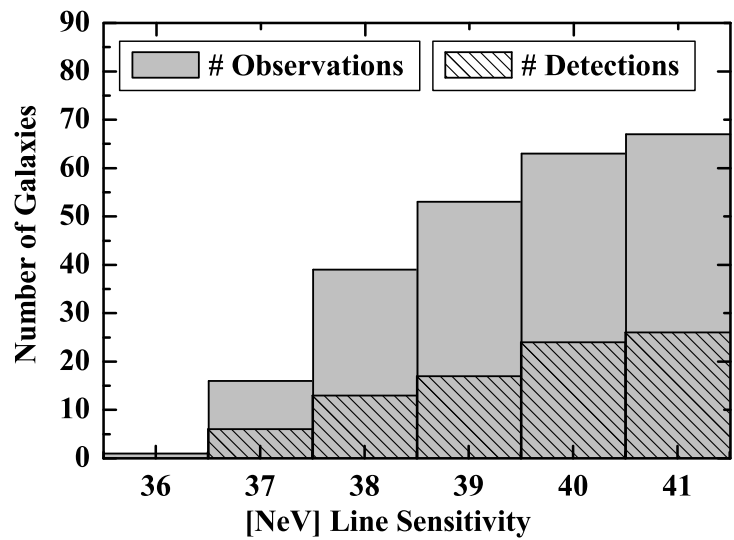}
\end{center}
\caption[]{ Detection Rate as a function of [NeV] Line Sensitivity.  We find a [NeV] detection rate of 39\%.   At a luminosity threshold of 10$^{38}$ erg s$^{-1}$ we find a detection rate of no more than 33\% implying that better sensitivity observations would not yield a higher detection rate.   }
\end{figure*}

In Figures 4 and 5 we examined the dependence of the AGN detection rate on Hubble Type and FIR luminosity.  For some galaxies the Hubble type could not be determined because the available observations lacked sufficient spatial resolution to determine the classification.  These galaxies are denoted by a question mark in Figure 4.  Though the sample is incomplete, prohibiting any significant statistical analysis, the detection rate is highest in mergers and lowest in early-type galaxies.  We note that Ho et al. (1997) find the majority of AGN (36 of 46) detected using broad H$\alpha$ were in early type (E-Sb) galaxies.  In fact, none of the objects with broad H$\alpha$ were in merging systems.  Similarly Nagar et al. (2002) find that 15 of 16 definitive AGN detections in LINERs were in massive elliptical galaxies. On the other hand, both the Ho et al. (1997) and the Nagar et al. (2002) samples stem from the Palomar survey of bright galaxies, preferentially excluding the most FIR-luminous galaxies, which are invariably mergers.  In addition to obscuration of the optical AGN signatures, this bias in sample selection may be partially responsible for the discrepancy in the AGN detection rate in mergers between the two samples.

\begin{figure*}[htbp]
\begin{center}
 \includegraphics[width=0.60\textwidth]{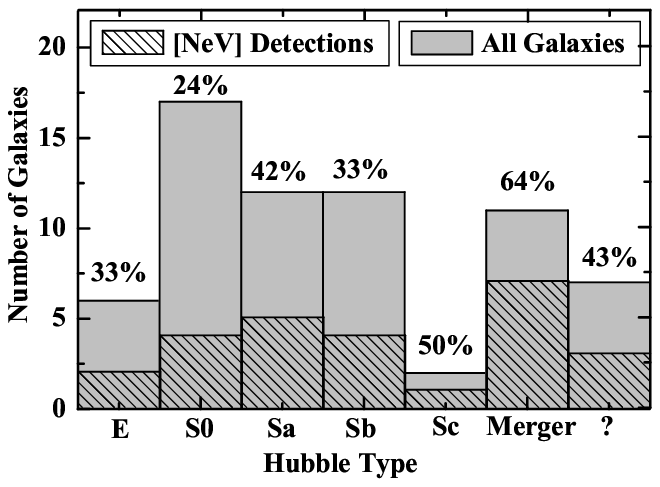}
\end{center}
\caption[]{ [NeV] Detection rate as a function of Hubble type.  For some galaxies the spatial resolution was insufficient to determine Hubble Type.  These galaxies are denoted by a question mark.  Mergers have the highest detection rate of all Hubble classes. }
\end{figure*}

\begin{figure*}[htbp]
\begin{center}
  \includegraphics[width=0.60\textwidth]{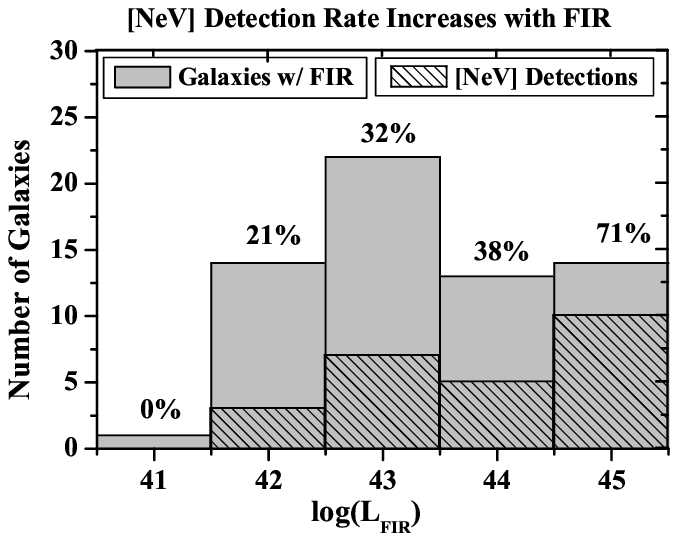}
\end{center}
\caption[]{ [NeV] Detection rate as a function of FIR luminosity.}
\end{figure*}

Figure 5 shows that the [NeV] AGN detection rate increases with L$_{FIR}$.  As pointed out in Section 2, our IR-bright population includes galaxies that are on average more distant than the IR-faint population.  This could induce a bias in that the IR-bright population in our sample can include a greater number of star-formation dominated galaxies than the IR-faint population.  However, the fraction of LINERs that are AGN increases with FIR luminosity, a result that is independent of any distance-induced bias. We note for Figure 5 that our sample is deficient in the most FIR luminous LINERs compared with the comprehensive Carrillo sample (see also Section 2).  The inclusion of more FIR luminous galaxies (with L$_{FIR}$ > 10$^{44}$ and 10$^{45}$ ergs s$^{-1}$ ) may alter the detection rates at the highest L$_{FIR}$.

 \section{Mid-IR vs. Optical, X-ray, and Radio AGN Diagnostics}

\subsection{Optical Observations} 

In the conventional picture of an AGN, the broad line region is thought to exist within a small region interior to a dusty molecular torus.  When viewed above the confines of the torus (face-on), Doppler broadened (FWHM exceeding 1000 km s$^{-1}$) Balmer emission lines are visible in the optical spectrum.  These broad permitted lines are the most widely used AGN indicators used in optical studies of LINERs.  The presence or absence of broad H$\alpha$ has been established in the literature for 84\% (56 of 67) of the galaxies in our sample and results in an optical detection rate of 30\%.  Broad H$\alpha$ results taken from the literature are given in Tables 5 and 6 along with the corresponding [NeV] detection information.  

As can be seen from Tables 5 and 6, there is very little consistency between the detection of [NeV] and the presence of Broad H$\alpha$ in the optical spectrum.  In fact 13 galaxies show [NeV] detections, but no broad H$\alpha$ line.  In addition, 11 show broad H$\alpha$ but no [NeV].   Using both IR and optical AGN detection diagnostics the combined comparative AGN detection rate is 54\% (30 of 56 galaxies). 
\begin{table*}
\fontsize{9pt}{10pt}\selectfont
\begin{center}
\caption{Multiwavelength Statistics: H97}
\begin{tabular}{lcccccccc}
\hline
\multicolumn{1}{c}{Galaxy} & \multicolumn{1}{c}{[NeV]} & Broad & Optical & X-ray & log(L$_X$) & X-ray & Radio & Radio\\

\multicolumn{1}{c}{Name} & \multicolumn{1}{c}{Detect?} & H$\alpha$? & Ref.
 & Detect? & erg s$^{-1}$ & Ref.& Detect? & Ref. \\

\multicolumn{1}{c}{(1)} & \multicolumn{1}{c}{(2)} & (3) & (4) & (5) & (6) & (7) & (8) & (9)\\ 

\hline
NGC1052  & Yes & Yes & A & AGN & 40.78 & G & $\cdots$ & $\cdots$\\
NGC1055  & No & No & A & nonAGN & $<$37.52 & H & No & S\\
NGC1961  & No & No & A & nonAGN & $\cdots$ & I & No & S\\
NGC266  & No & Yes & A & AGN & 40.88 & J & Yes & S\\
NGC2681  & No & Yes & A & AGN & 38.94 & G & No & S\\
NGC2787  & No & Yes & A & AGN & 38.30 & K & Yes & S\\
NGC2841  & No & No & A & nonAGN & 38.26 & K & Yes  & S\\
NGC3166  & Yes & No & A & $\cdots$ & $\cdots$ & $\cdots$ & No & S\\
NGC3169  & No & No & A & AGN & 41.41 & J & Yes  & S\\
NGC3190  & No & No & A & $\cdots$ & $\cdots$ & $\cdots$ & Yes & S\\
NGC3226  & No & Yes & A & AGN & 39.62 & G & Yes  & S\\
NGC3368  & Yes & No & A & nonAGN & 39.45 & L & No & S\\
NGC3507  & Yes & No & A & nonAGN & 38.31 & G & No & S\\
NGC3521  & Yes & No & A & $\cdots$ & $\cdots$ & $\cdots$ & $\cdots$ & $\cdots$\\
NGC3642  & Yes & Yes & A & $\cdots$ & $\cdots$ & $\cdots$ & No & S\\
NGC3884  & No & Yes & A & $\cdots$ & $\cdots$ & $\cdots$ & $\cdots$ & $\cdots$\\
NGC3998  & No & Yes & A & AGN & 41.67 & M & Yes & S\\
NGC4036  & Yes & Yes & A & $\cdots$ & $\cdots$ & $\cdots$ & No & S\\
NGC404  & No & No & A & nonAGN & 37.32 & L & No & S\\
NGC4143  & No & Yes & A & AGN & 40.04 & J & Yes  & S\\
NGC4203  & No & Yes & A & AGN & 40.08 & K & Yes  & S\\
NGC4261  & No & No & A & AGN & 40.65 & G & Yes & S\\
NGC4278  & Yes & Yes & A & AGN & 40.09 & K & Yes  & S\\
NGC4314  & No & No & A & nonAGN & 38.15 & L & No & S\\
NGC4394  & No & No & A & $\cdots$ & $\cdots$ & $\cdots$ & No & S\\
NGC4438  & Yes & Yes & A & nonAGN & 37.54 & N & No & S\\
NGC4450  & No & Yes & A & AGN & 40.35 & M & Yes  & S\\
NGC4457  & No & No & A & AGN & 39.22 & N & No & S\\
NGC4486  & No & No & A & AGN & 40.52 & L & Yes  & S\\
NGC4548  & No & No & A & nonAGN & 39.05 & N & Yes  & S\\
NGC4594  & No & No & A & AGN & 38.86 & K & $\cdots$ & $\cdots$\\
NGC4596  & No & No & A & nonAGN & 38.65 & G & No & S\\
NGC4736  & Yes & No & A & nonAGN & 38.72 & K & Yes  & S\\
NGC474  & No & No & A & $\cdots$ & $\cdots$ & $\cdots$ & No & S\\
NGC5005  & No & Yes & A & AGN & 39.49 & H & No & S\\
NGC5195  & No & No & A & nonAGN & 37.85 & K & No & S\\
NGC5353  & Yes & No & A & $\cdots$ & $\cdots$ & $\cdots$ & Yes & S\\
NGC5371  & Yes & No & A & $\cdots$ & $\cdots$ & $\cdots$ & No & S\\
NGC5850  & No & No & A & non-AGN & 39.83 & O & No & S\\
NGC5982  & No & No & A & $\cdots$ & $\cdots$ & $\cdots$ & No & S\\
NGC5985  & No & No & A & $\cdots$ & $\cdots$ & $\cdots$ & No & S\\
NGC6500  & No & No & A & AGN & 40.23 & L & Yes & S\\
NGC7626  & No & No & A & nonAGN & 41.51 & P & Yes & S\\
\hline 
\end{tabular}
\end{center}
\tablecomments{Columns Explanation:
Col(1): Common Source Names; 
Col(2): Galaxies with [NeV] Detected;
Col(3): Galaxies with Broad H$\alpha$ detected;
Col(4): Reference for Broad H$\alpha$;
Col(5): Galaxies with AGN-like nuclei in the X-rays;
Col(6): X-ray luminosity in erg s$^{-1}$
Col(7): Reference for X-ray morphology and luminosity. 
Col(8): Galaxies with compact flat spectrum radio cores.  
Col(9): Reference for radio morphology.  
}
\tablerefs{A - H97; B - Veron-Cetty \& Veron 2003; C - Goncalves, Veron-Cetty, \& Veron, 1999; D - Andreasian \& Khachikian 1987; E - Phillips 1983 F - Kewley et al. 2001; G - Gonzalez-Martin et al. 2006; H - Dudik et al. 2005; I - Colbert \& Ptak 2002; J - Terashima et al. 2003; K - Ho et al. 2001; L - Satyapal et al. 2004; M - Terashima et al. 2000; N - Satyapal et al. 2005; O - Georgantopoulos, Georgakakis, \& Koulouridis 2005; P - Philip et al. 2006; Q - Guainazzi, Matt, \& Perola, 2005; R - Teng et al. 2005; S - Nagar, Falcke \& Wilson 2005; T - Baan \& Klockner 2006}
\end{table*}

\begin{table*}
\begin{center}
\caption{Multiwavelength Statistics: V95}
\begin{tabular}{lcccccccc}
 \hline
\multicolumn{1}{c}{Galaxy} & \multicolumn{1}{c}{[NeV]} & Broad & Optical & X-ray & log(L$_X$) & X-ray & Radio & Radio\\

\multicolumn{1}{c}{Name} & \multicolumn{1}{c}{Detect?} & H$\alpha$? & Ref.
 & Detect? & erg s$^{-1}$ & Ref.& Detect? & Ref. \\

\multicolumn{1}{c}{(1)} & \multicolumn{1}{c}{(2)} & (3) & (4) & (5) & (6) & (7) & (8) & (9)\\ 

\hline

UGC556 & No & $\cdots$ & $\cdots$ & $\cdots$ & $\cdots$ & $\cdots$ & $\cdots$ & $\cdots$\\
IR01364-1042 & No & $\cdots$ & $\cdots$ & $\cdots$ & $\cdots$ & $\cdots$ & $\cdots$ & $\cdots$\\
NGC660 & Yes & No & A & nonAGN & 38.52 & H & No & S\\
IIIZw35 & No & No & B & $\cdots$ & $\cdots$ & $\cdots$ & AGN? & T\\
IR02438+2122 & No & $\cdots$ &  & $\cdots$ & $\cdots$ & $\cdots$ & AGN? & T\\
NGC1266 & No & $\cdots$ &  & $\cdots$ & $\cdots$ & $\cdots$ & $\cdots$ & $\cdots$\\
UGC5101 & Yes & Yes  & B & AGN & 40.89 & L & AGN & T\\
NGC4666 & Yes & $\cdots$ &  & nonAGN & $<$38.08 & H & $\cdots$ & $\cdots$\\
UGC8387 & Yes & $\cdots$ &  & $\cdots$ & $\cdots$ & $\cdots$ & $\cdots$ & $\cdots$\\
NGC5104 & Yes & $\cdots$ &  & $\cdots$ & $\cdots$ & $\cdots$ & $\cdots$ & $\cdots$\\
NGC5218 & No & Yes  & B & $\cdots$ & $\cdots$ & $\cdots$ & $\cdots$ & $\cdots$\\
Mrk273 & Yes & No & B & AGN & 42.18 & G & $\cdots$ & $\cdots$\\
Mrk848 & No & No & C & $\cdots$ & $\cdots$ & $\cdots$ & $\cdots$ & $\cdots$\\
NGC5953 & Yes & No & C & nonAGN & $<$38.69 & Q & $\cdots$ & $\cdots$\\
IR15335-0513 & Yes & $\cdots$ &  & $\cdots$ & $\cdots$ & $\cdots$ & $\cdots$ & $\cdots$\\
IR16164-0746 & Yes & $\cdots$ &  & $\cdots$ & $\cdots$ & $\cdots$ & $\cdots$ & $\cdots$\\
NGC6240 & Yes & No & D & AGN & 42.04 & G & $\cdots$ & $\cdots$\\
NGC6286 & Yes & $\cdots$ &  & $\cdots$ & $\cdots$ & $\cdots$ & No & T\\
ESO593-IG008 & Yes & $\cdots$ &  & $\cdots$ & $\cdots$ & $\cdots$ & $\cdots$ & $\cdots$\\
NGC7130 & Yes & No & E & AGN & 40.49 & G & $\cdots$ & $\cdots$\\
ZW453.062 & Yes & No & C & $\cdots$ & $\cdots$ & $\cdots$ & $\cdots$ & $\cdots$\\
NGC7591 & No & No & B & $\cdots$ & $\cdots$ & $\cdots$ & $\cdots$ & $\cdots$\\
IR23365+3604 & No & No & B & AGN & 41.52 & R & No & T\\
IR04259-0440 & No & No & F & $\cdots$ & $\cdots$ & $\cdots$ & $\cdots$ & $\cdots$\\
\hline
\end{tabular}
\end{center}

\tablecomments{Columns Explanation:
Col(1): Common Source Names; 
Col(2): Galaxies with [NeV] Detected;
Col(3): Galaxies with Broad H$\alpha$ detected;
Col(4): Reference for Broad H$\alpha$;
Col(5): Galaxies with AGN-like nuclei in the X-rays;
Col(6): X-ray luminosity in erg s$^{-1}$
Col(7): Reference for X-ray morphology and luminosity. 
Col(8): Galaxies with compact flat spectrum radio cores.  
Col(9): Reference for radio morphology.  
}
\tablerefs{A - H97; B - Veron-Cetty \& Veron 2003; C - Goncalves, Veron-Cetty, \& Veron, 1999; D - Andreasian \& Khachikian 1987; E - Phillips 1983 F - Kewley et al. 2001; G - Gonzalez-Martin et al. 2006; H - Dudik et al. 2005; I - Colbert \& Ptak 2002; J - Terashima et al. 2003; K - Ho et al. 2001; L - Satyapal et al. 2004; M - Terashima et al. 2000; N - Satyapal et al. 2005; O - Georgantopoulos, Georgakakis, \& Koulouridis 2005; P - Philip et al. 2006; Q - Guainazzi, Matt, \& Perola, 2005; R - Teng et al. 2005; S - Nagar, Falcke \& Wilson 2005; T - Baan \& Klockner 2006}

\end{table*}

\subsection{X-ray Observations} 
The standard model for AGNs includes an accretion disk that primarily emits in the optical and UV bands.  The hard X-ray (2-10keV) emission in AGNs is believed to result from inverse Compton scattering of these lower energy optical and UV accretion disk photons.  Because the hard X-rays (2-10keV) are not efficiently produced in normal star-forming regions, a hard nuclear X-ray point source coincident with the nucleus of a galaxy constitutes strong evidence for the presence of an AGN.  

We searched the literature for X-ray observations of all galaxies in our sample and list in Tables 5 and 6 the X-ray classification of the nucleus.  The vast majority of these classifications are based on the presence of a hard X-ray point source coincident with either the radio or 2MASS nucleus and log(L$_X$) $\geq$ 38 erg s$^{-1}$.  The resulting subsample consists of the 39 galaxies observed with {\it Chandra}, {\it XMM}, or {\it ASCA}.   We have excluded those galaxies observed by either {\it ROSAT} or {\it Einstein} since these lower-spatial resolution observations were limited to the soft X-ray band (0.2-4keV) so that the classifications of the X-ray sources are ambiguous and the luminosities could be contaminated or dominated by star forming regions.  However, even with this requirement, the absence of spectral X-ray data precludes the definitive determination of the nuclear source since a compact nuclear starburst with multiple X-ray binaries can mimic these signatures.  

The X-ray and IR detection rates in this case as well, are not consistent.  Sixteen galaxies classified as AGN in the X-rays show {\it no} evidence of a [NeV] line.   In addition, 7 galaxies have NeV detections but do not show evidence for an AGN in the X-rays.  The detection rate of galaxies observed in the hard X-ray band is 56\% (22 of 39).  Using both IR and X-ray AGN detection diagnostics, the combined AGN detection rate is 74\% (29 out of 39).  

We note that the available X-ray information is limited to primarily IR-faint galaxies since many of the luminous infrared galaxies in our sample have not been observed in the 2-10keV band.  The combined detection rate of 74\% may increase if more extensive X-ray observations of IR-bright galaxies are carried out. However we also note that some of the LINERs with AGN-like X-ray nuclei may be misclassified as such.  This is explored further in Section 5.5.

\subsection{Radio Observations} 
 A flat spectrum, compact nuclear radio source or nuclear, extended radio jet(s) are the two primary indicators of the presence of an active black hole in the radio (Nagar et al. 2001, 2002, 2005).    Very high spatial resolution observations are essential for uncovering radio jets in the nucleus of galaxies and such published observations are very rare.   Therefore we searched the literature for observations of either a flat spectrum nuclear radio source or evidence of radio jets in the nucleus of galaxies in our sample.  The majority of these observations are from the Nagar et al. (2005, hereafter N05), who conducted a high-resolution radio imaging survey of the majority of low luminosity Seyferts (LLAGNs) and LINERs in the H97 catalog of galaxies.  However, the majority of galaxies in the N05 survey are IR-faint and a subsequent search for observations of IR-bright LINERs yielded only 5 additional observations.   Based on the limited set of observations, the radio-based AGN detection rate is 46\%.  Tables 5 and 6 show that the IR observation and the radio observation agree in very few cases (42\%, 19 of 45 galaxies).  In fact the radio data does not agree very well with even the optical data (only 62\% of cases (27 of 43) agree).  Though data is limited, the radio observations seem to agree best with X-ray observations (75\%, 24 out of 32 cases).  

We note that X-ray binary sources are also known to display compact flat spectrum radio sources (Fender \& Belloni 2004) and compact starbursts can produce radio emission that mimics AGN spectra (Condon et al. 1991).  This could potentially result in misclassifications of some sources, which may provide some explanation for the inconsistencies seen here.  We also note that radio data for IR-bright LINERs could change the combined and overall detection rates significantly. Because Radio data is missing for IR-bright galaxies, we do not include these detection rates in the following analysis, however we have included the published data in Tables 4 and 5 for completeness.   

\subsection{Combined Multi-wavelength AGN Detection Rate in LINERs} 

The combined detection rate for galaxies with optical, X-ray and IR observations is at least 74\%, one of the highest AGN detection rates reported in the literature thus far.  The radio data has been excluded from this calculation because of insufficient data for IR-bright galaxies.   Inclusion of radio data will only increase this detection rate, as will more optical and X-ray observations of FIR luminous sources.

\subsection{Origin of Multi-Wavelength Inconsistencies}

As highlighted above, there are inconsistencies in the identification of AGN in our sample based on the diagnostic employed which warrants exploration.  In Figure 6, we plot the AGN detection rate for each diagnostic as a function of the host galaxy's far-IR luminosity (L$_{FIR}$).  This figure clearly shows that the optical AGN-detection rate decreases with L$_{FIR}$, while the [NeV] AGN detection rate increases with L$_{FIR}$.  Though data is limited at large L$_{FIR}$, the X-ray-based detection rate is independent of L$_{FIR}$.   Excess dust in LINERs with large L$_{FIR}$ can easily explain the broad H$\alpha$ deficiency in these galaxies.  In fact, near-infrared spectroscopic studies of infrared luminous galaxies show broad Pa$\alpha$ lines are detected in some IR-luminous galaxies, suggesting that the broad-line region is present, but attenuated by dust (Veilleux, Sander, \& Kim 1999).  

\begin{figure*}[htbp]
\begin{center}
\begin{tabular}{ccc}
\includegraphics[width=0.45\textwidth]{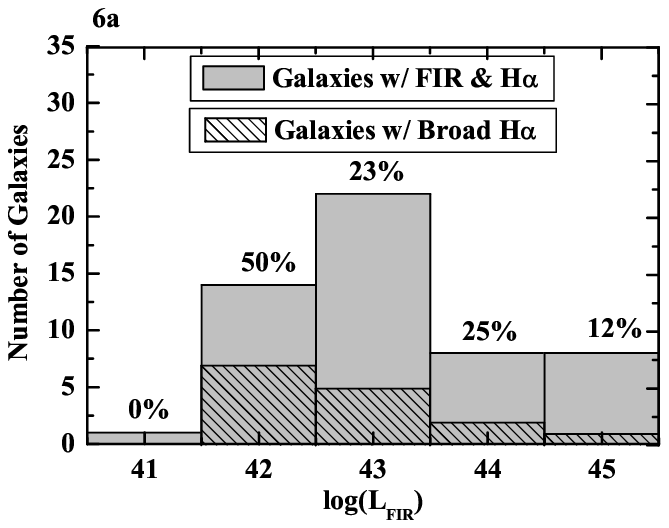} &
  \includegraphics[width=0.45\textwidth]{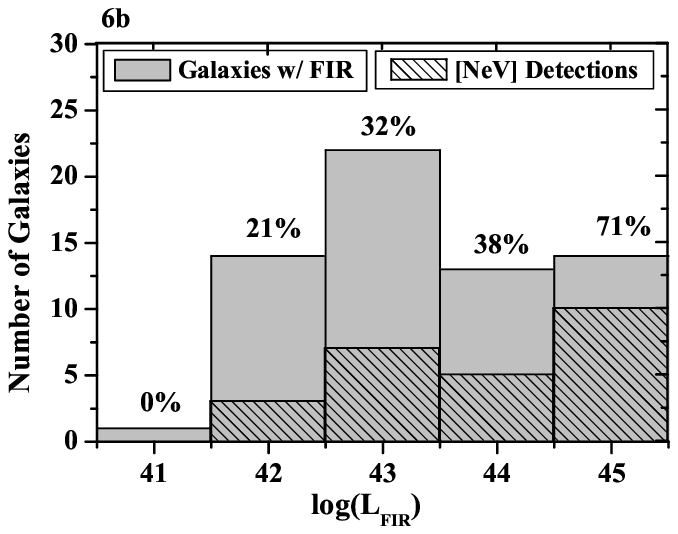} \\
\multicolumn{2}{c}{\includegraphics[width=0.45\textwidth]{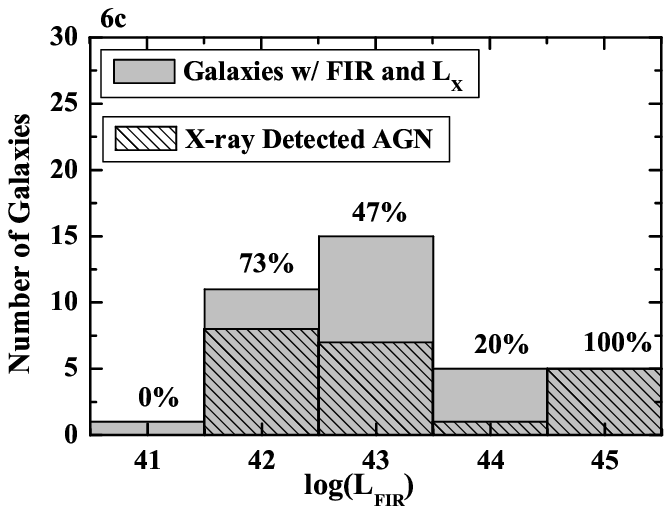}} \\
\end{tabular}
\end{center}
\caption[]{ Optical, IR, and X-ray detection rates as a function of far-IR luminosity.  As can be see from these plots, the optical AGN-detection rate decreases with L$_{FIR}$, while the [NeV] AGN detection rate increases with L$_{FIR}$.  On the other hand, the X-ray-based detection rate seems unrelated  to L$_{FIR}$. }
\end{figure*}

\begin{figure*}[htbp]
\begin{center}
\begin{tabular}{cc}
\multicolumn{2}{c}{\includegraphics[width=0.45\textwidth]{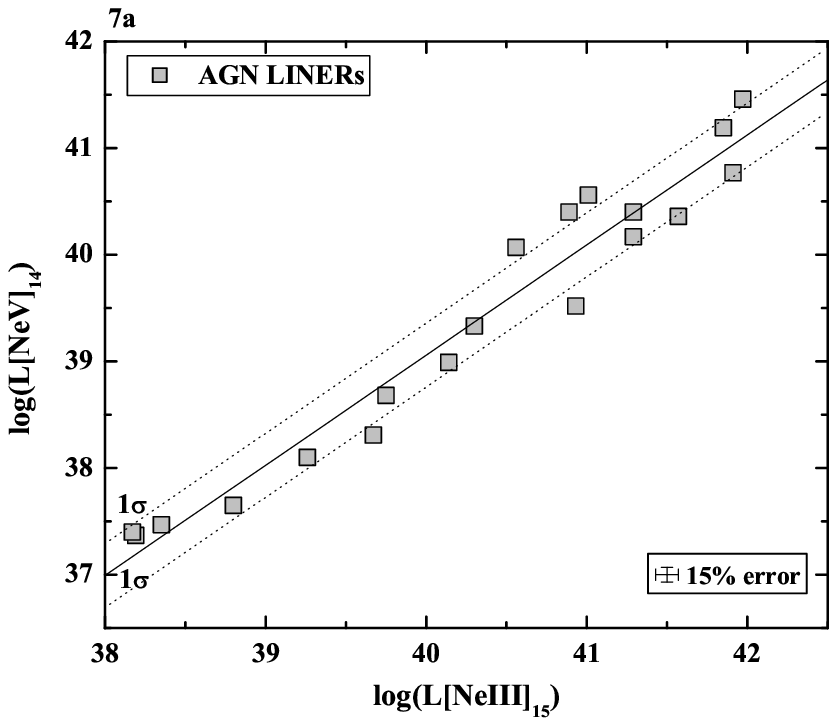}} \\
\includegraphics[width=0.45\textwidth]{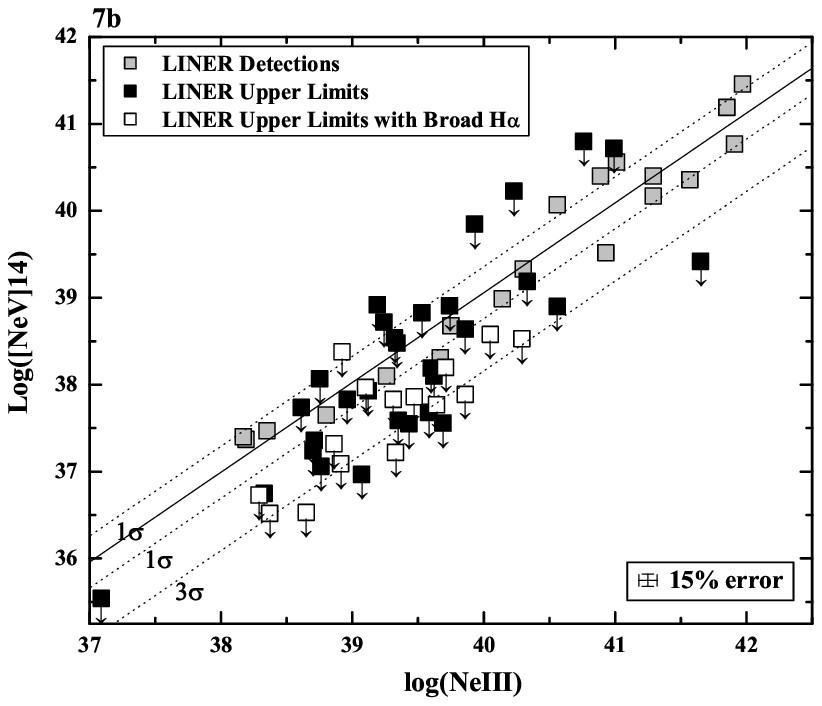} &
\includegraphics[width=0.45\textwidth]{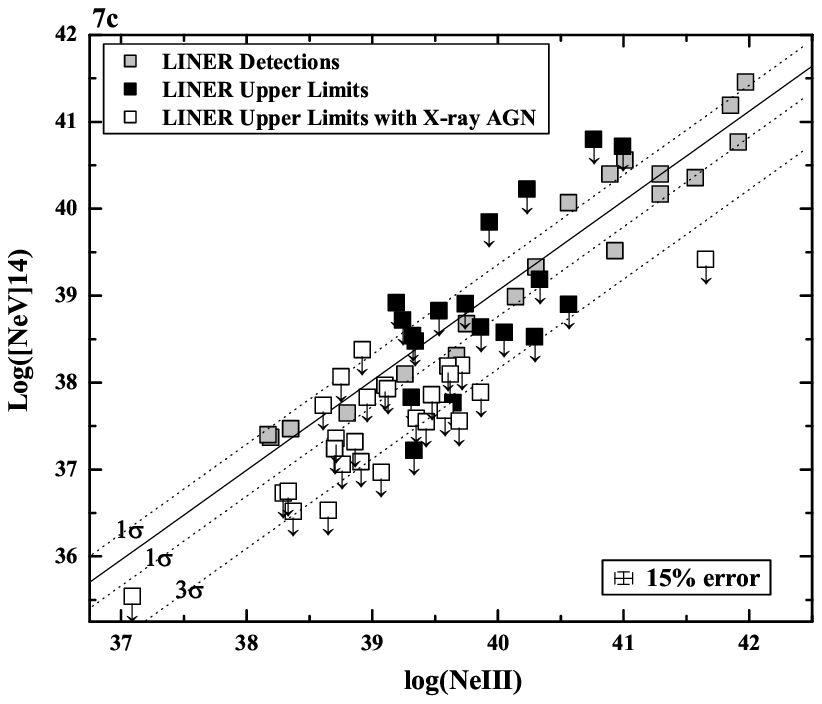} \\
 \end{tabular}
\end{center}
\caption[]{7a. log(L[NeV]$_{14}$) in erg s$^{-1}$ vs. log(L[NeIII]$_{15}$)in erg s$^{-1}$ for all of the AGN-LINERs for our sample, 7b and c.  log(L[NeV]$_{14}$) vs. log(L[NeIII]$_{15}$) for the detections and upper limits for our sample.  Open symbols in 7b \& c represent LINERs with AGN detections in the optical and X-rays respectively.  The error bars due to the 15\% calibration error on the LINER luminosities are roughly the same as the symbol size and could not be over-plotted. A representative error bar is plotted at the bottom right corner of each plot}
\end{figure*}

Some of the galaxies in our sample are reported to have broad H$\alpha$ in the literature but no [NeV] detections in their mid-IR spectra.  One explanation for this discrepancy is limited sensitivity in the {\it Spitzer} observations.  To test whether limited sensitivity was responsible for the missing [NeV], we estimated the expected [NeV] luminosity using the [NeIII] luminosity.  

The [NeV] 14$\mu$m luminosity is strongly correlated with the [NeIII] 15$\mu$m luminosity for all of the detections in our sample (see Figure 8). The best-fit linear relation for Figure 7 is:

\begin{equation}
\log(L_{\rm [NeV]14}) =  (1.03)\log(L_{\rm [NeIII]15}) - 2.26
\end{equation}

The dotted lines signify the 1 and 3 sigma deviations of fit.  The standard deviation is much larger than the calibration error of 15\% for all of the fluxes in the plot.  

In Figures 7b and 7c we add the 3$\sigma$ [NeV] upper limits to the [NeIII]-[NeV] correlation.  The open symbols for the upper limits represent those LINERs detected as AGN in either the optical (7b) or X-rays (7c).  As can be seen from Figures 7b and 7c, the vast majority of galaxies without [NeV] detections have upper limits that fall within the 3$\sigma$ deviation of the correlation.  This implies that the absence of [NeV] emission can be attributed to the limited sensitivity of the observations—one possible explanation for the multi-wavelength inconsistencies. Indeed, the sensitivity of the mid-IR observations (as noted above) as well as the sensitivity of X-ray observations could explain the discrepancy, since for the most part, these observations were part of large surveys searching for AGN rather than characterizing them.  

There are a few (3-5) upper limits that may be outliers to the correlation.  These upper limits have either broad H$\alpha$ or an AGN-like X-ray nucleus, however they fall below the three-sigma deviation of the correlation.  This may imply a [NeV] deficit in some LINERs, similar to that described by Ho (1999).   However, these upper-limits are so close to the three-sigma limit to the correlation, that deeper observations are needed to confirm this potential deficit and establish the nature of this subset of LINERs in our sample.  

\begin{figure*}[htbp]
\begin{center}
 \includegraphics[width=0.60\textwidth]{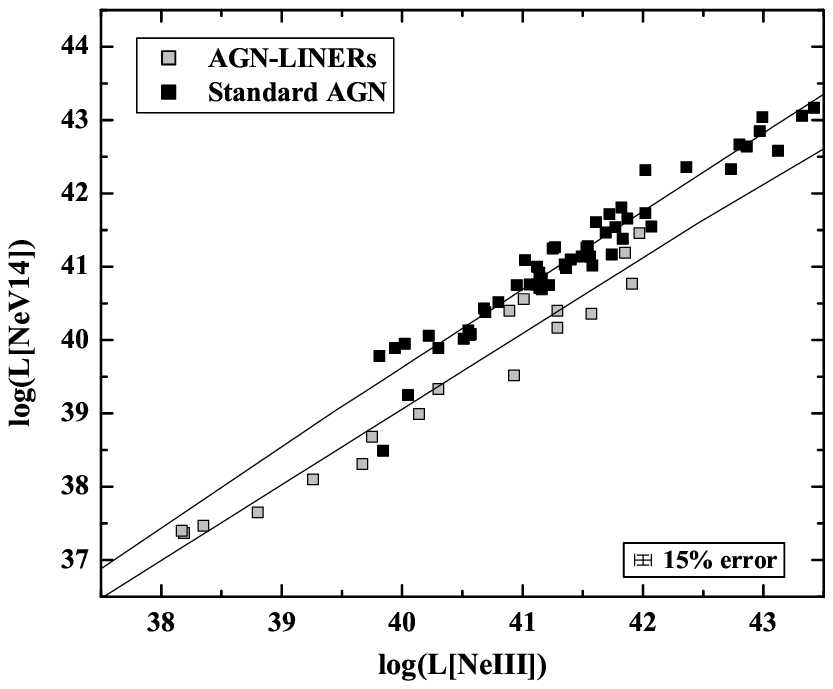} 
\end{center}
\caption[]{log(L[NeV]$_{14}$) in erg s$^{-1}$ vs. log(L[NeIII]$_{15}$)in erg s$^{-1}$ for all of the AGN-LINERs for our sample as a well as a sample of standard AGN from Gorjian et al. 2006.  We note that the slopes for both correlations are very similar, implying that the relation holds for both objects.   However the y-intercept for standard AGN is larger than that for LINERs, suggesting that either [NeV] 14$\mu$m luminosity is deficient or the [NeIII] is in excess in LINERs. The error bars due to the 15\% calibration error on the LINER luminosities are roughly the same as the symbol size and could not be over-plotted. A representative error bar is plotted at the bottom right corner of the plot}
\end{figure*}

We note that another possible cause for discrepancies between the multi-wavelength AGN identifications is misclassification of LINERs as AGN using optical and X-ray diagnostics.  Indeed, many of the reported broad H$\alpha$ lines taken from the literature have 1000 km/s $<$ FWHM $<$ 1200 km/s, very close to the line widths associated with the narrow-line region and which can also arise in ionizing shocks generated in starburst-driven winds.  In addition, as noted in Section 5.2 above, detailed X-ray spectral observations are needed to rule out X-ray binaries in a compact starburst as the source of the X-ray emission.  Very little spectral data is available for the galaxies in our sample, so it is possible that a subset of our X-ray classified AGN are in fact very compact star forming regions.

\section{How do AGN-LINERs compare with Standard AGN in the Mid-IR?. }

In a study of the nuclear spectral energy distribution (SED) of  7 low luminosity AGN-LINERs, Ho (1999) found that at least some LINERs lack the big blue bump--the optical/UV emission normally associated with emission from an optically thick, geometrically thin accretion disk (Lasota et al. 1996).  However since this discovery there has been some debate in the literature over whether this optical/UV deficit is present in some or all LINERs (e.g. Eracleous et al. 2008 and Ho et al. 2008 vs. Maoz et al. 2007) and, over whether this deficit is intrinsic in all objects or if it is due to extinction toward the optical/UV continuum (big blue bump).  In support of the former, Eracleous et al. (2008) find that this deficit is present in an IR-faint sample of Type 1 objects where obscuration {\it cannot by definition} be responsible for the missing big blue bump.  On the other hand, Satyapal et al. (2004) use ratios mid-IR spectral lines to compare the slope of the optical/UV continuum in IR-bright LINERs and standard AGN.  They find that the majority of AGN-LINERs in the sample have similar mid-IR line ratios to standard AGN, thereby suggesting that the optical UV spectrum is also very similar for these two classes.

The high ionization mid-IR line ratios can potentially probe the shape of the intrinsic ionizing SED illuminating the AGN narrow line region.  We emphasize however that such line ratios are dependent on a number of additional parameters including the ionization parameter and, for lower ionization lines, the contribution to the line emission from gas ionized by young stars.  Nonetheless, some insight can potentially be gained by comparing mid-IR line ratios between AGN-LINERs and standard AGN.  In Figure 8 we replot the [NeIII]15$\mu$m luminosity as a function of the [NeV]14$\mu$m luminosity for LINERs and the sample of predominantly standard AGNs from Gorjian et al. 2006 (hereafter G06).  G06 find a very tight correlation between L[NeIII]$_{15}$ and L[NeV]$_{14}$ in these objects.  Figure 8 shows that LINERs have [NeV]14$\mu$m/[NeIII]15$\mu$m ratios that are very different from standard AGN.  Of the two galaxies from G06 that overlap with the LINER slope, one (NGC 3079) is a LINER.  We have chosen [NeIII] 15$\mu$m and [NeV] 14$\mu$m since they are observed using the same {\it Spitzer} IRS module to avoid aperture effects that may result from taking ratios of lines from different sized apertures.  If distance-based aperture effects were inducing these results, we would expect the opposite effect: with the G06 sample (4-3000Mpc) showing smaller ratios than the LINERs (2.4-300Mpc). 

The best fit relation for the LINER population is given in Section 6, Equation 1.  The best fit relation for the G06 sample is given by:

\begin{equation}
\log(L_{\rm [NeV]14}) =  (1.06)\log(L_{\rm [NeIII]15}) - 2.99.
\end{equation}

The slopes for both correlations are very similar, implying that the relation holds for both objects.   However the y-intercept for standard AGNs is larger than for LINERs, implying either that the [NeV] 14$\mu$m luminosity is deficient in LINERs or that the [NeIII] is high.  A [NeV] deficiency may be due to an intrinsic optical/UV deficit in LINERS as discussed above (Ho 1999).  In addition, extinction toward the [NeV]14$\mu$m emitting region can also explain the deficiency (Dudik et al. 2007).  Finally, contamination by circum-nuclear star formation can also explain the relationship if  there is a strong contribution to the [NeIII] from star formation in LINERs.   These scenarios are explored further in Sections 6.1 and 6.2 below. 

\subsection{Do AGN-LINERs have a unique SED?} 

One could argue that the anomalous L[NeIII]$_{15}$ to L[NeV]$_{14}$ relationship in LINERs when compared to standard AGN suggests a different intrinsic SED in LINERs; one in which there is a UV deficit.  If this were the case, the [NeV] 24$\mu$m line flux should be similarly effected.  In Figure 9 we plot the [NeV] 24$\mu$m luminosity as a function of the [OIV] 26$\mu$m luminosity for both LINERs and standard AGN (AGNs taken from Dudik et al. 2007, Haas et al. 2005, and Sturm et al. 2002 which overlap with the G06 sample as well as AGN-ULIRGs from Farrah et al. 2007, hereafter DHSF sample).  This is meant to be analogous to the [NeV]14/[NeIII]15$\mu$m ratio, since [NeIII] and [OIV] are closest in ionization potential of all the detectable lines (41 and 55 eV respectively).  Again, the [NeV]24$\mu$m line and the [OIV]26$\mu$m line come from the same aperture, to avoid aperture effects.

\begin{figure*}[htbp]
\begin{center}
 \includegraphics[width=0.60\textwidth]{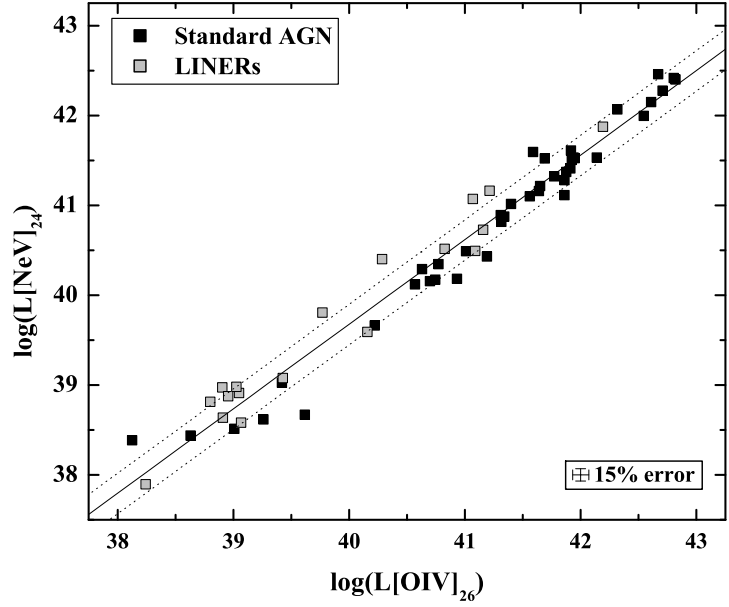} 
 \end{center}
\caption[]{The [NeV] 24$\mu$m luminosity to [OIV] 26$\mu$m luminosity for all of the detections in our sample as well as a sample of standard AGNs. We find that LINERs look very similar to standard AGN, suggesting that SED effects are not the cause of the discrepancy in Figure 8. The error bars due to the 15\% calibration error on the LINER luminosities are roughly the same as the symbol size and could not be over-plotted. A representative error bar is plotted at the bottom right corner of each plot}
\end{figure*}

As can be seen, both LINERs and standard AGN have very similar [NeV] 24$\mu$m/[OIV] 26$\mu$m line ratios.   Most of the LINER ratios fall well within the 1$\sigma$ deviation of the standard AGN correlation, strongly suggesting that the slope of the SEDs in these AGN-LINERs are very similar to those in standard AGN.  If LINERs with [NeV] do not have unique SEDs, what is responsible for the deficient [NeV]14$\mu$m emission in Figure 8?

\subsection{Evidence for Mid-IR extinction?}

Traditionally, ratios of infrared fine-structure transitions from like ions in the same ionization stage, but with different critical densities, have been used to provide abundance-independent density estimates for the ionized gas from which they arise.   However, a recent study of the [NeV]14$\mu$m/[NeV]24$\mu$m ratio in a sample of standard AGNs found that many AGN have [NeV] ratios well below the theoretical low density limit (Dudik et al. 2007).  Ruling out aperture effects and theoretical uncertainties, they argue that differential infrared extinction from dust in an obscuring torus may be responsible for deficient [NeV]14$\mu$m fluxes and therefore the anomalously low ratios.  If the LINERs in this sample suffer severe extinction toward the optical/UV band, one might expect the mid-IR line fluxes to be similarly effected.

\begin{figure*}[htbp]
\begin{center}
 \includegraphics[width=0.60\textwidth]{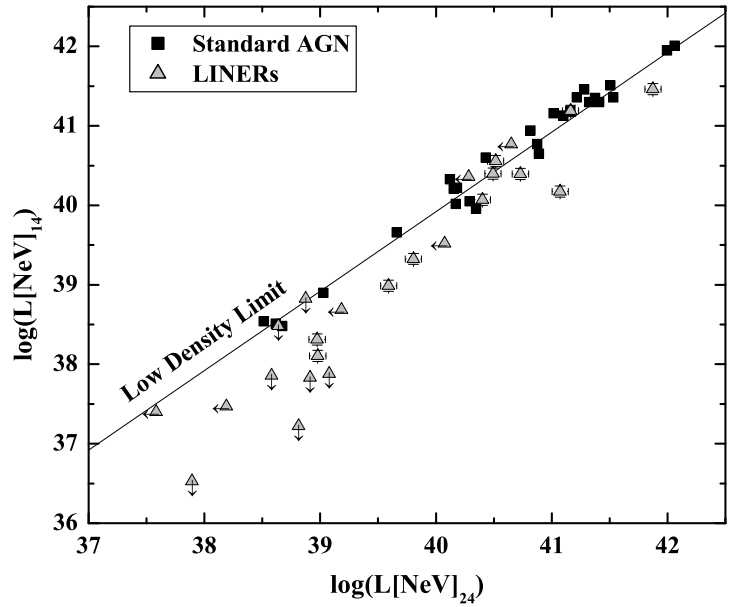} 
\end{center}
\caption[]{ The [NeV] 14$\mu$m luminsity to [NeV] 24$\mu$m luminosity for all of the detections in our sample as well as a sample of standard AGNs.  The majority of LINERs in our sample (15/26) have ratios well below the low density limit.  This may imply severe extinction toward the [NeV] 14$\mu$m emitting region in these galaxies. The error bars in this plot represent the calibration error of 15\%}
\end{figure*}

The low density limit for the [NeV]14$\mu$m/[NeV]24$\mu$m ratio is 0.83.  Table 7 lists the [NeV] ratio for all of the LINERs in our sample.  The majority of LINERs in the sample with [NeV] detections have ratios far below the theoretical limit.  Figure 10 shows the [NeV] 14$\mu$m luminosity vs. [NeV] 24$\mu$m luminosity for LINERs and a sample of standard AGNs.  Of the 26 LINERs with [NeV] detections, 15 have ratios that are below the low density limit to within the calibration error, while only 4 LINERs have ratios significantly above.   From Figure 10, it is clear that LINERs show [NeV] line flux ratios well below the low density limit, and significantly lower than those from standard AGNs from Dudik et al. 2007.   The average [NeV]14$\mu$m/[NeV]24$\mu$m ratio in LINERs where both lines are detected is 0.49.  The same ratio for our sample of standard AGNs is 0.98.  These results require that the extinction curve in LINER galaxies must be characterized by higher extinction at 14$\mu$m than at 24$\mu$m.

\begin{table}
\fontsize{9pt}{10pt}\selectfont
\begin{center}
\caption{NeV Line Flux Ratios}
\begin{tabular}{lccc}
\hline
\multicolumn{1}{c}{Galaxy} & \multicolumn{1}{c}{\underline [NeV]$_{14}$} & \multicolumn{1}{c}{Galaxy} & \multicolumn{1}{c}{\underline [NeV]$_{14}$}\\

\multicolumn{1}{c}{Name} & \multicolumn{1}{c}{[NeV]$_{24}$} & \multicolumn{1}{c}{Name} & \multicolumn{1}{c}{[NeV]$_{24}$}\\

\multicolumn{1}{c}{(1)} & \multicolumn{1}{c}{(2)} & \multicolumn{1}{c}{(3)} & \multicolumn{1}{c}{(4)} \\ 

\hline
NGC1052 &0.22 & UGC5101 & 1.06\\
NGC3166 &$<$0.70 & NGC4666 & $<$0.06\\
NGC3368 &$>$0.19 & UGC8387 & 0.13\\
NGC3507 &$>$0.69 & NGC5104 & 0.47\\
NGC3521 &$>$0.23 & Mrk273 & 0.39\\
NGC3642 &$<$0.08 & NGC5953 & 0.25\\
NGC4036 &$<$0.03 & IR15335$\cdots$ & $>$0.28\\
NGC4278 &$<$0.04 & IR16164$\cdots$ & 0.46\\
NGC4438 & $<$0.19 & NGC6240 & $>$1.31\\
NGC4736 & $>$0.66 & NGC6286 & 0.33\\
NGC5353 & $<$0.89 & ESO593$\cdots$ & $>$1.19\\
NGC5371 & 0.13 & NGC7130 & 1.10\\
NGC660 & $>$0.31 & ZW453.062 & 0.80\\
\hline
\end{tabular}
\end{center}
\tablecomments{Columns Explanation:
Col(1) and Col(3): Common Source Names; 
Col(2) and Col(4): [NeV]14$\mu$ / [NeV]24$\mu$ line flux ratio }

\end{table}

Figure 10 suggests that SED effects are not responsible for the low values of [NeV]14$\mu$m luminosity  seen in the [NeV]-emitting LINERs in Figures 8 and 10.  Rather they suggest that the 14 $\mu$m [NeV] emission is deficient, and that this deficiency is due to substantial extinction toward the [NeV] emitting region rather than an intrinsic optical/UV deficit.  In support of these findings, Desai et al. (2007) find that nearly all Ultraluminous Infrared Galaxies (ULIRGs) with LINER-like optical nuclei in their sample show silicate optical depth, $\tau$$_{9.7}$ in excess of 2.5, implying optical depths, A$_V$ $\geq$ 45.   At least for the luminous infrared galaxies plotted here, the [NeV] 14$\mu$m deficiency can easily be explained by such high columns of dust (see also Dudik et al. 2007).  We note that the data in Figures 8-10 is sparse, because there is currently very little [NeV] high-resolution data in the literature either for standard AGN or for LINERs.  [NeV] 24$\mu$m data and [OIV] data is particularly scarce.

\section{Conclusions}
 
Using the high resolution IRS modules on board Spitzer, we conducted an archival mid-IR spectroscopic study of 67 LINERs in order to search for mid-IR signatures of AGNs.  This is the most exhaustive mid-IR spectroscopic survey of LINERs to date and includes observations of the mid-IR [NeV] emission lines, the only robust spectroscopic diagnostic of an AGN in the mid-IR.  These results are compared with similar observations of standard Seyferts and quasars.  Our main results are as follows. 
\begin{enumerate}

\item  We detect the [NeV] 14 and/or 24 $\mu$m line in 26 of the 67 galaxies in our sample.  This yields a mid-IR AGN detection rate of 39\%.  

\item We find that many galaxies (23\%) show prominent [NeV] emission but no evidence for AGN activity in the optical.  This result emphasizes that optical observations fail at finding a significant fraction of AGN in LINERs.  This is particularly true in FIR luminous galaxies, the dominant and least studied population of LINERs.

\item The fraction of LINERs  with [NeV] emission increases with FIR luminosity, implying that large fraction of the dominant population of LINERs are AGN.

\item We find some  inconsistencies between [NeV] non-detections and either the presence of Broad H$\alpha$ or the presence of a hard X-ray point source.  This inconsistency is likely due to the limited sensitivity of the mid-IR observations and/or the misclassification of optical and X-ray sources. 

\item Using the combined X-ray, optical, and IR diagnostics, we find an AGN-detection rate in LINERs of 74\%, higher than previously reported.   

\item Contrary to previous studies, a comparison of the mid-IR line ratios in LINERs and standard AGN suggests that the ionizing SEDs in at least some LINERs are not significantly different from those of standard AGN.  

\item We find that the [NeV]14$\mu$m/ [NeV]24$\mu$m, line flux ratios in LINERs are significantly below the theoretical low density limit and substantially lower than the same ratios in standard AGN.  This result suggests that there is high extinction, even at mid-IR wavelengths,  toward the [NeV]14$\mu$m line emitting region in a significant number of LINERs.  

\end{enumerate}

\acknowledgements
We are very thankful to Devin Vega and Brain OÕHalloran for their invaluable help in the data analysis. We are also very grateful to Joe Weingartner, Luis Ho, and Jenny Greene for their enlightening and thoughtful comments.  Thanks also to Kartik Sheth, Daniel Dale, Eckert Sturm, and Henrik Spoon, who have clarified a number calibration questions relevant to this work. Finally we also are very grateful for the helpful comments from the referee, which significantly improved this paper.  This research has made use of the NASA/IPAC Extragalactic Database (NED) which is operated by the Jet Propulsion Laboratory, California Institute of Technology, under contract with the National Aeronautics and Space Administration.  SS gratefully acknowledges financial support from NASA grant NAG5-11432 and NAG03-4134X.  RPD gratefully acknowledges financial support from the NASA Graduate Student Research Program.

\end{document}